\documentclass[11pt]{article} 

\usepackage[utf8]{inputenc} 

\usepackage{geometry} 
\usepackage{graphicx} 
\usepackage[usenames,dvipsnames]{xcolor}
\usepackage{caption}
\usepackage{booktabs} 
\usepackage{array} 
\usepackage{paralist} 
\usepackage{verbatim} 
\usepackage{subfigure} 
\usepackage{url}
\usepackage{fancyhdr} 
\usepackage{sectsty}
\allsectionsfont{\sffamily\mdseries\upshape} 
\usepackage{color}

\usepackage[nottoc,notlof,notlot]{tocbibind} 
\usepackage[titles,subfigure]{tocloft} 

\usepackage{amsmath} 
\usepackage{amsthm} 
\usepackage{amssymb}

\title{ 
Minsky financial instability, interscale feedback, percolation and Marshall-Walras disequilibrium}

\author{Sorin Solomon, Nataša Golo, Hebrew University Jerusalem}
\graphicspath{{C:\Users\NATASA\Dropbox\AEL_paper\ }{D:/Minsky_paper/TeX/}{C:\Users\SORIN\Dropbox\AEL_paper\ }}

\begin{document}
\bibliographystyle{plain}
\maketitle

\abstract{ We study analytically and numerically Minsky instability as a combination of top-down, bottom-up and peer-to-peer positive feedback loops.
The peer-to-peer interactions are represented by the links of a network formed by the connections between firms; contagion leading to avalanches and percolation phase transitions propagating across these links.
The global parameter in the top-bottom -- bottom-up feedback loop is the interest rate.
Before the Minsky Moment, in the `Minsky loans accelerator' stage the relevant ``bottom" parameter representing the individual firms' micro-states, is the quantity of  loans.
After the Minsky Moment, in the `Minsky crisis accelerator' stage, the relevant `bottom' parameters are the number of ponzi units / quantity of failures / defaults.
We represent the top-bottom, bottom-up interactions on a plot similar to the Marshall-Walras diagram for quantity-price market equilibrium (where the interest rate is the analog of the price).
The Minsky instability is then simply emerging as a consequence of the fixed point (the intersection of the supply and demand curves) being unstable (repulsive).
In the presence of network effects, one obtains more than one fixed point and a few dynamic regimes (phases).
We describe them and their implications for understanding, predicting and steering economic instability.}

\section{Requisiteness of the agent based and network approach }
\label{sec:Introduction}
\subsection{Background on Minsky theory of instability}
\label{subsec:Minsky background}
Minsky's theory of instability has as a central element the procyclical self-reinforcing feedback loop between the individual behavior of economic players and the state of the economic system as a whole. This explains the fact that in spite of his brilliance and in spite of the personal admiration by many of his peers, for a long time, Minsky's work could not be absorbed within the mainstream economic theory \cite{Wray 2011a}. Indeed there are a series of crucial features in the Minsky instability theory which are incompatible with the neo-classical framework:
\begin{enumerate}
\item the relationship between the individual and the macroscopic behavior cannot be expressed in a framework where both the system and the individuals are fused into a representative agent by simply ignoring aggregation concerns.  
\item the positive feedback loop between the system as such and the agents composing it, leads to a dynamics which instead of approaching equilibrium, as assumed by the neo-classical approach, runs away from it. Departures from the equilibrium, instead of being corrected by the forces within the economic system, are, in Minsky's theory, exacerbated until a Minsky moment takes place and from then on, new divergence from equilibrium takes place in the opposite direction. Thus, as observed by Minsky \cite{Minsky 1975}, in a capitalist economy, stability is destabilizing and thus impossible.
\item Minsky divided the companies into 3 classes:
\begin{itemize}
\item[-] Hedge - companies that are entitled to enough cash flow to completely pay their debts including their interest,
\item[-] Speculative - companies that are only entitled to enough cash flow to serve their interest payments,
\item[-] Ponzi - companies that do are not entitled to enough cash flow to pay interest on their loans.
\end{itemize}
This classification (see below for their formal definitions in Eqs. \ref{eq1}-\ref{eq4} and the associated text) requires an explicit representation of the companies and their dynamics in terms of a multi-agent system in which the processes take place by long chains of action and reaction between the various players. In fact the dynamic motion of companies forth and back between Hedge, Speculative, Ponzi and failed positions is at the very heart of the Minsky instability.
\end{enumerate}

Thus Minsky had to construct his thinking on the basis of previous work that offered alternatives to the homogeneous, equilibrium, representative agent \cite{Kirman 1992}: 
\begin{itemize}
\item[a.] Keynes' expectation dynamics under uncertainty .
\item[b.] Hayek's emphasis of the role of individuals in driving macroeconomic behavior.
\item[c.] Schumpeter's theory of innovation followed by creative destruction \cite{Schumpeter 1936}\cite{Biondi 2008}.
\end{itemize}

Minsky exposed his early views in a book entitled John Maynard Keynes \cite{Minsky 1975}. Rather than attempting therein a scientific biography of Keynes, he used the Keynesian conceptual framework to emphasize the non-equilibrium, dynamical character of the economy and in particular the crucial role of uncertainty and individual subjective expectations in inducing fast changes in prices, values, production and employment.
Following Keynes \cite{Keynes 1937}, Minsky endeavored to avoid in this way the difficulties that Hayek feared in deducing the dynamics of the economic system out of the behavior of individual agents. Hayek believed that in the absence of detailed information at the single agent and single transaction level, it would be impossible to deduce and even less to control, the emergence of market prices and dynamics.  One critical issue in this debate has been coordination: Hayek renewed the classical wisdom that coordination of individual plans is impossible for a central planner, leaving coordination only to market(s), while Keynes and others argue against this received wisdom.

In Hayek, the only collective dimension is performed through the market, while Keynes introduced non-market coordination by governmental action among others. Keynes' choice was to represent the macroscopic dynamics independently of an explicit microeconomic foundation. This was as a natural choice as the physicists' choice (e.g. \cite{Carnot 1890}) to use macroscopic thermodynamic concepts at a time when their atomic-molecular foundations were not yet available, neither empirically nor methodologically. Obviously, this comes at a price: the non-equilibrium aspects and the detailed description of the space and time transitions between various regimes are outside the reach of such a macro theory. Yet, as in the case of Carnot who devised thermal engines based only on macroscopic facts (e.g. that increasing temperature increases pressure) \cite{Keynes 1937}  was able to devise monetary and fiscal policies by which the economy regulators could steer the economic system out of depression and into full employment. He also described how these interventions can cause the growth or the transformation of the entire system. Thus, for crude macro-policy interventions in acute crisis conditions, Keynes' attempts to circumvent the issue of micro-economic foundations was justified. Hayek's despair of achieving a meaningful aggregation with the experimental and theoretical means of his time was justified as well.

Unfortunately this left the economic field largely at the mercy of naive aggregation in terms of representative agents which merely substituted the emergent complex collective dynamics by a copy of the mechanical behavior of a microeconomic ideal agent (additionally defaced by the assumptions of rationality, perfect knowledge or even perfect foreknowledge (cf. rational expectations theory \cite{Lucas 1972})).

Minsky was left in the distinguished, but very limited, company of a minority that recognized bounded rationality \cite{Simon 1997}, non-equilibrium \cite{Schumpeter 1928} and heterogeneity \cite{Kirman 1992}. Most of the field continued to believe in Say's wealth conservation, Walras equilibrium pricing, money neutrality and to ignore the role of debt financed investment.  As recognized already by Minsky's teacher  \cite{Schumpeter 1936}, and by an increasing tide in the current literature (see e.g. \cite{Biondi 2008}) the debt financed investment is a crucial factor in the capitalist economy. Unfortunately the mainstream economy had no place for Minsky's disequilibrium ideas and no way to internalize this concepts: in the neoclassical frame of mind loans are just transfering the funds from individuals with less need for cash to individuals with more need for cash and are irrelevant at the macroeconomic level.  One of the ambitions of the present paper is to express those Schumpeter-Minsky instability ideas in a format familiar from the study of equilibrium economic systems. 

\subsection{Role of solvable agent based models in the study of unstable economic systems}
\label{subsec:ABM}
The novel capabilities to gather data at the resolution of individual agents and individual transactions as well as the statistical mechanics, field theory and complexity tools to extract non-trivial collective dynamics out of large sets of individual interactive agents, allow students at the present time to go beyond the positions that deny, in principle, the possibility for the regulators to steer the macroeconomic behavior. The original position expressed by Hayek, that there is no hope of reliably deducing from the myriads of microscopic acts a meaningful macroscopic collective dynamics, can be now re-evaluated. In the present paper we propose a way to re-address, from a positive perspective, the emergence of the macroeconomic collective phenomena from microeconomic individual interactions.
We argue that the feature that can guide and give coherence and (to a degree) reproducibility to the connection between the micro and macro scales is autocatalyticity:  the non-trivial structure of feedback loops that amplify the microeconomic individual events to systemic changes and reshape in turn individual behavior, according to the macroeconomic state. 

In the last decade, the research bridging between the microeconomic basis and the emergent macroeconomic phenomena led to significant breakthroughs beyond the naive assumption that a collective of microscopic agents would behave as a single representative agent with properties similar to the typical individual. This has been discussed in the book ``Microscopic Simulation of Financial Markets. From Investor Behavior to Market Phenomena" \cite{LLS 2000}. However, initially the impression was created that the agent based models are tractable only by numerical methods, as can be seen from the comment of Harry Markowitz \cite{Mark 2000}: 
\begin{quotation}
 ``{\it Microscopic Simulation of Financial Markets} points us towards the future of financial economics. If we restrict ourselves to models which can be solved analytically, we will be modelling for our mutual entertainment, not to maximize explanatory or predictive power."  \end{quotation}
In fact the name “agent based modeling” became identified  in the literature (mistakenly as we argue below) with computer simulations. The implicit hope in such a position was that using agent based models, which have all the necessary details at the microscopic level, the mechanisms and processes that govern the emergence of macroeconomic phenomena will reveal themselves in a direct and self-evident way. It might have been the case even that the mysterious invisible hand (as well as its trembling in market frictions and failures) would emerge from the multitude of individual actions and agents coordinated by market(s) and become visible in their simulations' visualization and / or in their theoretical analysis. In particular, the access by modern electronic communication and information processing to the individual acts composing the global economic activity might have circumvented the difficulties that Hayek perceived at his time as forbidding the possibility to access and analyze the immense number of economic acts that compose the systemic economic dynamics.
Those hopes were based on the success of similar scenarios played out during the last century in various other fields and notably in Physics. In the past it had been customary in many disciplines to formalize a collective of many similar objects in terms of a “mean field” or a “representative agent” characterized by the average of their individual properties and behavior. In reality such collectives may possess completely new properties and behavior than their components. In turn, they often constitute the elementary objects of a higher level of organization. The representative agent, mean field, continuum, linear way of thinking missed the higher level/order effects that are responsible for the emergence of life from chemistry, conscience from life, society from conscious individuals, etc \cite{Holland 1992}, \cite{Lovelock 1979}, \cite{Gell-Mann 1994}, \cite{Prigogine 1997}. Understanding this connection between the elementary objects of one science and the collective phenomena overarching them allowed scientists to achieve many syntheses and insights and to overcome the obstacles that kept the classical sciences as isolated sub-cultures.
Thus the possibility to go both empirically and theoretically to the individual economic agent level seemed a great opportunity to repeat in the economics field the great successes that physics achieved to understand the emergence of macroscopic phenomena and even to describe the conditions in which the macroscopic systems switch from one regime to another (e.g. phase transitions between solid and liquid or liquid and gas). A corresponding success in economics would be the understanding of the abrupt transition from the euphoria state to the panic state with a Minsky moment as the point of phase transition between the two.Richard Roll, former president of the American Finance Association and one of the leading research authorities in finance, gave a very optimistic evaluation of these prospects emerging from \cite{Mark 2000}:
 \begin{quotation}
``This book contains the first fully comprehensive treatment of microscopic simulation in Finance. The authors make a compelling case that this technique originally used in physics to solve otherwise intractable problems, is destined to become a standard tool in finance."
 \end{quotation}
However it is not guaranteed that a method which worked well in the physical sciences applies with no change to the social, biological and human sciences. According to the Popperian paradigm, a scientific method has, on the one hand, to be able to make falsifiable statements: make nontrivial predictions that can be confronted with the empirical data. And on the other hand, according to Popper, ``science is the art of systematic simplification": scientific understanding means reducing the explanations of a phenomenon to a limited number of premises (Occam's razor). The question is, whether this kind of oversimplification does not compromise the possibility to both understand phenomena of interest, and make reliable predictions.
In the same way in which economists, witnessing a real life situation, might interpret it along very divergent narratives, the mere representation of a system in all its `micro' details in the computer will not reveal automatically its salient features. In fact, a model with too many parameters to tune can predict anything -- and thus nothing. If one has myriads of microscopic features it is very difficult to know which one is responsible for the macroscopic effects. By representing exactly in the computer a system, which one doesn't understand, one ends up with two systems that one doesn't understand: the original one and the computer model of it.
  
So, in order to validate complex, non-equilibrium theories such as Minsky's, one has to look beyond computer simulations of agent based models. Minsky's ideas suggest a generic way out of this dilemma:
one should keep in the models only those individual features that are directly involved in the amplification of micro to macro. In the sequel, various models will be introduced; these will include only microscopic features involved in the process of amplification of microscopic individual events to macroscopic collective features. In this way, one still obtains the macroeconomic effects and the capability of making macroeconomic predictions applicable to policy recommendations, while not getting cluttered by the microscopic noise and the myriad of parameters related to it. 

Thus the present paper, beyond the effort in describing in a precise way the Minsky's financial instability scenario and its implications, has an additional methodological ambition: to emulate the success in the physics and the mathematics literature of treating multi-agent percolation processes by analytical methods \cite{Stauffer 1985}, \cite{Grimmett 1999}. In fact our main tool will be the crucial relation Eq. \ref{Nfailed of rho} that relates analytically the density of susceptible agents in a network to the size of the contiguous clusters that they form.

However, the relation Eq. \ref{Nfailed of rho} by no means reduces the agent based network model to a continuous one:
as in  \cite{Stauffer 1985}, \cite{Grimmett 1999},  the discrete agent character will continue to show up in the large fluctuations in many aspects affecting the predictions and policy recommendations pertinent to steering the economic system. The discrete character of the agents is amplified in the phase transition parameter ranges to large variability in:  
\begin{itemize}
\item[-] the fractal geometrical distribution of failures within the network of companies;
\item[-] the intermittent fluctuation in the time sequence of failures; 
\item[-] the large non-self averaging variability between different realizations of the same or very similar systems.
\end{itemize}
In the following sections we will describe in detail the application of autocatalytic processes to economic systems. 

\section{Autocatalytic feedback mechanisms applied to economic systems}
\label{sec:AFL}
\subsection{Background on Autocatalytic mechanisms}
\label{sebsec:AFL background}
In modern terminology, Minsky's proposal for understanding the instability and the crises of the complex capitalist economy is to identify and characterize the autocatalytic loops that destabilize it. In particular Minsky identified positive feedback loops that act between the global system level and its components at various scales. These autocatalytic loops are the filters that sort out the individual level events that may trigger a systemic catastrophe from those destined to drown in the noise of local, short lived perturbations. This implies that most of phenomena that make it to the macroscopic/systemic level do present, and are fuelled by, some kind of autocatalytic positive feedback loop. This fact was noticed in the past in many occasions and disciplines but in the absence of specific mechanisms that could explain it, it was often dismissed as incompatible with the ethos and ideology of scientific thinking.  
The ubiquity of such occurrences in so many fields (markets, economies, social organization, life, ecology, conscience, creativity) suggests an equally generic paradigm: in order to understand the emergence of collective behaviors in macroscopic systems, one has to find among the myriads of interactions and rules that govern their microscopic components, the ones that have the capability to generate autocatalytic feedback loops. It is only a behavior associated with such a positive feedback and autocatalytic loop that has the chance to be amplified to the macroscopic scale and to govern the system's global dynamics. 

In a series of models starting from heterogeneous interacting agents it has been found, using analytical, simulation and empirical methods, that indeed such positive feedback loops lead to the emergence of adaptive collective objects that change completely the dynamics of the system \cite{Levy 1994}, \cite{LLS 2000}, \cite{Shnerb 2000}, \cite{Yaari 2009}. Thus, by generalizing the auto-catalysis concept one was able to explain how random microscopic elements may self-organize spontaneously into highly resilient localized collective objects and change dramatically the naively expected behavior of the system as a whole. Examples of such generalized autocatalytic mechanisms are: iterative contagion of neighbors (or business partners), proliferation (of successful entities under appropriate conditions), generation by such entities of the very conditions that produce them (or makes them grow), interactions between various aggregation levels within the system (individual events contributing to the general mood that encourage the further occurrence of such events) \cite{Dover 2009}, \cite{Malcai 1999}, \cite{Yaari 2008}.

One is led to the conclusion that the generic criterion that separates phenomena doomed to remain local and buried in the noise from phenomena destined to take over the system, is their positive feedback potential in its many guises and forms. In order to understand, predict and steer systemic changes one has to discover, identify and characterize the particular feedback loops that sustain and amplify them. These positive feedback mechanisms select just those  sustainable emergent collective structures that possess such self-sustaining properties \cite{Nowak 2010}, \cite{Cantono 2010}.
The autocatalytic processes are responsible for many of the sudden changes that threaten the climate, the environment ecology, the social order, or the economic stability around the world. The emergence of new elements that trigger new autocatalytic loops in the system may highly and rapidly destabilize the current state of the system. The new elements may be attributed to external causes but often they are the unavoidable result of the intrinsic instability of the system \cite{Biondi 2008}. 
\begin{paragraph}{Reflexivity}
As mentioned above, the positive feedback / self-reinforcing loop -- or the autocatalytic feedback loop as we will name it from now on -- are not new concepts; one could write an entire monograph about their origins, history and the forms which they took at different times and places. One aspect, which became pre-eminent due to its relevance in the context of the latest global crises, is reflexivity. Reflexivity is a feedback loop between cause and effect in systems of self-conscious individuals. While the reflexivity concept has been known more recently in the context of evolutionary economics \cite{Nelson 1982} and of rational expectations theory \cite{Akerlof 2009}, \cite{Lodhia 2005} its roots are very old: The principle of reflexivity was perhaps first enunciated by William Thomas \cite{Thomas 1923}, \cite{Thomas 1928}: ``the situations that men define as true become true for them."
A particular aspect of reflexivity is the self-fulfilling prophecy -- a prediction /prophecy that causes itself to become true, due to the positive feedback between belief and behavior.  Karl Popper called the self-fulfilling prophecy the Oedipus effect \cite{Popper 1976}:
\begin{quotation}
``One of the ideas I had discussed in The Poverty of Historicism was the influence of a prediction upon the event predicted. I had called this the `Oedipus effect', because the oracle played a most important role in the sequence of events which led to the fulfillment of its prophecy. \ldots For a time I thought that the existence of the Oedipus effect distinguished the social from the natural sciences. But in biology, too --- even in molecular biology --- expectations often play a role in bringing about what has been expected."
\end{quotation}

George Soros is an active promoter of the relevance of reflexivity to economics. For Soros \cite{Soros 1987}, if traders believe that prices will fall, they will sell -- thus driving down prices, whereas if they believe prices will rise, they will buy -- thereby driving prices up. As exposed in \cite{Soros 2008}, the central idea in his conceptual framework is
\begin{quotation}
 ``that social events have a different structure from natural phenomena. In natural phenomena there is a causal chain that links one set of facts directly with the next. In human affairs the course of events is more complicated. Not only facts are involved but also the participants' views and the interplay between them enters into the causal chain. There is a two-way connection between the facts and opinions prevailing at any moment in time: on the one hand participants seek to understand the situation (which includes both facts and opinions); on the other, they seek to influence the situation (which again includes both facts and opinions). The interplay between the cognitive and manipulative functions intrudes into the causal chain so that the chain does not lead directly from one set of facts to the next but reflects and affects the participants' views. Since those views do not correspond to the facts, they introduce an element of [social] uncertainty into the course of events that is absent from natural phenomena. That element of uncertainty affects both the facts and the participants' views."
\end{quotation}

\cite{Merton 1968} describes the emergence of a typical bank run: One day, a large number of customers come to the bank at once -- the exact reason is never made clear,  it could be a large statistical fluctuation. Customers, seeing so many others at the bank, begin to worry. False rumors spread that something is wrong with the bank and more customers rush to the bank to try to get some of their money out while they still can. The number of customers at the bank increases which in turn fuels the false rumors of the bank's insolvency and upcoming failure, causing more customers to come and try to withdraw their money. The rumor of insolvency caused a sudden demand of withdrawal of too many customers, which could not be answered, causing the bank to become insolvent and declare bankruptcy. The rumoured prediction of a collapse led to its own fulfillment.
\end{paragraph}

\subsection{Minsky's scenario and the role of ponzi units}
\label{subsec:Ponzi}
Following Keynes, Minsky made an important point that the expectations and their dynamics are the strongest economic incentives and a driver for business cycles. Expectations play a central role in the precipitation of the crisis, as well as in the creation of the conditions which characterize financial fragility. According to Minsky's theory of financial instability:
\begin{quotation}
``Stability -- even of an expansion -- is destabilizing in that more adventuresome financing of investment pays off to the leaders, and others follow." \cite{Minsky 1975}
\end{quotation}
In other words, during a prolonged period of prosperity, the positive expectations which initially were resulting in a self-fulfilling prophecy (``success breeds daring"), at some point start (exogenously or endogenously) to be perceived -- again in a self-fulfilling way (rationally justified or not) -- as irrationally exuberant \cite{Shiller 2006}. Expectations start then to diverge from the manipulative capabilities of individuals and therefore result in some initial occurrences of failures and bankruptcies. Such individual occurrences very rapidly lead to the change in expectations from positive to negative, triggering the precipitation of a crisis.  The realization that the exuberance is irrational is not based on the microscopic experience of each of the individuals but rather on a global view of the system which is not available directly to the individuals. This explains the great delay and the unpredictable timing of the switch between the exuberant and panic moods and consequently the dramatic and sudden characteristics of the switch -- the `Minsky moment'.
In the year preceding the latest crisis, ``the adventuresome financing of investment" led to unprecedented levels of leverage. 

Following Minsky, \cite{Biondi 2013} classifies economic entities into Hedge, Speculative or Ponzi finance positions using a cash-based financial analysis. He assumes that net cash flows from operations (cash earnings) fully cover both the interest charges and principal repayments in hedge positions, but only the interest charges in speculative positions, while they do not even fully cover the interest charges in ponzi positions. 
Accordingly, one can say that there was a multitude of companies (especially financial institutions) that, in search for increasing profits, became `speculative units' before the crisis. Their financial solvability depended on the possibility to collateralise assets or refinance positions to cover principal repayments, because their net cash flows from operations did not enable such repayment. More seriously, those financial companies encounter, therefore, the risk of becoming `ponzi companies':  companies which cannot service the interest on their loans from their inflows from operations (cash earnings or earnings thereafter), being dependent on collateralisation, leveraging and market-dependent operations even to satisfy interest charges.
Occasionally we shall call such companies in short {\bf `ponzi'} (we offer our apologies to the reader and to Mr. Ponzi  \cite{Ponzi 1935} for this crude short of hand). 

Formally we define as ponzi a company which cannot pay the interest on its debt from its cash earnings (earnings, hereafter):
\begin{equation}
\label{eq1}
\textit{earnings} - \textit{debt} \times \textit{interest rate} < 0.
\end{equation}
It would be useful in the next sections to use Eq. \ref{eq1} to define the resilience $r(n)$ of a company $n$ as the earnings to debt ratio. 
\begin{equation}
\label{eq2}
\textit{resilience}=r(n)=\frac{\textit{earnings}(n)}{\textit{debt}(n)}.
\end{equation}
With this definition, a company becomes a `ponzi unit' at any moment if the interest rate increases to such an extent that it exceeds the company's resilience:
\begin{equation}
\label{ponzi def}
r(n)<i=\textit{interest rate}.
\end{equation}
Thus the difference between the value of the resilience of a company and the current interest rate (`distance to ponzi status'),
\begin{equation}
\label{eq4}
DP(n)=r(n)-i
\end{equation}
is a crucial indicator of the failure susceptibility of a company to changes in the global mood and in particular in the interest rate.

The role of the Eqs. \ref{ponzi def} and \ref{eq4} is to connect the discrete qualitative conceptualization by  the original Minsky classification with the real life quantitative continuous heterogeneous parameterization of companies in terms of their resilience.  While the companies' resiliences take naturally values in a continuous probability distribution, their comparison with the current interest rate, Eq. \ref{ponzi def}, provides a sharp criterion which separates in a discrete way the class of ponzi companies from the rest.  Of course, this criterion is well defined at each given time. However, as described in detail in the next sections, the dynamics of the interest rate (depending in turn on the individual companies' dynamics through Eq. \ref{default of i}) is continuously moving companies into and out of the ponzi status as part of the `Minsky accelerator' dynamics.

In the present paper we will occasionally distinguish between ponzi companies and failed companies.
We will elaborate on this distinction as we progress in defining increasingly realistic models.
The distinction is meant to reflect the empirical fact that many ponzi companies are not necessarily recognized as such by the other companies nor by the system as such.
A well know example of this was the Madoff scheme where a ponzi unit acted for many years and it was treated as a perfectly healthy company.
Only when external factors affected the company adversely, did its ponzi status become known, which influenced in turn the status of its creditors and the status of the system as such.
Thus, in addition to the Minsky classification Hedge, Speculative and Ponzi, we introduce the `failed', status which is a ponzi which is recognized and treated as such. In the same way as the ponzi status, the failed status may be reversible.
Typically we will consider that a ponzi becomes a failed unit by contagion, when one of its partners has failed.
Thus the differentiation between a simple ponzi and failed one is relevant in the network case when each company (node) has only a limited number of partners (nodes directly connected to it, as described in Sections \ref{sec:Percolation} and
\ref{sec:Minsky on network}).
  
In the case in which there are no network effects, as in   Sections \ref{sec:Marshall Walras} and  \ref{sec:Minsky accelerator}, one assumes that all companies are connected, and
there will be no effective difference between the ponzi and failed statuses: a company will become failed as soon as it becomes ponzi.
Thus in the non-network models of Sections \ref{sec:Marshall Walras} and \ref{sec:Minsky accelerator}, the debt-deflation phase is described in terms of a simple  feedback loop: the increase in the interest rate causes many ponzi companies to fail which leads to an increase in the interest rate which in turn causes more speculative companies to become ponzi and fail, thus closing the feedback loop.

In the network based  `Minsky accelerator percolation model' described in Section \ref{sec:Minsky accelerator},  the ponzi status implies that the company is `susceptible". More precisely, the company will fail iff any of its partners fails'. 
Thus in the percolation model one differentiates between the `susceptible' or ponzi status of a company and the `contaminated'  or failed status, in which the company is recognized as ponzi by the system.

 By `failure' we do not necessarily mean bankruptcy, but rather serious distress and missed payments, i.e. a failure of a company to pay its debt obligations (default) which leads to sanctions by creditors and in particular to credit limitations by banks. The actions in the case of default could be the restructuring of the debt (extension of the due date and haircut for the debt repayment and the modification of the interest rate), or sometimes, the liquidation of all the assets. The mathematical relations connecting the interest rate to the number of ponzi and failed companies and quantifying thereby the Minsky accelerator will be introduced and discussed in Sections \ref{sec:Minsky accelerator} and \ref{sec:Percolation}.

The `ponzi unit' concept is crucial to Minsky's theory of instability:
\begin{quotation}
 ``An increase in the ratio of ponzi finance, so that it is no longer a rare event, is an indicator that the fragility of the financial structure is in danger zone for a debt-deflation."                   
                                                                                                  Minsky (1986)
\end{quotation}
This Minsky vision came to happen during the great recession only too often: companies that increased their debt without increasing their earnings placed themselves in danger of becoming ponzi at the slightest increase in the interest rate.  This has been central to the global financial crisis and its explanation (and suggested regulatory reforms) \cite{Wray 2012}, \cite{Wray 2013}. Companies which incurred large leverage decreased their resilience and had difficulties to withstand the credit crunch. The interaction between the resilience in the macroeconomic context and the resilience of the microeconomic agent will be discussed in the subsequent sections.
In particular in Figure \ref{fig:phaserho} we cast our phase diagram describing the system susceptibility to a Minsky instability in terms of the density (fraction) of ponzi companies within the economy.

In addition to the usual Minsky crisis accelerator introduced in Section \ref{sec:Minsky accelerator}, we will discuss in the Section \ref{subsec:Exuberance}  the Minsky loans accelerator that characterizes the  `exuberance phase'.

In the present paper we will formalize and study in detail the Minsky financial accelerator feedback loop by showing that there exists generically a critical fraction or density of ponzi companies above which the system becomes unstable. Congenial efforts to formalize the effect of over-leveraging on the global financial crisis and the network effects in the dynamics of systemic risk were offered in many recent works \cite{Biondi 2005}, \cite{Stein 2011}, \cite{Adrian 2012}. In particular a model capturing these effects in a non-equilibrium context was presented in \cite{Biondi 2012}. An agent-based model of socio-economic interaction-based phenomena and expectation formation (positive, negative or neutral) has been developed and studied in \cite{Hohnisch 2005}.  In that model, the swings between various market moods are explained in terms of the influence that the peer groups are exercising on each of their members. The approach of the collective as an entity has been advocated also in \cite{Cantono 2010}, \cite{Cantono 2012}, \cite{Biondi 2005}, \cite{Biondi 2010}. 

\subsection{Connecting the Minsky Accelerator to the Marshall-Walras formalism}
\label{subsec:MW intro}
The price formation in neoclassical economics seems very distant or even contradictory to the kind of Minsky accelerator dynamics described in the previous section. Yet, in the following section, we will show that one can find formal connections which in turn might illuminate the kind of dynamics that precedes the Minsky moment and which takes, in the first place, the system away from equilibrium. We will treat the Minsky accelerator in a form that makes it quite similar to the Marshall-Walras formalism. More specifically, in the basic supply-demand diagram, on the y axis, instead of the price we will use the interest rate.
The x axis will be different before and after the Minsky moment:
\begin{itemize}
\item[-] before the Minsky moment on the x axis we will plot the quantity of loans
(Figures \ref{fig:Walras Conv},  \ref{fig:increasing returns converge}, \ref{fig:fast increasing returns});
\item[-] after the Minsky moment the x axis will represent the number of ponzi defaults (Figures \ref{fig:convergence},  \ref{fig:divergence}).
\end{itemize}
These formal parallels might seem artificial (especially the second) but they will emerge naturally from the
mathematical processing of quite realistic assumptions.

This application of neoclassical concepts to Minsky's ideas encounters two main obstacles that turn out to be
more idiosyncratic than real:
\begin{itemize}
\item[-] using Marshall-Walras to describe run-away from equilibrium, rather then convergence to equilibrium;
\item[-] treating, in the Marshall-Walras formalism, the amount of loans and the loan defaults as a production quantity.
\end{itemize}
Both items above seem at first sight to be in contradiction to the very basis of the neoclassical thinking:
\begin{itemize}
\item[-] the very motivation for Walras to invent the `tatonnement' procedure and the quite similar market adjustments procedure by Marshall, was to substantiate Adam Smith's insight that markets (prices and quantities) are brought to equilibrium by an `invisible hand'.
Using it to demonstrate crisis and divergence from equilibrium seems the very opposite. It also seemingly goes against the `common sense' view  which 
suggests that the prices of the goods which are produced in excess will fall
and the prices of the good for which demand is greater then supply, will rise.
\item[-] in the neoclassical thinking, money is not production commodity (only exchanged: Say's law) and even when new money is injected in an economy the reaction is (`in the long term') neutral (rational expectations).
\end{itemize}

However one should not get the impression that there are no precedents in the economic literature to the two `departures' above:
\begin{itemize}
\item[-] for using the market mechanisms to obtain run-away from free-market equilibrium:
\begin{enumerate}
\item the possibility, and in fact likelihood, that market equilibria can be unstable has been established by Sonnenschein-Mantel-Arrow-Debreu \cite{Arrow 1954}. 
within the neoclassical framework;
\item even the rational expectations idea has been introduced with the help of an example \cite{Muth 1961};
of a divergence of the Walras procedure, which in fact happens for a very wide range of parameters;
\item Veblen has introduced ideas in which the increase in the price increases demand, while
\item the existence of economies with increasing returns (producing more costs less per product) has long been recognized (see for example \cite{Arthur 1994});
\item the social positive feedbacks leading to herding and allowing demand
to make large excursions out of equilibrium have been considered (see for example \cite{Kirman 2010} \cite{LLS 2000} \cite{Solomon 2000}).
\end{enumerate}
\item[-] For treating money or loans as a product, there was significant resistance because the conservation
of money through the transactions within an economy has been one of the most efficient instruments to obtain useful theorems and models, starting with Say, through Hicks' model and within the rational expectations theories. However, the use of Walras-like price-quantity diagram for interest rates and loans supply has been made in the past.
For instance \cite{Keen 2011} has emphasized that in addition to income earned by selling goods and services (which primarily finances consumption of goods and services and thus by and large conserves money), there exists the supply of money by banks in the context of entrepreneurial debt (which primarily finances investment) and in the context of rising ponzi debt (which primarily finances the purchase at increasing prices of existing assets).
\end{itemize}

A more significant departure from the neoclassical pattern is our analysis of the `Minsky crisis accelerator' that takes place after the Minsky moment. The evolution of the ponzi density and the interest rate during this period occupies the bulk of the present paper. 
We will show that their dynamics mathematically parallels the Marshall-Walras feedback loop where, in order to find the equilibrium state, one equates:   
\begin{itemize}
\item[-] the price per product offered by the suppliers as a function of the quantity of supplied products, with
\item[-] the price per product for which the demand by the clients equals the same quantity of products. 
\end{itemize}
The model based on the feedback loop between interest rate and ponzi failures will assume not only the autocatalytic loop between
\begin{itemize}
\item[-] the bottom-up  regulation (the influence of the quantity of ponzi companies on the interest rate offered by the lenders) and
\item[-] the top-down feedback (the influence of the interest rate on the number of ponzi companies),
\end{itemize}
but it will also take into account the peer-to-peer interaction in terms of network effects.
This will lead to a nontrivial configuration of stable and unstable fixed points Figure \ref{fig:solutions}
and to a correspondingly complex phase diagrams as shown in Figures \ref{fig:phasediag} and \ref{fig:phaserho}.

\section{The Marshall-Walras price formation process applied to the interest rate and loans volume}
\label{sec:Marshall Walras}
\subsection{The Marshall-Walras equilibrium in a loans market with decreasing returns}
\label{subsec:decreasing}
In neoclassical market equilibrium analysis, the formalism for finding fixed (equilibrium) states and establishing whether they are stable has been belabored for a long time, starting with Marshall and Walras and brought to mathematical perfection by Arrow-Debreu \cite{Galor 2007}. 
In our case we are looking for equilibria (and the dynamics leading towards or away from them) which result from the dynamics of the interest rate. 
In the period preceding a Minsky moment, the relevant interaction is between the interest rate, $i$,  and the amount of loans outstanding, $N_{loans}$. In the period following the Minsky moment, the relevant interplay is between the number of ponzis, $N_{ponzi}$, and the interest rate, $i$. 

Let us first apply to the debt or loans market the Marshall-Walras method for price formation. We will discuss later its limitations and their important implications. The crucial assumptions are that:
\begin{itemize}
\item[-] The amount  of loans demanded by the debtors, $N_{loans}$, is a decreasing function of the interest rate, $i$, that they have to pay:  
\begin{equation}
\label{N loans of i}
N_{loans} (i)=({i}/{k})^{-\mu}
\end{equation}
where $k$ and $\mu$ are constants. 
This equation Eq. \ref{N loans of i} implies that the number of loans increases as the interest rate decreases.
This connection is also at the basis of the Hansen-Hicks IS-LM model of credit (money) supply.
More precisely, the IS part of the model assumes that as the interest rate decreases, even investments with modest returns become lucrative
because they can still give returns higher than the interest rate:
one will do better by investing in a productive business than in keeping the money in the bank.
Conversely, by borrowing money at very low interest one will gain even if one invests it in a modestly lucrative business.
Thus the number of investments eligible for loans increases as the interest rate decreases.
As the increase in investments leads to an increase in production (GDP) which establishes the neo-classical IS inverse connection between the 
interest rate and the GDP.
\item[-] The interest rate, $i$, that the banks are charging is an increasing function of the amount of loans, $N_{loans}$, they agree to supply. This is the famous law of decreasing returns. The argument is that the banks are charging an increasing price if their lending capacity is stretched up to or even beyond their limits, thus incurring a higher risk of default.  We will discuss the realism of this assumption later. For the moment, we adopt this position and assume for definiteness that: 
\begin{equation}
\label{i of Nloans}
i ( N_{loans} ) =i_{0}N_{loans}^{\alpha}
\end{equation}
where $i_0$ and $\alpha$ are constants.
\end{itemize}

The power laws in Eqs. \ref{N loans of i} and \ref{i of Nloans} have both empirical and theoretical justifications:
\begin{itemize}
\item[-] The square root rule of influence (corresponding to $\alpha = 1/2$) has been argued in social psychology by Wallacher and Nowak \cite{Nowak 1998}.
\item[-] While the square root impact of orders on prices (corresponding to $\alpha = 1/2$) has been argued by Farmer at al, see \cite{Farmer 2006}.
\item[-] A power law is the only functional form that is scale invariant. 
If the dynamics is not strongly influenced by other scales, it is natural to assume a power law behavior. 
In particular, the scale invariance can be connected in the context of pricing with the fact that the price behavior should be invariant to the monetary unit in which it is measured (a form of `money neutrality').
\end{itemize}
However the results in the present paper are more general then the details of the power functional forms involved.

\begin{figure}[ht]
\centering
\subfigure [The macroscopic state of the system is expressed in terms of credit availability,  parameterized by the interest rate, $i$. The top-down part of the regulatory feedback loop is related to Eq. \ref{N iter}:  an increase in $i$ leads to a decrease in the demanded quantity of loans, $N_{loans}$. The decrease in the demand, $N_{loans}$, leads, according to Eq. \ref{i iter}, to a decrease in the price, $i$. In this case the resulting feedback loop has a stabilizing effect on the loans market, as shown in Figure  \ref{fig:Walras Conv}.] {\includegraphics[scale=0.3]{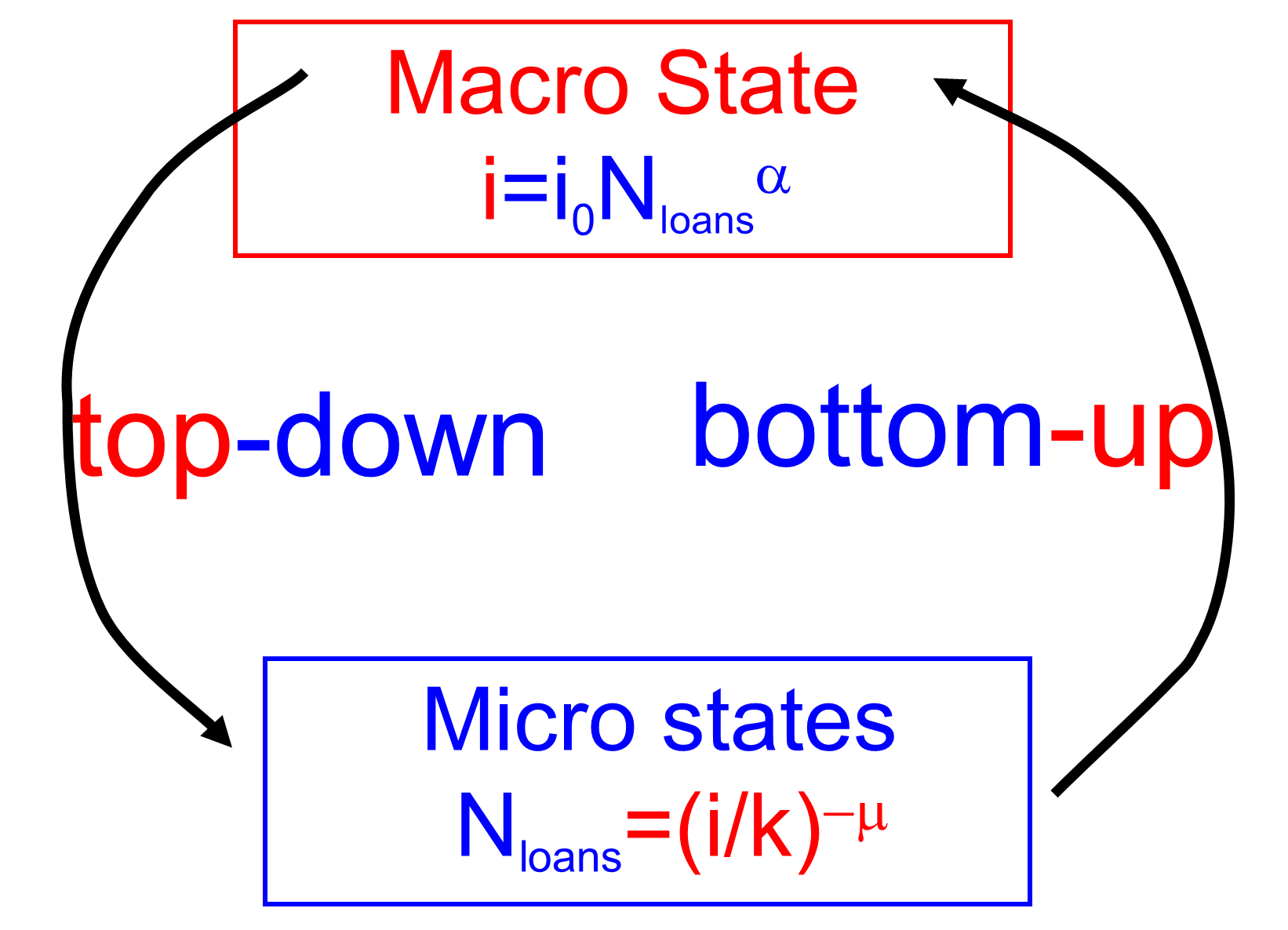} \label{fig: Walras Loop}}
\hfil
\subfigure[Application of the Walras procedure shown in Figure \ref{fig: Walras Loop} equalizing supply and demand for the case where the product quantity is the amount of loans, $N_{loans}$, and the price is the interest rate, $i$. The intersection point defines the equilibrium values $N_{fix}$ and $i_{fix}$. For the case $\alpha \mu < 1$ the iterative application of the  Eqs. \ref{N loans of i} and \ref{i of Nloans} converges to the equilibrium point irrespective of the initial value. Graphically this consists in iteratively drawing vertical arrows to obtain the new $i_{t+1}$  from the previous $N_{t}$ and horizontal arrows to obtain the next $N_{t+1}$.]{\includegraphics[scale=0.3]{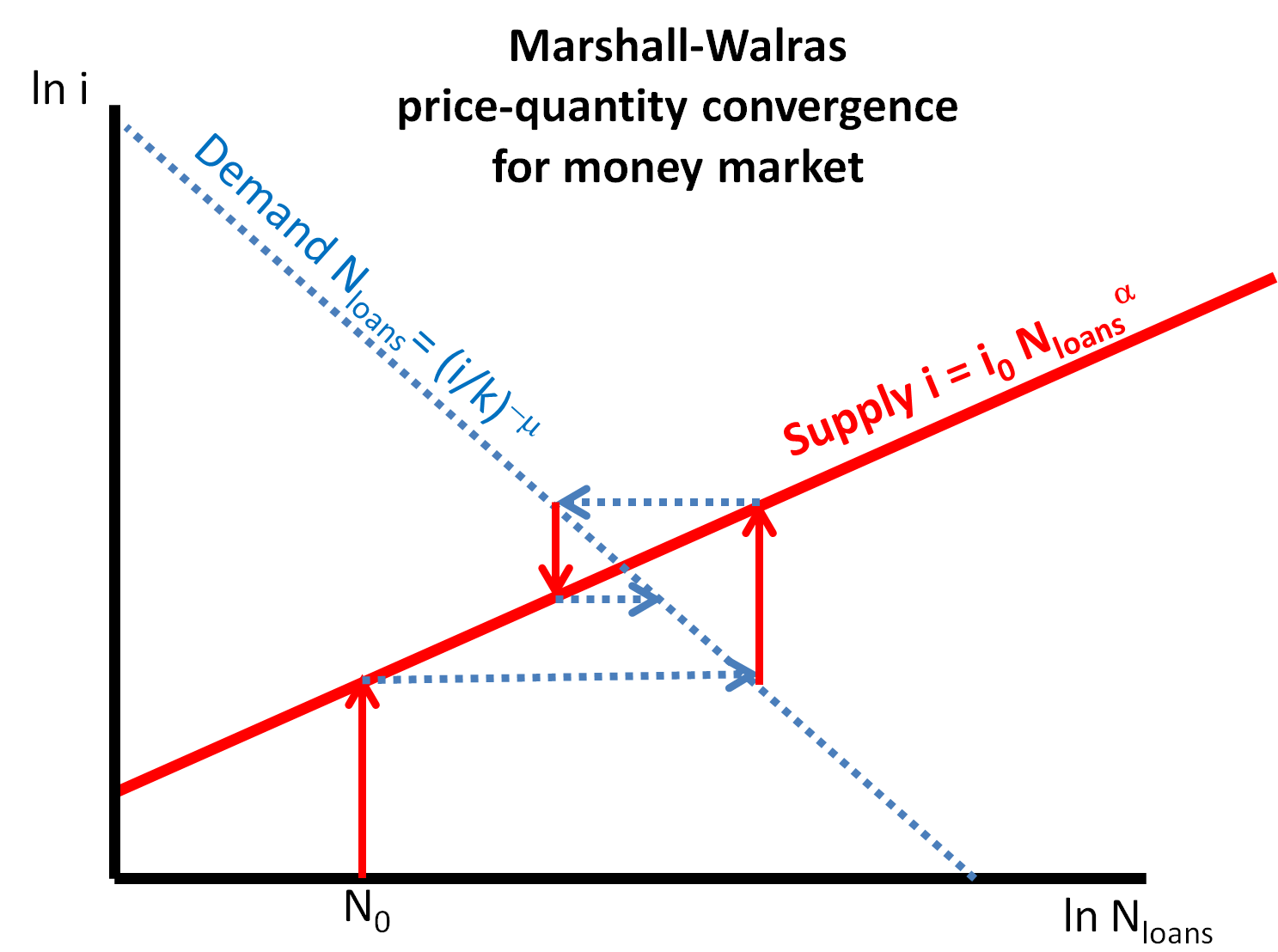} \label{fig:Walras Conv}}
\caption{\small Figure (\ref{fig: Walras Loop}, the Marshall-Walras convergence loop representing the interactions in the case of a loans market with decreasing returns. Figure \ref{fig:Walras Conv} represents the case when the parameters are such 
$ | \alpha \mu | < 1$ so that the $i$ and $N$ coevolution leads to convergence. In the Appendix \ref{sec:appendix decreasing returns} examples of both the convergent and the divergent coevolution are shown.}
\label{fig: Walras Loop and Conv}
\end{figure}

Marshall and Walras differed in the interpretation of the iterative procedure that may converge to the equilibrium point.
While for Walras the iterative `tatonnement' (`tweaking', `groping') towards an equilibrium price, $i_{fix}$,  and an equilibrium level of production, $N_{fix}$, took place in virtual time $t$, Marshall considered the approach of the equilibrium as a genuine process in real time, $t$, in which the supplier (in our case, the lender) reacted to the excess demand (in our case, of loans) by modifying the price (in our case, the interest rate) while the client (in our case, the debtor) reacted to the new price by modifying the quantity demanded:
\begin{equation}
\label{N iter}
N_{loans}=N_{t}=({i_t}/{k})^{-\mu}
\end{equation}
\begin{equation}
\label{i iter}
i_{t+1}=i_{0}N_{t}^{\alpha}.
\end{equation}
The equations \ref{N iter} and \ref{i iter} define a `top-down' -- `bottom-up' feedback loop as shown in Figure \ref{fig: Walras Loop}.

Starting from a certain initial interest rate, $i_0$,   and an initial quantity of loans,  $N_0$, one triggers a `tatonnement' chain reaction:
\begin{equation}
\label{process}
N_0  \xrightarrow{Eq. \ref{i iter}} i_1 \xrightarrow{Eq. \ref{N iter}} N_1 \xrightarrow{Eq. \ref{i iter}} i_{2} \cdots N_t  \xrightarrow{Eq. \ref{i iter}} i_{t+1} \xrightarrow{Eq. \ref{N iter}} N_{t+1} \cdots
\end{equation}
This iterative process  is represented graphically in Figure \ref{fig:Walras Conv}:
\begin{itemize}
\item[-] each step $N_t \xrightarrow{Eq. \ref{i iter}} i_{t+1}$ is represented as a vertical arrow with $x$ coordinate equal to $N_t $ and ending on the curve $i ( N_{loans} )$ given by Eq. \ref{i of Nloans}; 
\item[-]  each step $i_{t} \xrightarrow{Eq. \ref{N iter}} N_{t}$ is represented as a horizontal arrow with $y$ coordinate equal to $i_t $ and ending on the curve $N_{loans} ( i )$ given by Eq. \ref{N loans of i}. 
\end{itemize}
The equilibrium interest rate and loans quantity are then obtained as the common solution of the Eqs. \ref{N loans of i}
and \ref{i of Nloans}, as seen in Figure \ref{fig:Walras Conv}:
\begin{eqnarray}
\label{Walras solution i}
i_{fix}&=&( i_{0} k^{\alpha \mu} )^{1/(1+\alpha \mu )}\\
\label{Walras solution N}
N_{fix}&= &(k/i_{0})^{\mu / (1+ \alpha \mu )}.
\end{eqnarray}
In the Appendix \ref{sec:appendix decreasing returns} the exact evolution of $N_t$ during this process is deduced:
\begin{equation}
\label{N of t}
N_t =  N_{fix} [{N_0}/{N_{fix}}]^{{(-\alpha \mu )}^t}.
\end{equation}
According to Eq. \ref{N of t}, for $\alpha \mu < 1$, since the exponent ${( -\alpha \mu )}^t \rightarrow 0 $ vanishes for $t \rightarrow \infty$, $N_t$ converges to $N_{fix}$.
Thus, the Marshall-Walras process, Eq. \ref{process}, visualized in Figure \ref{fig:Walras Conv}, leads to the evolution of the number of loans, $N_t$, given by Eq. \ref{N of t}, which converges to the stable fixed point, given in Eq. \ref{Walras solution N}.

For $\alpha \mu > 1$, $N_t$ and $i$ both diverge to infinity, somewhat similar to the fears of the banks after the Lehman crash in September 2008. Yet,
 the convergence may still be achieved by modifying the details of the procedure. For instance, one could introduce smaller steps (say consisting of a fraction, $s$ ) instead of the full adjustment steps that overshoot the fixed point:
\begin{eqnarray}
\label{N small}
N_{t}&=&s ({i_t}/{k})^{-\mu} + (1-s) N_{t-1}\\
\label{i small}
i_{t+1}&= &s ( i_{0}N_{t}^{\alpha}) + (1-s) i_{t}
\end{eqnarray}
Such solutions have been proposed in the past: \cite{Kaldor 1934} (see especially pages 133-135) suggested that the adaptation of $N_{t}$ and $i_{t}$ takes place in small (individual transactions) steps (price/production stickiness) while \cite{Muth 1961} assumed that the changes are performed by agents with foreknowledge or `rational expectations'.
Another possibility could be:
\begin{eqnarray}
i_{t+1}&=&i_t+s(N_t-N_{t-1})^{\alpha} \label{eqA5ii}.
\end{eqnarray}
The models Eqs. \ref{N small} - \ref{eqA5ii} are genuinely time-independent Markov processes, unlike Eq. \ref{i iter} where the value of the interest rate at the beginning of the process, $i_0$,  is explicitly remembered throughout the process.
 However, this is not a problem for our models: they represent human reactions to specific events and moments, so singling out the initial value of the variables, $i_0$  , $N_0$, is quite natural.
 E.g. the Minsky moment  is definitely a memorable point of reference which the agents are quite naturally likely to remember throughout the crisis evolution.
Thus, we will not pursue the models of the type Eqs. \ref{N small} - \ref{eqA5ii} here and will rather concentrate on models of the type Eqs. \ref{N iter}, \ref{i iter}.

The analysis in the present subsection is a paradigm which we will repeat in the next subsections and following sections, introducing increasingly realistic conditions. As envisaged in the general market equilibrium analysis of Sonnenschein-Mantel-Debreu, in the systems that we will consider,  the generalization of the curves Eqs. \ref{N iter} and \ref{i iter} may have more (or less) than one intersection (common $(N_{fix}, i_{fix})$ solution of Eqs. \ref{N loans of i}, \ref{i of Nloans}) and neither a unique equilibrium interest rate, $i_{fix}$, nor even that the process stops at a given amount of loans, $N_{fix}$, will be guaranteed. A quite involved phase diagram will eventually emerge -- Figure \ref{fig:phasediag}.

\subsection{The Marshall-Walras equilibrium in loans market with increasing returns}
\label{subsec:increasing}

In the previous section it was assumed that $\alpha >0$. This corresponds essentially the law of decreasing returns applied to the money market. It represents the assumption that the interest rate, $i$, charged by the creditors increases with the quantity of loans, $N$, that they supply.
While the law of decreasing returns is often invoked in the neoclassical literature starting with Mills and Ricardo \cite{Ricardo 1996}, it applies less and less in the current economic conditions \cite{Arthur 1994}. While it was more difficult to produce more apples from the same land area or to extract more coal from a mine that was already in use for some time, it is often easier to produce a computer or a copy of a program after one has already produced many units of it.  However, the decreasing returns assumption is still often invoked in order to ensure market equilibrium, as indeed turned out to be the case in  Figure \ref{fig:Walras Conv}.

\begin{figure}
\centering
\subfigure[The top-down part of the regulatory feedback loop is related to Eq. \ref{N iter}: namely, an increase in $i$ (top) leads to a decrease in the demanded quantity of loans quantity, $N$ (bottom). The bottom-up effect is defined by Eq. \ref{i iter increasing}: a decrease in $N$ leads to a further increase in $i$. The difference in the thickness of the arrows in the drawing implies that this interest rate regulations is \emph{slower} than the increase in loan demand ($\alpha \mu< 1 $), and so the resulting feedback loop has a stabilizing effect on the loans market.]{\includegraphics[scale=0.3]{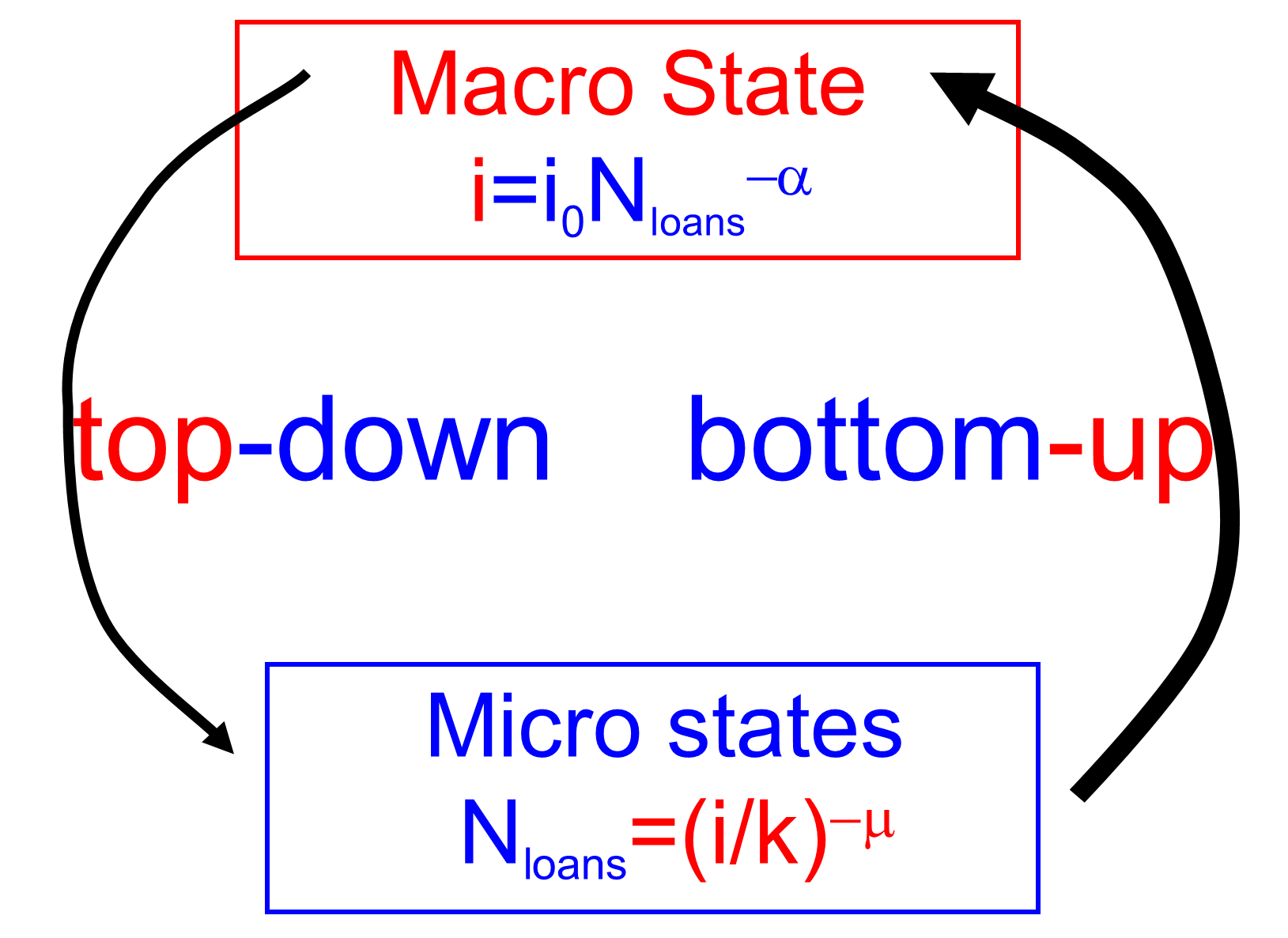} \label{fig:WalrasLoans}}
\hfil
\subfigure[Coevolution of the parameters in the loop Figure \ref{fig:WalrasLoans}: supply of loans $N$ and demand $i$. For the case $\alpha \mu < 1$, the process converges to the equilibrium point ($N_{fix}, i_{fix}$), marked by the intersection of the curves Eqs. \ref{N loans of i}, \ref{i increasing returns}. The convergence occurs irrespective of the initial value of $N_0$.  Graphically such an iterative convergence process \ref{process increase} is represented by alternatively drawing vertical arrows to obtain the new $i$  from the previous $N$ and horizontal arrows to obtain the next $N$ from the current $i$.]{\includegraphics[scale=0.3]{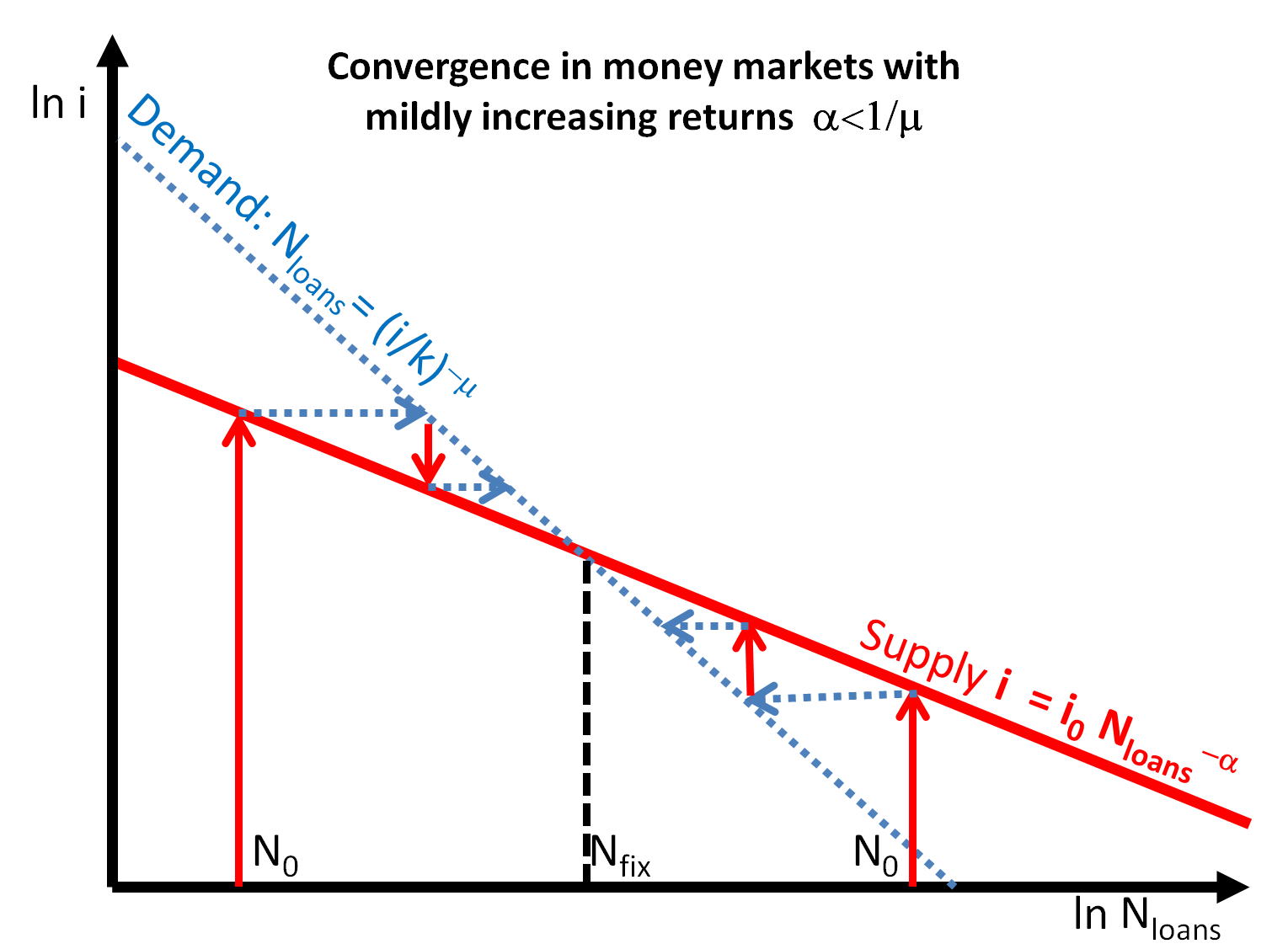} \label{fig:increasing returns converge}}
\caption{\small The Marshall-Walras procedure for increasing returns to scale and slow top-down feedback, i.e. when the cost of loans (interest rate, $i$) is a slowly decreasing function of the quantity of loans supplied, $N$.}
\label{fig:WalrasLoans Increasing Converge}
\end{figure}

As in many other cases, in the case of loans, modern conditions are conducive to a law of {\bf increasing returns}:
a bank that gave already many loans, has a lot of assets (the loans and their interest) which are also diversified over many debtors.
Thus it is well insured against occasional creditor defaults. Therefore, it can afford to allocate more loans to more clients at  lower interest rates. On the contrary, a bank with few loans has less assets and less diversification and has to charge a higher interest rate in order to protect itself against occasional defaults.
Increasing returns has been recognized in the last decades both as a daily occurrence in the empirical economic reality and also as a significant factor determining the departures of the real life from neoclassical economic theory \cite{Arthur 1994}.
Moreover as detailed in \cite{Biondi 2005} bank entities differ from markets in a way that makes the banking sector one of the sectors most likely to be affected by the factors leading to economies of scale because of its specific economic organization and specific features: 
\begin{itemize}
\item[-]  economies of knowledge (learning by doing, aggregating information from scale and variety of operations)
\item[-] overheads split across various clients (leading to diminishing average and marginal costs)
\item[-] implicit public guarantees for bigger entities (too big too fail), etc.
\end{itemize}
Thus, instead of Eq. \ref{i of Nloans} one is lead to study the case of a supply curve where the interest rate, $i$, decreases with the quantity of loans, $N$:
\begin{equation}
\label{i increasing returns}
i ( N_{loans} ) =i_{0}N_{loans}^{-\alpha}
\end{equation}
as exemplified in Figure \ref{fig:increasing returns converge}.
In the case of increasing returns, Eqs. \ref{N iter} and \ref{i iter} become:
\begin{eqnarray*}
N_{ t}&=&({i_t}/{k})^{-\mu} \\
\label{i iter increasing}
i_{t+1}&=&i_{0}N_{t}^{-\alpha}.
\end{eqnarray*}
The process of their coevolution, iteratively over time, is:
\begin{equation}
\label{process increase}
N_0  \xrightarrow{Eq. \ref{i iter increasing}} i_1 \xrightarrow{Eq. \ref{N iter}} N_1 \xrightarrow{Eq. \ref{i iter increasing}} i_{2} \cdots N_t  \xrightarrow{Eq. \ref{i iter increasing}} i_{t+1} \xrightarrow{Eq. \ref{N iter}} N_{t+1} \cdots
\end{equation}
The amount of outstanding loans at any one time is (see Appendix \ref{sec:appendix increasing returns} for the derivation):
\begin{equation}
\label{N of t negative}
N_t =  N_{fix} [{N_0}/{N_{fix}}]^{{( \alpha \mu )}^t},
\end{equation}
where the fixed point is defined as the intersection between the two curves, Eqs. \ref{i increasing returns} and \ref{N loans of i}:
\begin{eqnarray}
\label{increase solution N}
N_{fix}&= &(k/i_{0})^{\mu / (1- \alpha \mu )}\\
\label{increase solution i}
i_{fix}&=&( i_{0} / k^{\alpha \mu} )^{1/(1-\alpha \mu )}.
\end{eqnarray}
As opposed to the decreasing returns case (Eqs. \ref{Walras solution i}, \ref{Walras solution N}) the convergence condition  $ \alpha \mu < 1$ cannot be circumvented by small step modifications of the type given in Eqs. \ref{N small}, \ref{i small}. Indeed, if $ \alpha \mu < 1$  the Eq. \ref{N of t negative} converges in the $t \rightarrow \infty$  limit to $N_{fix}$ since the exponent ${(\alpha \mu )}^t \rightarrow 0 $.
However for $ \alpha \mu > 1$, instead of $N_t$ oscilating between  $0$ and $\infty$ , as in the previous case Eq. \ref{N of t} where the exponent oscillated between $+ \infty$ and $- \infty$,  one has now a monotonic behavior because ${(\alpha \mu )}^t \rightarrow + \infty $.
Thus, if the initial point $N_0$ is less then $N_{fix}$, one has in Eq \ref{N of t negative} a quantity less then $1$ at the power $\infty$ which converges to $0$.
On the contrary, if $N_0> N_{fix}$, then one has  in Eq. \ref{N of t negative} a quantity larger then $1$ at the power $\infty$ which diverges to $\infty$.

Thus we conclude that a genuine instability occurs at $N_{fix}$ if the interest rate offered by the banks decreases with the volume of loans Eq. \ref{i iter increasing} faster  than the decrease in the interest rate sufficient to insure an increase in the loans demand ($\alpha\mu > 1$), Eq. \ref{N iter}. This case is analyzed in the next subsection. 
The difference between the stable and unstable loan market with increasing returns can be best understood graphically by comparing Figures \ref{fig:increasing returns converge} and \ref{fig:fast increasing returns}, where, on a double logarithmic scale graph:
\begin{itemize}
\item[-]the function $i(N)=i_{0}N^{-\alpha}$  is represented  by a straight line of slope $-\alpha$; 
\item[-] while the function $N(i)=(i/{k})^{-\mu} $ is represented by a straight line of slope $- 1/ \mu$ because $N$ corresponds to the horizontal  x axis while $i$ is measured on the vertical y axis.
\end{itemize}
The evolution of the iterative process $N(i_t)=(i_t/{k})^{-\mu} $, $i_{t+1}(N_t)=i_{0}{N_t}^{-\alpha}$ is represented by arrows. Following the arrows in Figures \ref{fig:increasing returns converge} and \ref{fig:fast increasing returns}, one can see that:
\begin{itemize}
\item[-] if $-\alpha > -1/\mu$,  i.e. the slope of $N(i)$ is steeper then the slope of $i(N)$ the process converges to the fixed point; 
\item[-] if, on the contrary, the slope of $i(N)$ is steeper then the slope of $N(i)$, as in Figure \ref{fig:fast increasing returns}, then the fixed point is unstable (repulsive).

\end{itemize}

The generic condition for the convergence and stability of a fix point of a discrete dynamical system is (cf. \cite{Galor 2007}):
\begin{equation}
\label{divergence condition}
\left| \frac{\partial i}{\partial N}  \right| 
\left|\frac{\partial N}{\partial i} \right| <1.
\end{equation}
We will use this inequality in the rest of this paper and especially in obtaining the phase diagrams in the non-linear case with multiple fixed points given by Figs. \ref{fig:solutions}, \ref{fig:phasediag}; not only does this criterion help establish the direction of the iterative process in the neighbourhood of the fixed points, but it also helps identify the character and the evolution of the process in the entire parameter intervals between the fixed points. In fact they correspond to the various phases (stable or unstable) of the system.

The formal convergence condition Eq. \ref{divergence condition} is visually and more intuitively enforced in the various diagrams
of the type shown in Figure \ref{fig:increasing returns converge}, by just following the arrows that represent graphically the iterations
of the process \ref{process increase}: horizontal arrows bring the process from $i_t$ to the corresponding $N_t$ while the vertical arrows advance the process from $N_t$ to $i_{t+1}$. 

More specifically, in Figure \ref{fig:increasing returns converge} the chains of arrows are pointing towards the stable fixed point, indicating that the process is converging towards it irrespective of on which side of it one starts. On the contrary, in Figure \ref{fig:fast increasing returns} the chains of arrows point away from unstable fixed point 
such that starting slightly above or below the fixed point, the distance to it increases as the process advances.
Automatically this means that the entire interval between two fixed points
(e.g. Figure \ref{fig:solutions}) has a uniform behavior -- it constitutes one phase of the system; irrespective of where one starts within that interval, the process will converge towards the stable fixed point and run away from the unstable fixed point. 

\subsection{The Rational of Irrational Exuberance}
\label{subsec:Exuberance}

\subsubsection{The Loans Accelerator}
\label{subsubsec:Loans}
The Minsky vision of intrinsic instability has been perceived by both supporters and adversaries as incompatible with the mainstream neoclassical tradition. 
As in many other cases (Keynes, Schumpeter, etc.), the incompatibility is mainly in the minds of the researchers rather than in the actual mathematical hard core or the methodology.
The instability is clearly singled out as the default rather then the outlier in the Sonnenschein-Mantel-Arrow-Debreu analyses. Concentrating on the convergent cases is only a particular choice which was (too) often made in the past. In the present sub-section we will use a  Marshall-Walras neoclassical-like analysis to substantiate Minsky's point that instability is a natural condition for a capitalist regime.   
In fact to obtain it, one only has to consider the $ \alpha \mu > 1$ case  in the analysis of the previous sub-section.

In this case, the process Eq. \ref{process increase} diverges because in Eq. \ref{N of t negative}
the exponent, ${( \alpha \mu )}^t$, instead of vanishing for $t \rightarrow \infty$, diverges ${( \alpha \mu )}^t \rightarrow \infty$.
Thus 
\begin{itemize}
\item[-] for an initial ${N_0} > {N_{fix}}$, Eq. \ref{N of t negative} implies   $N_t \rightarrow \infty$. Of course for a finite number of agents $N_{total}$,
the process \ref{process increase} will rather stop at $N_t = N_{total}$ 
\footnote{As discussed below, the finiteness of $N_{total}$ may bring upon the Minsky moment:
the Minsky loan accelerator is relying on the expectations that one may continue indefinitely to make
loan-financed investments and pay the interests from the earnings. The finiteness of the economy means that at some stage
the investments would exceed the capacity of the market to buy and will not give the returns necessary to pay the interest on the loans. This will push some of the most aggressive investors in a Ponzi position and trigger the Minsky moment.}.
\item[-] for an initial ${N_0} < {N_{fix}}$, Eq. \ref{N of t negative} implies   $N_t \rightarrow 0$ .
\end{itemize}
\begin{figure}
\centering
\hfil
\subfigure[Minsky Loan Accelerator for a loan market with fast increasing returns (fat top-bottom arrow, i.e. fast decrease of the charged $i$ with the supplied loan quantity $N$). Similarly to Figure \ref{fig:WalrasLoans}, the top-down part of the regulatory feedback loop is related to Eq. \ref{N iter}: a decrease in $i$ leads to an increase in the demanded quantity of loans, $N$. The increase in $N$ leads, according to Eq. \ref{i iter increasing}, to a further decrease in $i$. 
If the decrease in the charged interest rate is so fast that it exceeds the decrease necessary to increase the demand for loans $(\alpha \mu > 1)$, this   has a destabilizing effect on the loans market and results in a divergent feedback loop.]{\includegraphics[scale=0.3]{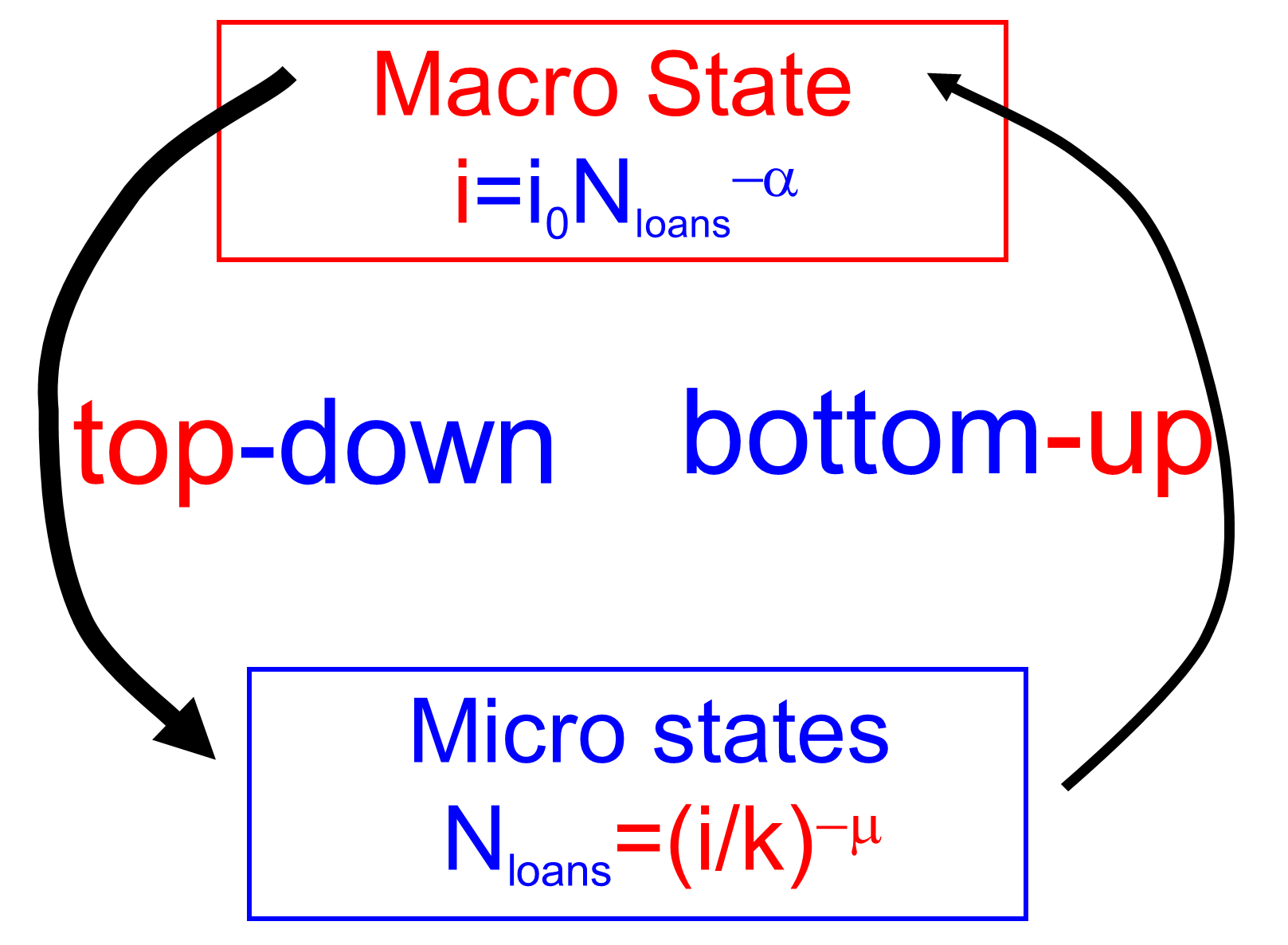} \label{fig:WalrasLoansFast}}
\hfil
\subfigure[The Marshall-Walras procedure for the supply and demand of loans, Eq. \ref{i increasing returns}, in the case of fast increasing returns, i.e. $\alpha > 1/\mu $.
Graphically the iterative process \ref{process increase} is represented by alternatively drawing vertical arrows to obtain the new $i_t$  from the previous $N_{t-1}$ and horizontal arrows to obtain the next $N_t$ from the current $i_t$.
 The process diverges from the unstable fixed point marked by the intersection 
$(N_{fix}, i_{fix})$ of the curves Eqs. \ref{N loans of i}, \ref{i increasing returns}.
In the limit as $t \rightarrow \infty$ for an initial loans quantity $N_0 < N_{fix}$, 
the quantity of loans shrinks to $N_t \rightarrow 0$. 
In contrast, for an initial loan quantity $N_0 > N_{fix}$, the quantity of loans increases $N_t \rightarrow \infty$ until this `bubble' is stopped by a Minsky moment.  ]{\includegraphics[scale=0.3]{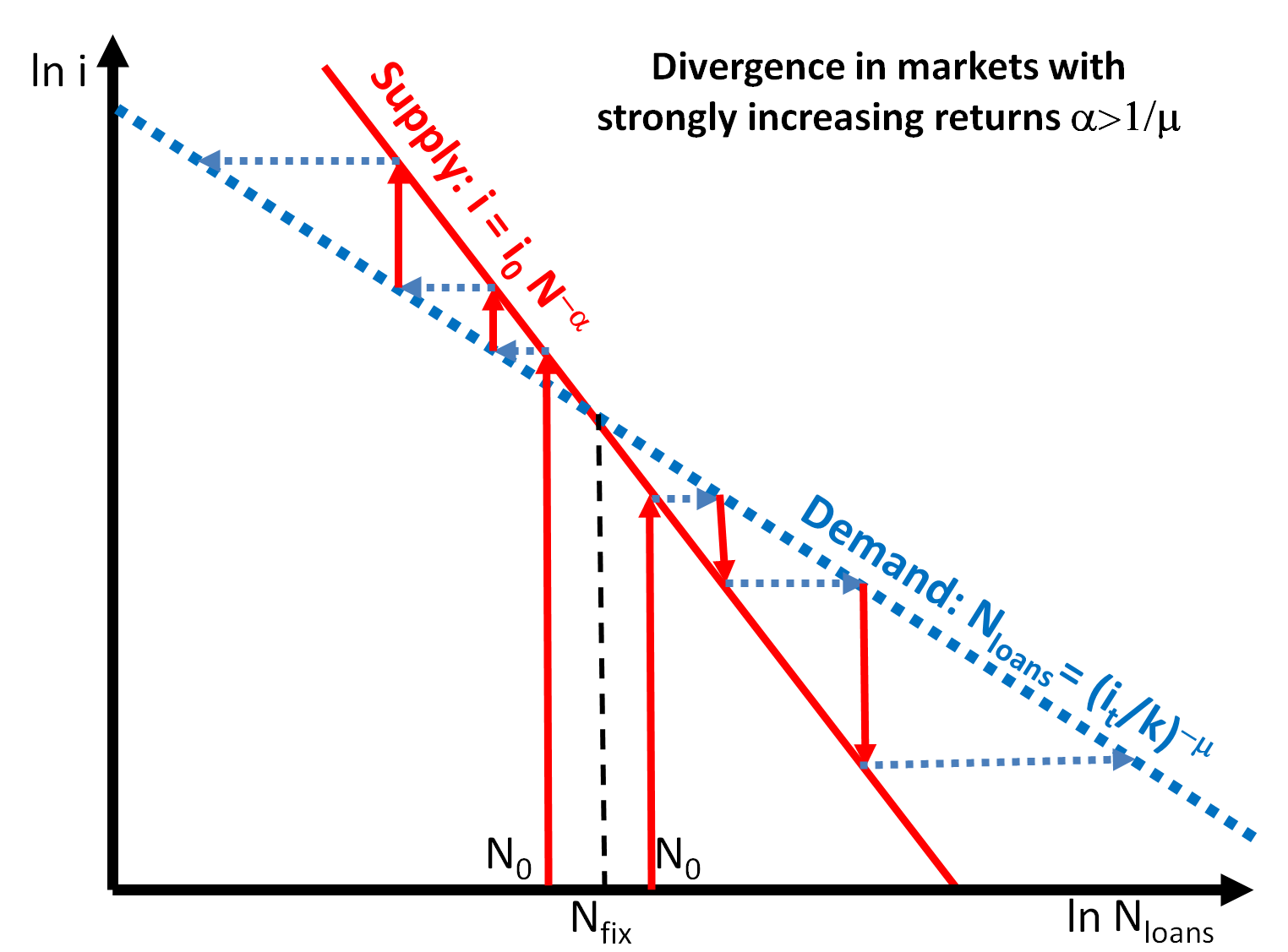} \label{fig:fast increasing returns}}
\label{fig:WalrasLoans with fast increasing returns}
\caption{\small Minsky Loan Accelerator for a loan market with fast increasing returns to scale of the supply of loans (lower interest rate) and a graphical representation of the coevolution of the parameters $i$ and $N$. The condition that determines the difference between the slow and the fast increasing returns is the product $\alpha \mu$ of the interscale feedback parameter $\alpha$ and the heterogeneity coefficient $\mu$, which, in the slow returns markets, is less than and, in the fast returns market, is greater then 1.}
\end{figure}
Thus, in the case $ \alpha \mu > 1$, the fixed point $(N_{fix}, i_{fix})$ becomes repulsive.
Graphically this is seen in Figure \ref{fig:fast increasing returns}, where, starting at  points $N_0$ very close to (slightly above or slightly below) $N_{fix}$,  leads to $N_t$ (and $i_t$) evolving in opposite directions, towards $0$ or $\infty$, respectively.
Figure \ref{fig:WalrasLoansFast} illustrates visually the autocatalytic feedback loop responsible for this instability in which the top-down feedback is dominant (fat top-down arrow).

The implications for the loans market are disastrous: 
banks that dare to lend in an aggressive way large quantities at low interest rate take over the market, while banks or financial sectors that take a more conservative position are squeezed out of the market.
It becomes rational, and in fact unavoidable, for the creditors to have an as risky as possible policy just in order to avoid marginalization. In turn the debtors are encouraged to borrow larger  and larger amounts at lower and lower interest rates.

However, as seen above and in the next sections, the formal expression of the run-away analysis 
has many common points with the neo-classical curve-crossing techniques used to find stable equilibria.

In the pre-crisis mode, the system is so successful that it slips into a state of exuberance which feeds back upon itself and increases exponentially the debt financed investment.  \cite{Greenspan 1996}, \cite{Shiller 2006} have called this exuberance `irrational'. However this is a matter of point of view: from the individual profit-seeking (capitalistic, in Minsky's word) point of view this behavior is rational in as far as it maximizes one's personal profits. However, for all-knowing agents who in particular would know the tragedy of the commons and the contents of the present paper, it should be clear that their behavior is likely to end in a collective loss. 

Some commentators made Greenspan himself and the Federal Reserve Bank (Fed) responsible for the bubble by having kept interest rates too low and thereby encouraging the dangerous exuberance.  Indeed, if loans with low interest $(i=\textit{interest payments}/\textit{loan})$ are available, and the assets that can be bought with them have a high enough resilience: $r=earnings/loan$, which exceeds the interest rate: $r>i$, the bottom line is that in this regime: $earnings>\textit{interest payments}$. Thus borrowing in order to buy those assets looks from an myopic individual point of view (and in absence of understanding of the emergent collective dynamics) safe and profitable and thus rational! 

This Minsky loans accelerator addresses a perennial puzzle faced by the mainstream neo-classical economics: the business cycle fluctuations. Not only the neo-classical models predict a crisis-free steady state but in fact they assume equilibrium as their main conceptual basis. The only way to admit some level of fluctuations is to attribute it to exogenous shocks that temporarily take the economy out of its stable equilibrium state \cite{Kydland 1982}. According to those models, following such a shock, the invisible hand of the efficient market gently brings the system back to its equilibrium. On the opposite side, Minsky's position is that crises are intrinsic to capitalism: the quite visible hand of debt financed investment is leading the capitalist system out of equilibrium. As seen below, the system eventually brings itself to a state in which the slightest noise is amplified to a systemic crisis. This "Minsky moment"  is described in the next subsection.

\subsubsection{The Minsky Moment}
\label{subsubsec:MinskyMoment}

As shown in the previous subsection, banks that would NOT lend with decreasing interest rates and debtors that would NOT take increasingly leveraged positions would in fact be the ones to be punished by lower profits, lower growth rates and market share loss. In order to give more loans, the banks will have to lend not only to solid highly promising companies (which at some stage become over-leveraged themselves), but also to units which are only riding on the positive-feedback expectations loop. Moreover, it is rational that the companies will adopt very large leverage positions that given the current interest rate are affordable at their current level of earnings. In fact this in itself leads to a version of the Minsky self-referential feedback loop: as the asset (stock market / real estate market) prices increase, so do the borrowing capabilities of the units holding them. This provides those units with more collateral for borrowing more on the same assets (which have now increased market value). The new loans received by those units  may then go back reinvested in stock market / real estate, reinforcing the loop as long as the market price increases.

In fact, in order for the bank to continue the increase in its volume of lending, it has to find new borrowers. When all the good borrowers already have a loan, the bank has to lower its lending standards to capture new borrowers who were previously shut out of the credit market \cite{Adrian 2010}. In an infinite economy and in the absence of noise this can be shown to be sustainable \cite{LLS 1995}.
However in the real world, one eventually rediscovers the  `Herbert Stein's Law': 
\begin{quotation}
``If something cannot go on forever, it will stop." \cite{Krugman 2010}.   
\end{quotation}
In the present case the positive feedback loop is broken once there is even a small downward fluctuation of the rate $r$ of growth of the assets $n$ bought from the loan.
This can in particular be brought upon by the finiteness of $N_{total}$: the markets expand indefinitely to keep up with the ever increasing (loan financed) investments.
 Once the earnings from these investments $r \times loan$ cannot cover the interest payments on their debt $i \times loan$  , those units become ponzi and -- when discovered -- fail.  This is the Minsky moment (Paul McCulley coined the term `Minsky moment' to describe the 1998 Russian financial crisis). Following the Minsky moment, the previously lax credit policies are tightened, interest rates are increased $i_{new}>i_{old}$ (Eq. \ref{default of i} below) which brings even mildly speculative (but until then viable, $r>i_{old}$) companies into ponzi positions ($r<i_{new}$) and eventual failure. In turn, this feeds back onto tightening further credit availability in the system and closing the feedback loop. 

There is always some level of external noise that may  cause the system to depart dramatically from that specified by these equations. As long as the leverage of companies is below a certain threshold, the noise is not sufficient to induce failures. However, as the leverage increases, the slightest fluctuation in the interest rate or the earnings may lead some of the borrowers into failure. This is the Minsky moment. The first wave of failures, induced by the noise, is then amplified systematically by the feedback between the number of failures and the interest rate; in order to avoid becoming ponzi, companies then have to deleverage. This increases the interest rate, which in turn forces more companies to deleverage or become ponzi.

\begin{figure}
\centering
\includegraphics[scale=0.35]{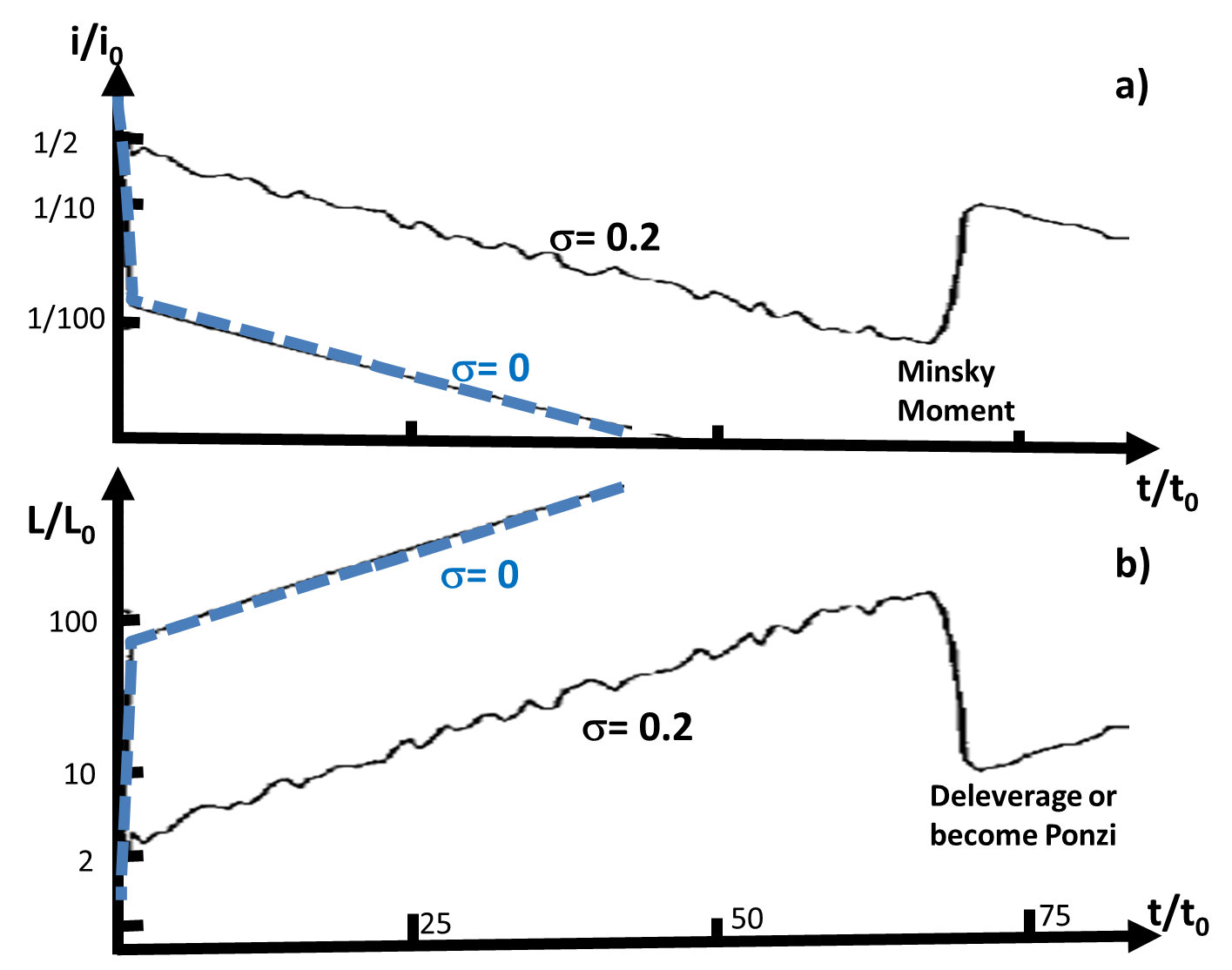}
\caption{\small The Minsky moment adapted from the LLS \cite{LLS 1995} model. \\
The original LLS model  \cite{Levy 1994} \cite{LLS 1995} \cite{LLS 2000} described the burst of a bubble where the fraction, $f$, of investment in the risky assets increased indefinitely approaching 100 $ \% $ of the individuals' wealth. This in the present context corresponds to the sustainable leverage diverging to arbitrary large values of the order $L \sim {L_0} / i $ (cf. the interrupted line in Figure \ref{fig:lls}(b).
 By sustainable average one means the leverage which is still consistent with the company not being ponzi. For instance, $L_0$ is the initially sustainable leverage  $L_0 = earnings / (i_0  \ assets)$. \\
It was shown in LLS that while in the absence of noise ($\sigma =0$) this regime may continue indefinitely, in the presence of noise the bubble bursts when $ L >> 1/ {\sigma}^2 $ i.e. in our case when $i << {\sigma} ^2$.
In the present paper, the noise appears as a intrinsic consequence of the  granularity of the agent based model. \\
One may wonder if, in order to avoid the Minsky moment, one should minimize the noise amplitude, $\sigma$.
The answer is that reducing $\sigma$ does indeed allow the system to pursue the exuberance for a longer period and tolerate without defaults higher leverages $L$.
However, this would only delay the Minsky Moment, not eliminate it.
Moreover it would greatly increase the severity of the crisis once it is triggered.
For instance, the presence of a $\sigma = 0.2$ noise keeps $i$ (Figure \ref{fig:lls}(a)) continuous line) almost two orders of magnitude larger than it would be in its absence $\sigma = 0$ (Figure \ref{fig:lls} (a) interrupted line). Consequently the jump in the interest rate at the Minsky Moment is, for $\sigma =0.2$,  from order of $i \sim O(1 \% )$ to order of $i \sim O(10 \% )$  (cf. Figure \ref{fig:lls} (a)). 
Cf. Figure \ref{fig:lls} (b) this corresponds to a jump of one order of magnitude in the sustainable  leverage $L$.  In turn this means that all the companies with leverages in between the old and new sustainable leverage values are suddenly thrown into the ponzi category unless they have enough cash reserves to deleverage instantly.
}
\label{fig:lls}
\end{figure}

Figure \ref{fig:lls} adapts into the present context the analysis of \cite{LLS 1995} of the role of noise and finite size in triggering the Minsky moment.
In the figure it is shown that in the total absence of noise the risk taking collective behavior may continue to infinity: the interrupted lines, corresponding to 0 noise, show that the interest rate can continue to decay to arbitrary small values which in turn makes it possible (and profitable) for borrowers to take increasingly large leverages. By contrast, the slightest noise is bound to trigger sooner or later a catastrophic Minsky moment. In fact smaller the noise, later is the crisis and more catastrophic its consequences. Trying to avert the burst of the bubble in those conditions, achieves only its delay at the price of making it more severe. In the end assessment, it is better to have an economy with uncontroled noise, rather then one with uncontrolable crashes. 

The dynamics following the Minsky Moment can be analysed with methods mathematically identical with the one used in the previous sections. This will be detailed in the next section. However, it will turn out that in order to obtain realistic results one has to include the microscopic granularity of the agents and the network effects of their interactions. This will be the subject of the rest of the paper.
\section{A Marshall-Walras-like procedure for the interest rate vs ponzi quantity: Minsky crisis accelerator}
\label{sec:Minsky accelerator}

In neoclassical thinking the interest rate, $i$, is the driving factor toward equilibrium in the debt market: increasing the interest rate is expected to inhibit risk taking and borrowing when they exceed a certain limit. One sees that in Minsky's scenario, beyond a certain "critical" point increasing the interest rate has an opposite procyclical effect of triggering the crisis. 
In the previous section we have considered the period leading up to the creation of a large amount of
unwarranted, unsecured, low interest loans. Of course, the way to stop the `Minsky loans accelerator' would be to increase the interest rate. 

However, this would render a further part of debtors to be ponzi, leading to a `Minsky moment' which marks the end of the `Minsky lending accelerator' and the start of the `Minsky crisis accelerator': ponzi failures decrease the credit availability in the system and thus increase the interest rate, $i$, which in turn increases the number, $N_{ponzi}$, of companies $n$ forced into the ponzi status Eq. \ref{ponzi def}. Below we explain, justify, formalize and analayse the components of this feeback loop. 

This `Minsky crisis accelerator' is believed nowadays to be one of the main mechanisms behind the propagation of the current economic crisis [Wray 2011b] although its mathematical formulation is under fierce debate \cite{Keen 1995}, \cite{Keen 2012}, \cite{Eggertsson 2012}, \cite{Krugman 2012}. 
As it will turn out, the analysis of the Minsky crisis accelerator reduces to a  similar mathematical formalism as the Minsky lending accelerator, except that instead of the quantity of loans, $N_{loans}$, it will be the number of ponzi companies, $N_{ponzi}$, that closes the feedback loop with the interest rate, $i$. 

Let us define in detail the formal framework.
Earlier measurements by \cite{Takayasu 2000} indicate that both debts and earnings, i.e. the denominator and the numerator in Eq. \ref{ponzi def}, are distributed according to power laws. In fact this has been connected with the Pareto wealth distribution power law   \cite{Klass 2006}. Therefore it is reasonable to assume that the resilience of company $n$: 
\begin{center}
$r(n)=earnings(n)/debt(n) <i$,
\end{center}
 also follows a power law distribution with a heterogeneity exponent $1/\beta$:
\begin{equation}
\label{eq10}
r(n)=k  n^{1 / \beta}
\end{equation}
where $k$ and $\beta$ are empirically fixed parameters. 

Inverting Eq. \ref{eq10} allows us to obtain that the number of ponzi companies for a given interest rate, $i$, i.e. the number of companies that have resilience $r(n)<i$.
  More precisely, if at a certain time, $t$, the interest rate becomes $i_t$, this will bring the current number $N_t =N_{ponzi}(t)$  of ponzi companies to: 
\begin{equation}
\label{default of i}
N_{t}=({i_t} / {k})^{\beta}.
\end{equation}          

\begin{figure}
\centering
\subfigure[The Minsky top-down bottom-up feedback loop. The macroscopic state of the system in terms of the credit availability is parametrized by the interest rate $i$. The states of the individual companies are characterized by their ponzi or non-ponzi status. The top-down part of the Minsky crisis accelerator feedback loop is related to Eq. \ref{ponzi def}: an increase in $i$ will switch the status of some of the companies from non-ponzi to ponzi. In turn, this increase in the number of ponzi companies may induce, by the bottom-up part of the Minsky accelerator loop Eq. \ref{default of i}, a further increase in the interest rate. For $\alpha \beta <1$ the process leads to a stable fixed point.]{\includegraphics[scale=0.3]{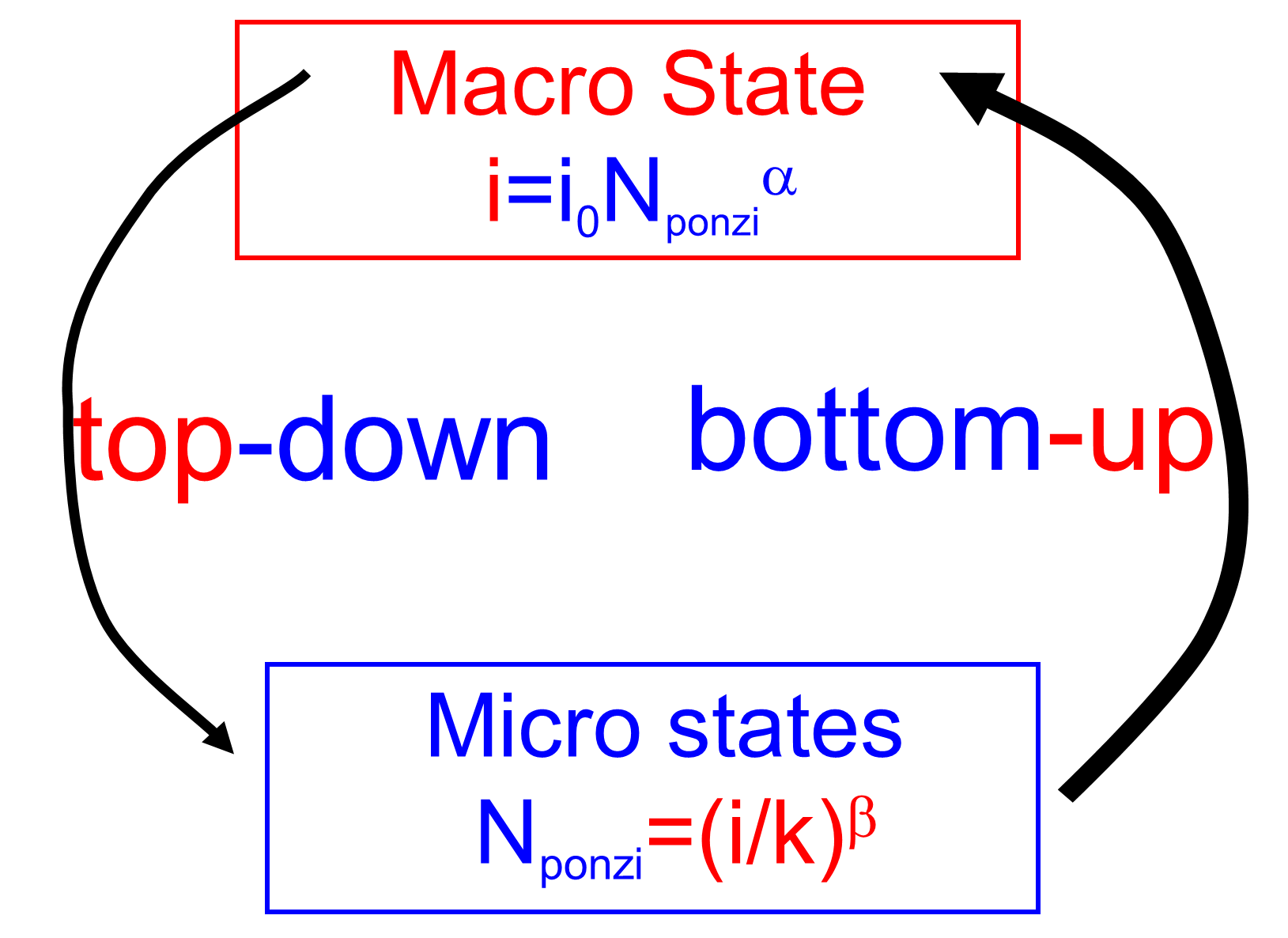} \label{fig:Ponzi Loop Conv}}
\hfil
\subfigure[The graph represents the system of Eqs. \ref{default of i}  and \ref{i of default}. 
It allows us to represent graphically  the iterative process \ref{process loan diverge}:
one starts from the initial number of ponzis (failures) $N_0$ on the $x$ axis and moves on the vertical full arrow keeping $N= N_0$ until one intersects the full line $i = i_0 {N_{ponzi}}^{\alpha}$.
 The intersection point defines the new interest rate $i_1$. 
From the point $(N_0, i_1)$ one moves on the horizontal dotted arrow keeping $i=i_1$ till one intersects the dotted line $N_{ponzi} = (i/k)^{\beta}$. The intersection defines the new $N_1$.
And so on in general:
one moves on vertical arrows with fixed $N_t$ to intersect $i = i_0 {N_{ponzi}}^{\alpha}$ and thus find $i_{t+1}$ and 
then one moves on horizontal arrows with fixed $i_{t+1}$ to intersect $N_{ponzi} = (i/k)^{\beta}$ and thus find $N_{t+1}$.
One sees that the process converges as long as on the graph the slope of $N_{ponzi}(i)$ is steeper then $i(N_{ponzi})$ i.e. when
$\alpha \beta < 1$.]{\includegraphics[scale=0.3]{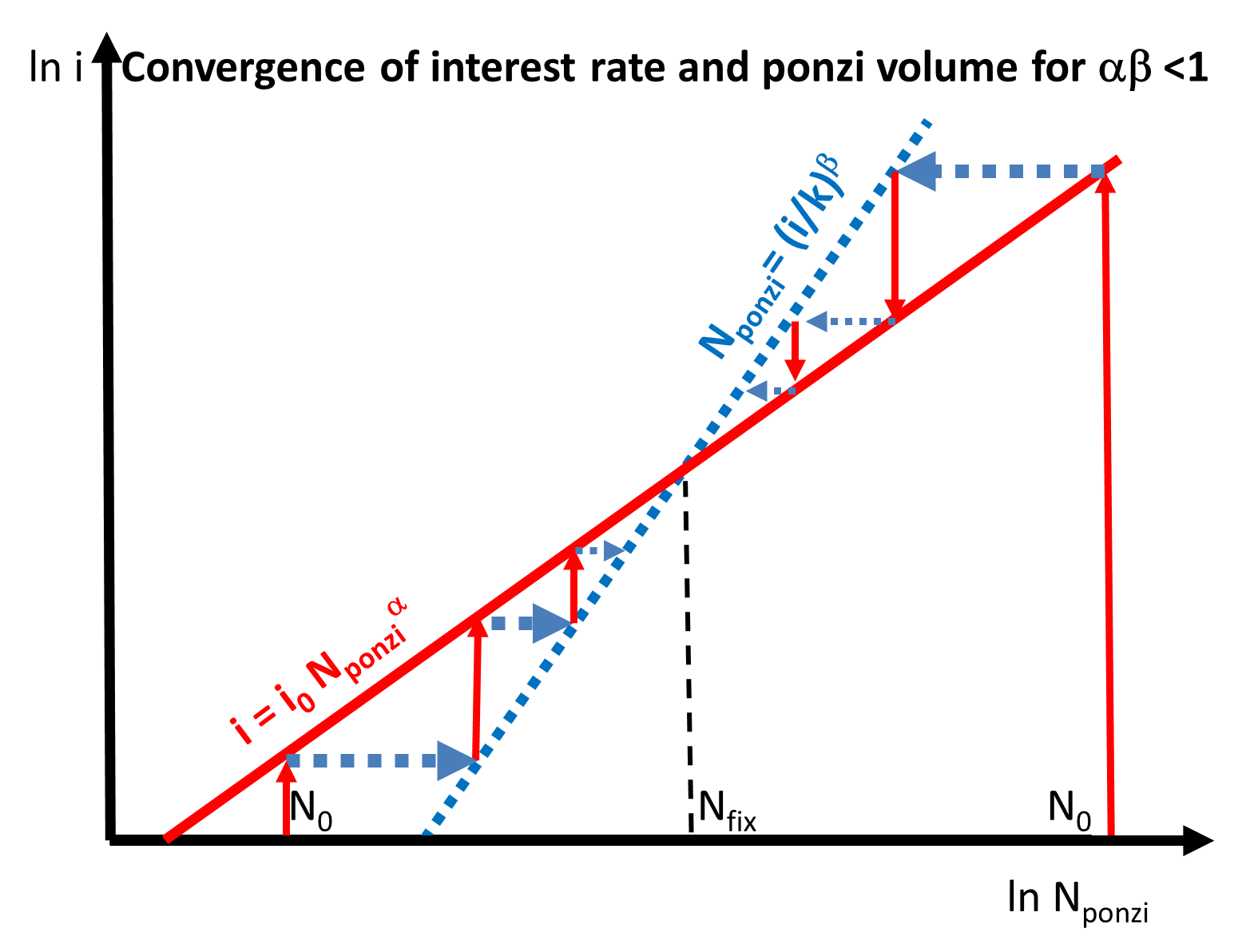}\label{fig:convergence}}
\caption{\small The Minsky crisis feedback loop represented by the system of Eqs. \ref{default of i} and \ref{i of default}. Note that (only) when the slope of the curve $N_{ponzi}(i)$, which is $1/\beta$, exceeds the slope $\alpha$ of the curve $i (N_{ponzi})$ (i.e. $\alpha \beta < 1$), does this process converge. The divergent case ($\alpha \beta >1$) is illustrated in Figure \ref{fig:Ponzi loop with divergence}. It is interesting to note that in the $\alpha \beta < 1$ there exists the possibility that a very large exogenous shock $N_0 > N_{fix}$ can be partially reversed by the system healing itself. This is explained intuitively by the existence of companies with very large resilience which even after the increase in the interest rate following the shock, remain viable (not Ponzi). Such companies even when forced momentarily into failure by the external shock would return to paying their interest as soon as the external shock is absorbed.}
\label{fig:Ponzi loop with convergence}
\end{figure}

On a double logarithmic scale in Figure \ref{fig:convergence} (where $N$ is represented on the $x$ axis, and $i$ on the $y$ axis) $N(i)$, as defined by Eq. \ref{default of i}, is represented  by a straight line of slope $1/ \beta$. 
In this section we assume for simplicity that a company defaults as soon as it becomes ponzi $r(n)<i$.\footnote{ In the next sections we will significantly modify this extreme assumption.}   
In the Minsky scenario, these $N_t$ defaults will cause credit to shrink and consequently the interest rate, $i_{t+1}$, to go up. The following reasons lead us to expect this effect:
\begin{itemize}
\item[1] Banks will get increasingly worried about lending, because of the increasing danger of companies failing. 
\item[2] The debt left un-served by those companies that failed is now leaving their creditors short of cash. Consequently the creditors in turn may not be able to pay their own debts.
\item[3] The liquidation of the collaterals used to guarantee the failed companies' debt will lead to a devaluation of the value of similar guarantees held or posted by other companies in the system \cite{Delli Gatti 2008}.
\end{itemize}

For definiteness we assume that the increase in the interest rate, $i$, will depend on the current number of defaults as a power law. More precisely, if the number of ponzis  at time $t$ is $N_t$ this will induce an interest rate:
\begin{equation}
\label{i of default}
i_{t+1}=i_0  N_t^{\alpha}.
\end{equation}
On a double logarithmic scale in Figure \ref{fig:convergence} the function $i_{t+1} (N_t)$ is represented by a straight line of slope $\alpha$.
Thus, as in the illustration of the previous iterative process Eq. \ref{fig:Walras Conv}, we have in Figure \ref{fig:convergence} on the same graph two lines with different roles: 
\begin{itemize}
\item[-]  the line representing Eq. \ref{default of i} is always and only used to obtain $N_t$ for a given $i_t$ while 
\item[-] the line representing Eq. \ref{i of default} is to be used only to obtain  $i_{t+1}$ for a given $N_t$.  
\end{itemize}

As in the case of Figure \ref{fig:Walras Conv}, this is indicated by the arrows in Figure \ref{fig:convergence}:
\begin{itemize}
\item[-]  horizontal arrows will be used to obtain $N_t$ for a given $i_t$ while 
\item[-]  vertical arrows will be used to obtain $i_{t+1}$ for a given $N_t$.  
\end{itemize}
The initial conditions for the iterative process are characterized by:
\begin{itemize}
\item[-]  the state of the system before the shock as expressed by the initial interest rate $i_0$, and
\item[-]  the strength of the shock as expressed by the number of companies $N_0$ initially knocked down into failure by it.
\end{itemize}

Following the Eqs. \ref{default of i} and \ref{i of default}, and assuming the occurrence of an exogenous shock producing an initial number of ponzi  $N_0$ companies, at a given initial interest rate $i_0$, the unit iteration cycle is:
\begin{equation}
\label{eq15}
N_t \xrightarrow{Eq. \ref{i of default}} i_{t+1} \xrightarrow{Eq. \ref{default of i}} N_{t+1}.
\end{equation}
One can further represent the entire iterative process, where the given initial shock of $N_0$ ponzi companies triggers a chain reaction:
\begin{equation}
\label{process loan diverge}
N_0  \xrightarrow{Eq. \ref{i of default}} i_1 \xrightarrow{Eq. \ref{default of i}} N_1 \xrightarrow{Eq. \ref{i of default}} i_{2} \cdots N_t  \xrightarrow{Eq. \ref{i of default}} i_{t+1} \xrightarrow{Eq. \ref{default of i}} N_{t+1} \cdots
\end{equation}

The main questions that such iterative process poses are: 
\begin{itemize}
\item[-] is the process leading to an increasing sequence of $i_t$ and $N_t$ 
 (corresponding to a crisis)
 and if so, 
\item[-] is the increasing sequence converging to a finite value (limited `mini'-crisis) or diverging to a systemic crisis (Minsky financial accelerator unleashed)?
\end{itemize}

Thus the parameter ranges of stability vs. crisis are determined by the initial values   $(i_0$ and $N_0 )$ 
and especially their position with respect to the fixed points where curves $N_{ponzi}(i)$ and  $i(N_{ponzi})$ intersect. In the present section (non-network case), there is only one intersection: the common solution of the Eqs. \ref{default of i} and \ref{i of default} after imposing stationarity $i_{t+1}= i_t$:
\begin{equation}
\label{eq17}
N_{fix}=({i_0} / {k})^{\beta/(1-\alpha \beta)}
\end{equation}
\begin{equation}
\label{eq18}
i_{fix}=\left(i_0 / k^{\alpha \beta} \right)^{1/(1-\alpha \beta)}.
\end{equation}

\begin{figure}
\centering
\subfigure[The top-down / bottom-up feedback in the Minsky accelerator.
The macroscopic state of the system in terms of the credit availability,  parameterized by the interest rate $i$.  The states of the individual companies are characterized by their ponzi or non-ponzi status. The top down part of the Minsky accelerator feedback loop is related to the Eq. \ref{ponzi def}: an increase in $i$ will switch the status of some of the companies from non-ponzi to ponzi. In turn, this increase in the number of ponzi companies may induce by the bottom-up part of the Minsky accelerator loop a further increase in the interest rate (Eq. \ref{default of i}).]{\includegraphics[scale=0.3]{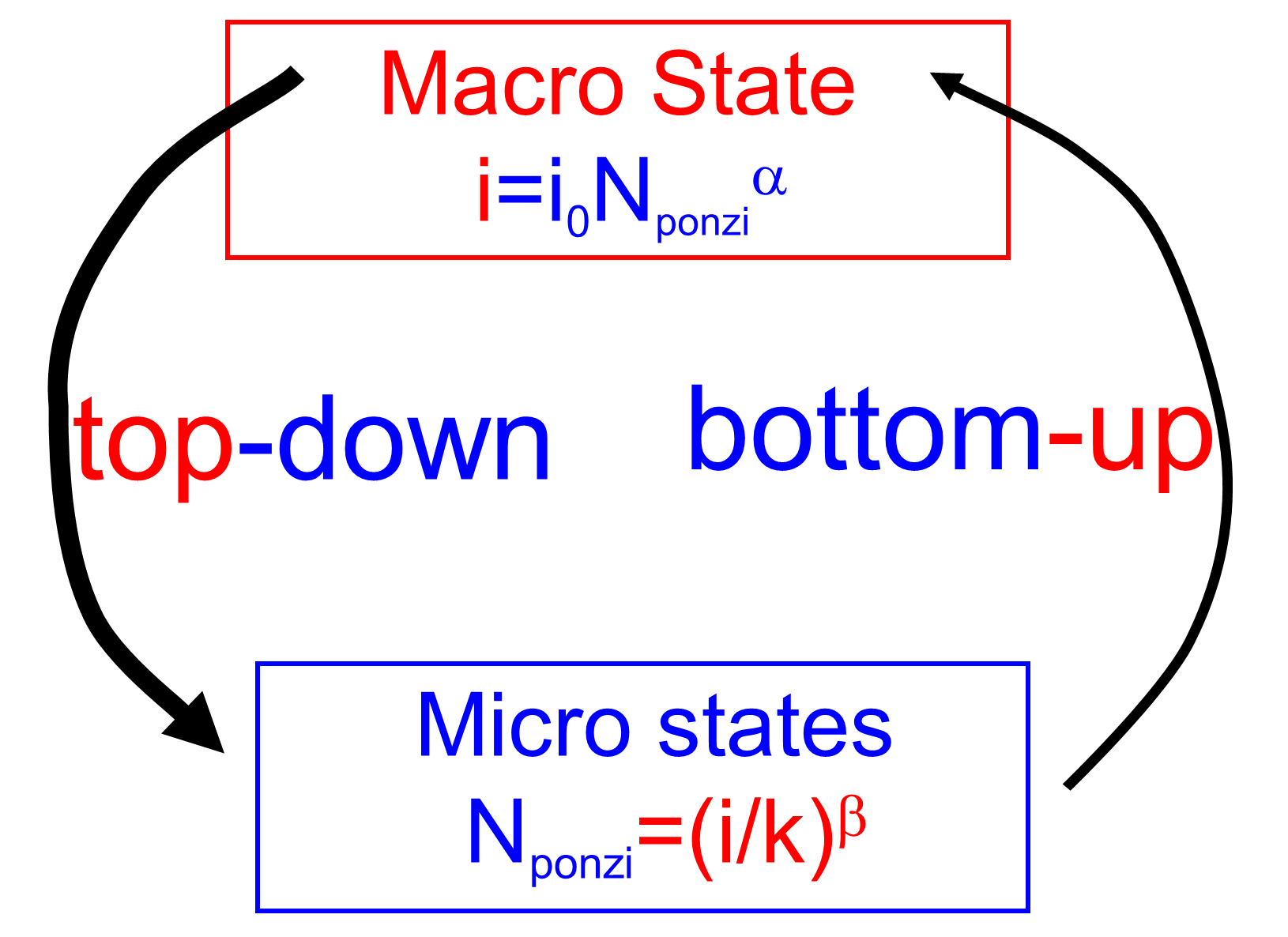} \label{fig:Ponzi Loop}}
\hfil
\subfigure[ The graph shows the coevolution between the interest rate $(i)$ and the number of ponzi companies / failures $(N)$ in the Minsky accelerator. The slope of the red solid line $(\alpha)$ is steeper than the slope of the blue dashed line $(1/\beta)$, which means that in each iteration step the number of ponzi $N$ and the interest rate $i$ diverge from the fixed point (graphically represented by the crossing of the blue and the red lines). Whether $N$ and $i$ will have a positive increment during their coevolution  (which would lead to a systemic crisis) or a negative one (stable system), it depends on whether $N_0 > N_{fix}$ or the opposite. ]{\includegraphics[scale=0.3]{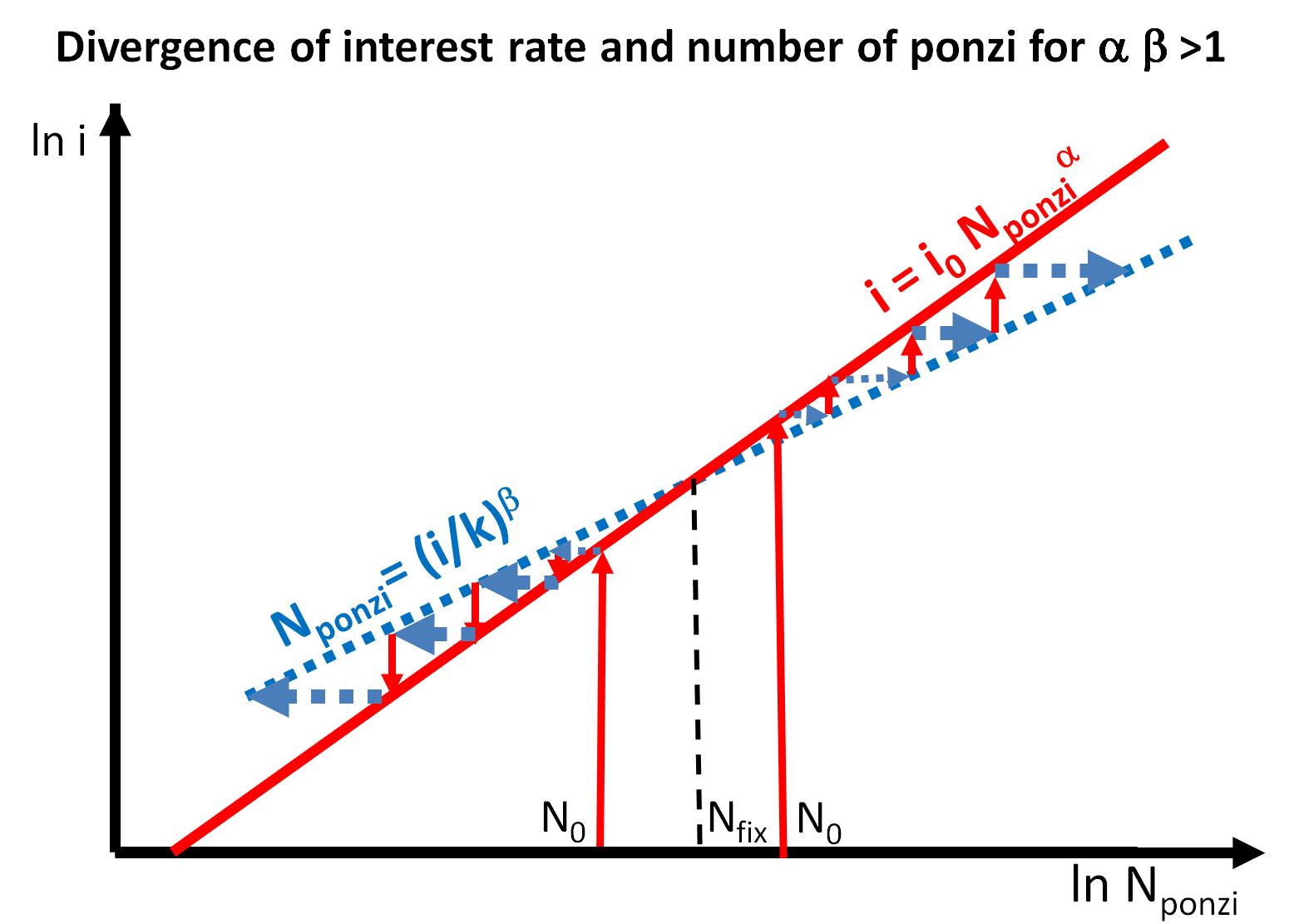} \label{fig:divergence}}
\caption{\small Illustration of the Minsky accelerator loop which assumes that $\alpha \beta >1$ and as a result of it, the iteration process \ref{process loan diverge}, graphically represented in Figure \ref{fig:divergence}, diverges. For $\alpha \beta >1$ and the initial number of failures $N_0>N_{fix}$ the process leads to a macroscopic chain of  failures:  the Minsky crisis accelerator.\\
Note that while passing from the case $\alpha \beta <1$ to the case $\alpha \beta >1$ has dramatic consequences for the system, in terms of the basic model assumptions the differences between the 2 cases are minor: a slight increase in the dependence of the interest rate on the number of failures or a slight change in the distribution of resiliences within the system can throw the system from the converging regime $\alpha \beta <1$ to the Minsky Instability Accelerator regime $\alpha \beta >1$. }
\label{fig:Ponzi loop with divergence}
\end{figure}

As detailed in the Appendix \ref{sec:appendix2}, the time evolution of the quantity of loans in the process  Eqs. \ref{default of i}, \ref{i of default}, \ref{process loan diverge} shown in Figures \ref{fig:Ponzi Loop}, \ref{fig:convergence}  is represented mathematically by:

\begin{equation}
\label{N of t defaults}
N_t =  N_{fix} [{N_0}/{N_{fix}}]^{{( \alpha \beta )}^t}
\end{equation}

This means that in general for $\alpha \beta < 1$ the exponent ${{( \alpha \beta )}^t} \rightarrow 0$ for $t \rightarrow \infty$ and  the fixed point \ref{eq17},  \ref{eq18} is stable: 
\begin{itemize}
\item[-] starting with a smaller number of ponzi companies, $N_0 < N_{fix} $, will lead to a limited crisis that will stop at $(N_{fix}, i_{fix})$. 
\item[-] starting with a severe economic state where $N_0 > N_{fix}$ the system will heal itself: $N_t$ and $i_t$ will iteratively shrink reaching eventually the same stable point  $(i_{fix}, N_{fix})$. 
This assumes that as soon as the interest rate falls sufficiently to make a ponzi company non-ponzi any adverse systemic effects that its ponzi status had (say, on the interest rate) are erased. 
\end{itemize}

For $\alpha \beta > 1$, (illustrated in Figure \ref{fig:divergence}) and visualized conceptually in Figure \ref{fig:Ponzi Loop}  the situation reverses:
the exponent  ${{( \alpha \beta )}^t} \rightarrow \infty$ for $t \rightarrow \infty$ and the fixed point \ref{eq17},  \ref{eq18} is unstable:
\begin{itemize}
\item[-] starting below the fixed point $N_0 <  N_{fix}$ will lead to further shrinking in the number of ponzi companies and in the interest rate.
\item[-] if the system is brought (endogenously or exogenously) above the fixed point $N_0 > N_{fix}$ it will enter in a Minsky financial accelerator and $(N_{t}, i_{t})$ will further increase until complete economic collapse or until exogenous measures -- e.g. forcing exogenously the interest rate, $i$, (and / or the failures) down -- stop the process. 
\end{itemize}

For instance, central banks in major countries can and have forced down the interest rate (and expanded inter-bank credit availability) artificially by creating fiat money and lending it at very low interest rate to banks. That may break (or even reverse) the feedback loop by forcing the red resilience trajectory below the resilience line of companies and borrowers. The Fed did it immediately while the European Central Bank (ECB) did it eventually, under the pressure of the markets on the sovereign bonds of Greece, Spain and Italy.

This section has dealt with the non-network view of the Minsky lending and crisis accelerators.
We are now turning towards casting the basic Minsky ideas into a more realistic agent based network model.
The main new ingredient will be that not all ponzi companies will default or fail.
Rather the ponzi condition will make a company susceptible of being recognized as such and denied credit.
The actual switch from the susceptible to the failed status will depend on the interactions of each ponzi with its trade partners. 
More specifically we will assume that a ponzi company will openly fail only when one of the companies directly connected to it (by a network link) defaults. 
To prepare this synthesis we first describe in the next section the propagation of distress by contagion on a network.
We will follow therein the formalism of social and market percolation as described in [Solomon 2000] and [Goldenberg 2000]. 

\section{Propagation of distress by contagion. Crisis Percolation across networks of companies}
\label{sec:Percolation}
\subsection{The relevance of contagion and percolation in financial networks }
\label{subsec:Percolation intro}

In order to express formally the feedback loops that enforce the propagation of distress between companies we will use the mathematical concept of percolation in a widened -- dynamical -- sense, better adapted to Minsky's non-equilibrium thought. In its usual sense, the mathematical term of {\bf percolation} describes the conditions in which a  population of `nodes' connected by binary links is capable of forming macroscopic connected clusters. However, in the present paper we will use a more dynamical version of this concept, as has been introduced under the names of social percolation in \cite{Solomon 2000} and market percolation in \cite{Goldenberg 2000}, \cite{Jager 2011a}, \cite{Jager 2011b}, \cite{Jager 2010}. In this version, the stress is less on the question of the static existence of macroscopic clusters, but more on the `contagion' process that sweeps across the population, creating  macroscopic clusters of `contaminated' agents. This allows, in turn, feedback between the growth of the contaminated cluster and processes that it influences \cite{Solomon 2000} \cite{Cantono 2010}, \cite{Cantono 2012}, \cite{Kindler 2013} which can, in their turn, feedback on the growth of the cluster. 

Percolation models are agent based models in which the agents are the nodes of a network and their interactions are the contagion via the network edges. Contrary to common belief, the computer simulation of the evolution of the individual states of the agents is neither the only nor the most illuminating way to extract information about such systems. Quite to the contrary, the formula in Eq. \ref{Nfailed of rho}, which we deduce below in a toy setting, and which predicts the size of the contagion avalanche as a function of the density of susceptible agents, is only the simplest example of a wealth of analytic results that are not only more precise but also more informative than the direct simulation of the system. 

We start by describing the `market percolation' mechanism using the `ponzi' concept introduced in Section \ref{subsec:Ponzi}; this will be relevant for the network extension of the Minsky accelerator.

In the simplest non-network Minsky model, we considered a situation in which any company that becomes ponzi is immediately identified as such by the economic system and loses its capability to get further loans. In the absence of credit, such a company becomes incapable of continuing its normal activity and so, until the conditions change (e.g. new funds or better loan conditions become available), it has to freeze its activities. The model dynamics consisted then on companies either becoming ponzi (i.e. failing) or recovering from being ponzi. Once a company is ponzi, its failure was considered immediately known to the entire system and the consequences of its status were immediately enforced. In this sense the model adopted the neoclassical assumption that all the agents in the system have perfect and immediate knowledge of the system state, including the state (failed or not) of all other agents.

However, reality is often different. Consequently, we will consider here a different kind of situation and its model, in which the exact financial positions of companies is not known to all, especially not to potential creditors, and a ponzi company would fail openly only if a specific event uncovers (highlights or brings public attention to) its problems. For instance, it took the distress in much of its environment to uncover the fact that the Madoff company was ponzi and so to trigger its failure, even though it had been in the ponzi position for many years previously.

One may think of many ways by which a ponzi company may openly fail, but definitely its immediate trade and credit associates are very likely to have an important role in this. For instance, as long as the company's banks or suppliers are in very good shape, they will not make problems in financing the company's further purchases, operations and investments. Only when the bank notices either that the company's clients are themselves in trouble or have failed or that one of its suppliers is not in a position (due to its own distress) to provide supplies on credit, may the bank start have a closer look at the company and, noticing its ponzi status, stop its line of credit.  Only then will the ponzi company start failing to meet its obligations and thus its distress will become public knowledge. Reciprocally, only then will its ponzi status have a negative impact  upon its environment (e.g. contributing to the crunch in the availability of credit, an increase in the interest rate etc.). In particular, other companies, including banks, in its environment will be more reluctant to lend to one another and may well ask higher interest when they do.

It thus becomes clear that the mechanisms by which ponzi companies are known openly to have `failed' rather than only potentially in trouble, are contagion-like mechanisms \cite{Solomon 2000}, \cite{Weisbuch 2000}.
Once this is recognized, the theory of percolation becomes relevant \cite{Stauffer 1985}.
In particular, the contagion rules that were used in \cite{Solomon 2000}, \cite{Weisbuch 2000}, \cite{Stauffer 1985} can be re-expressed in the Minsky context as follows:
\begin{itemize}
\item[-] A ponzi (`susceptible') company that is connected by either business or credit ties only with non-ponzi (hedge or speculative) partners will be safe against failure. 
\item[-] On the contrary, a ponzi company that is part of a large connected cluster of ponzi companies (`ponzi percolation cluster' from now on)
 will be very exposed: once one of the members of the cluster fails (say by an exogenous shock, or an internal whistle-blower), its ponzi partners will be uncovered too and then their partners' partners and so forth. The cascade will eventually reach the entire cluster, including this company.
\end{itemize}
Thus, in such a model where recognizing the `failed' status of ponzi companies depends on a mechanism of contagion by partners, the ponzi percolation clusters are a measure of the size of the crises that individual events can trigger.

Fortunately, the dependence of the sizes of connected cluster on the density of susceptible (ponzi) nodes in a network has been studied extensively for the last 40 years by statistical physicists and mathematicians. In particular, in a very wide range of network geometries there exists a critical density $\rho_{C}$ of ponzi companies, which separates the range where the clusters are all microscopically localized from the range in which a giant cluster spans across the entire system. Thus, for a low density (fraction of the population) of ponzis, the ponzi clusters have small sizes and the contagion avalanches will be limited in size. By increasing the number (and thus the density) of ponzi companies, these clusters gradually become larger and the gaps between them filled. 

\begin{figure}
\centering
\subfigure[The evolution of the contagion avalanche in the case where the noise circle (marked by a bold blue circle surrounding it) is not ponzi. The circles representing ponzi units are filled (with yellow). 
Each cluster of ponzi nodes is marked by a line sourounding it.
In this particular case only the indicated seven red circles  (each marked inside with the time of its contagion) fail and the rate of contamination is slow: one agent per time period. At the end of the process not only most of the agents but also most of the ponzi have not failed. ]{\includegraphics[scale=0.35]{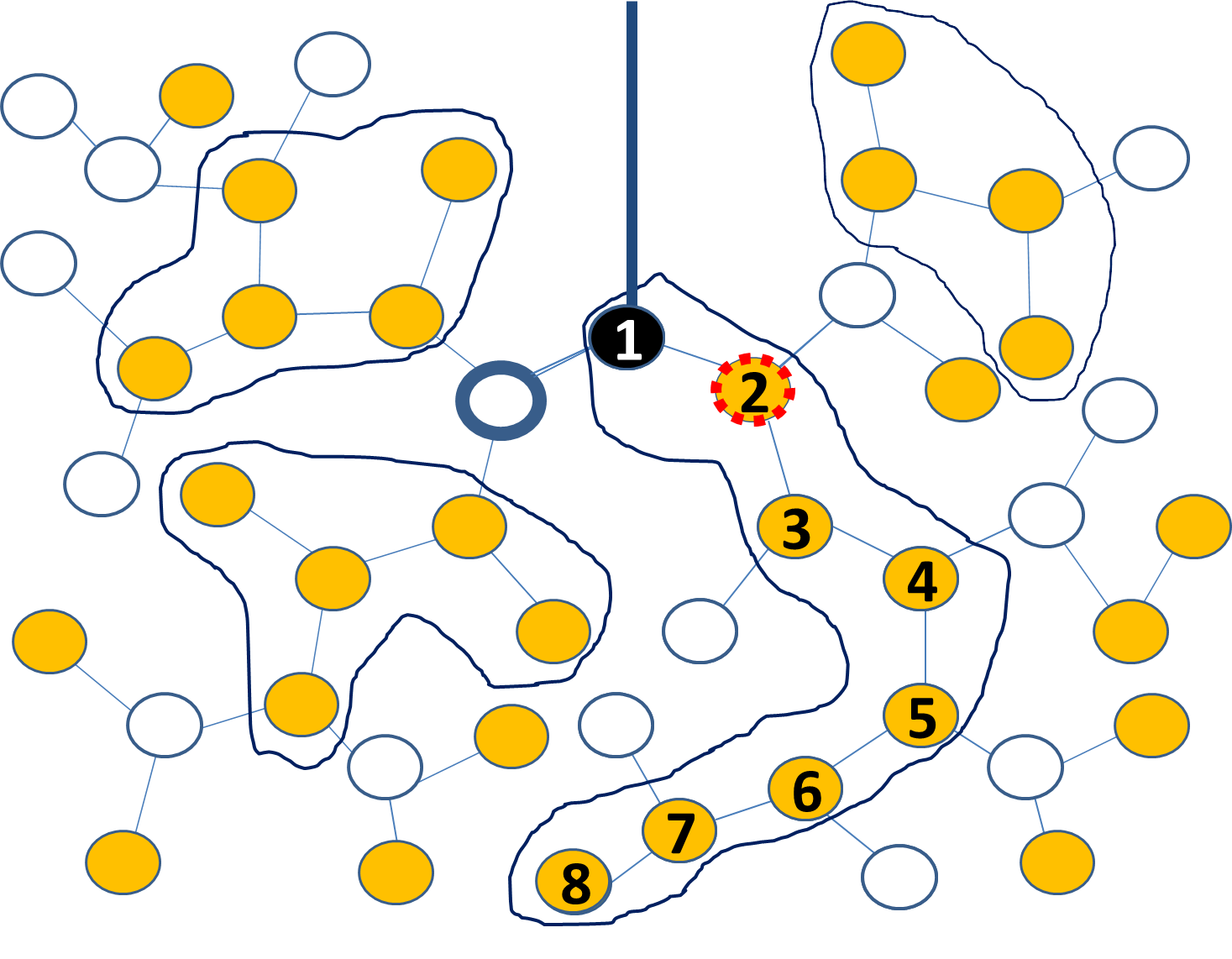} \label{fig:net1}}
\hfil
\subfigure[This time we assume that the `noise circle' (surrounded by the bold blue circle) is ponzi.
Consequently, the ponzi clusters change significantly: three of them are united in one single cluster.
This time, the evolution of the contagion avalanche starting from the node marked by 1 leads to a failure avalanche involving a
large fraction of the ponzi. One sees that the rate of contamination shows large time fluctuations. More specifically, the time series representing the number of contaminations at each time step is now 2,3,5,5,1,1,1,0.]{\includegraphics[scale=0.35]{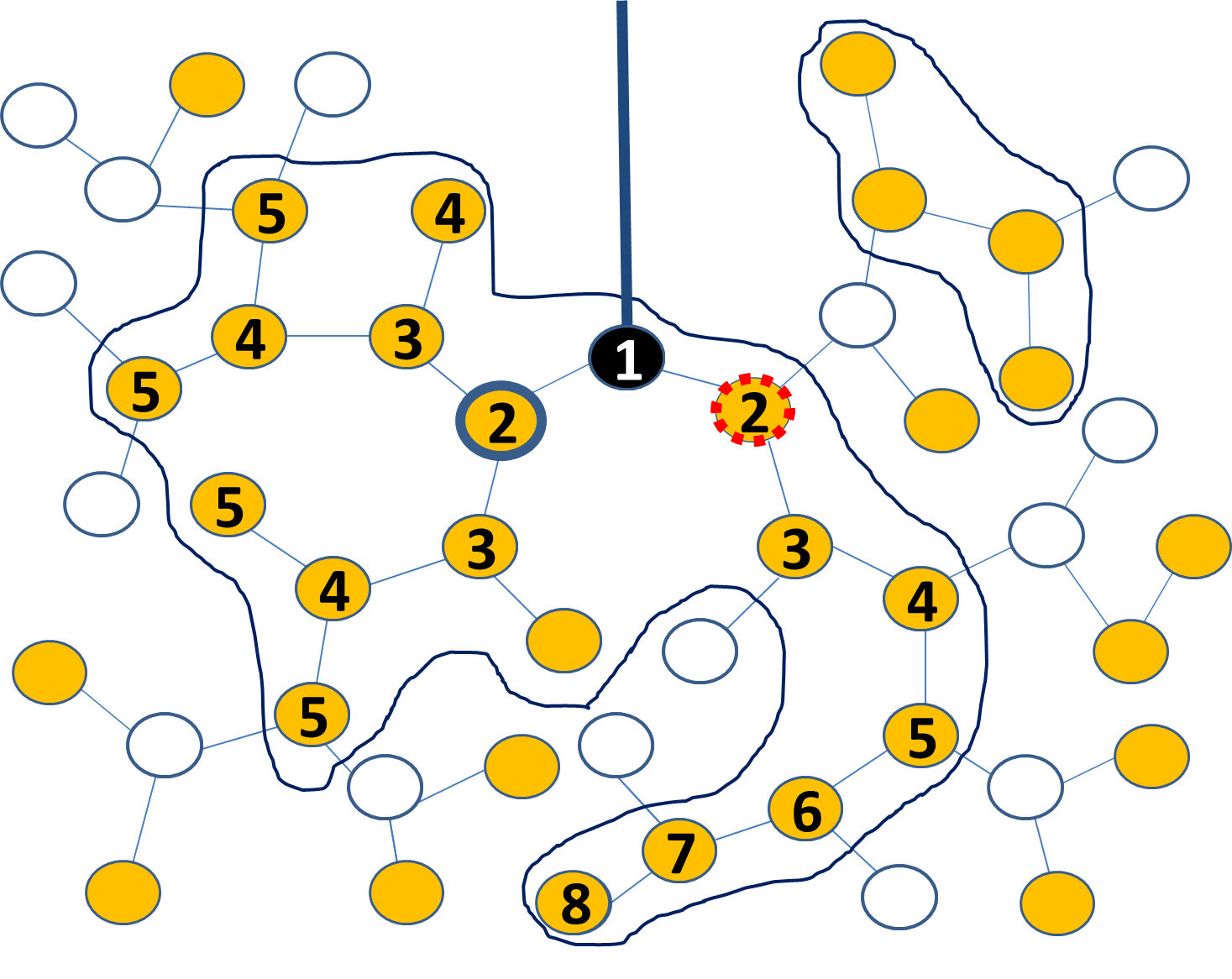} \label{fig:net2}}
\caption{\small Figures \ref{fig:net1} and \ref{fig:net2} reveal the emergence of macroscopic fluctuations around the critical point of a percolation phase transition. The ponzi companies are marked by full circles while the non-ponzi are marked by empty circles.
The ponzi clusters are shown surrounded by frontiers to facilitate their identification as disjoint one from the other.
One assumes that at time 1 the node marked by 1 fails. A failures contagion avalanche is initiated. 
The contagion process cannot cross the cluster frontiers.
To visualize the time history of the contagion avalanche, inside the circle representing each contaminated node the time step at which it failed is shown. The small difference between these figures demonstrates that there can be a large difference between very similar systems. Note that in the case when the second neighbor (surrounded by an interrupted circle) of the initially failed ponzi (marked by 1) would also not be ponzi, then the process would stop as soon as it, starts without leaving any significant trace. This figure is an illustration of the very steep character of the percolation transition.
}
\label{fig:Nat}
\end{figure}

Thus, the most relevant property of the percolation model is the existence of a phase transition: as the density of ponzi companies, $\rho_{ponzi}$, comes close to the critical density, $\rho_{C}$, even a minute increase in $\rho_{ponzi}$ can lead to dramatic increase in the cluster and thus avalanche size. The smallest noise can cause the small clusters to fuse into a giant cluster that has the length and width of the size of the entire system. 
Figure \ref{fig:Nat} shows how a noise consisting of making one single company to be ponzi can completely change the situation and create, from a few small disconnected clusters, a giant cluster that contains most of the ponzi companies in the system. The main point is that around the critical density, $\rho_{C}$, of ponzi companies the dynamics is very sensitive to small changes in their density, $\rho_{ponzi}$, within the population. This will have crucial implications for the Minsky accelerator acting on networks of companies.

A process  like Minsky's, rather than affecting locally a small number of ponzis, acquires systemic size and, consequently, the capability of eliciting significant feedback from the whole system. For instance, the dynamics introduced in \cite{Solomon 2000}, \cite{Weisbuch 2000} can be interpreted as assuming that, after a large contagion avalanche of failures, the macroeconomic policy may be tuned so as to suppress the number of ponzis  companies which are the ones susceptible to contagion. This lead in \cite{Solomon 2000}, \cite{Weisbuch 2000} to the system self-organizing  and self-tuning itself to a state that keeps $\rho_{ponzi}$ close to the critical density of ponzi companies, $\rho_{C}$. According to Eq. \ref{Nfailed of rho}, in this range of $\rho_{ponzi}$, the size of the largest cluster changes between finite to infinite (system size) values, with the consequence that the dynamics is very sensitive to the smallest change and leads to macroscopic fluctuations. Thus, paradoxically, an effort to limit instability by reacting to past large crises can sometimes lead, if it is not carried out to the last implications \cite{Solomon 2000}, \cite{Weisbuch 2000}, to a regime where large fluctuations become the rule.
In the present paper the source of instability will be different but with similar effects: the reaction of the system to
the failures of ponzi companies will be to reduce the availability of credit generally, in particular by increasing the interest rate.
This will increase the density of ponzis and thus the size of the default avalanches. 
Thus, instead of the system stabilizing, its reaction to defaults will only keep increasing the number of ponzi companies and thus the number of failures.

\subsection{The Market Percolation transition to Systemic Crisis}
\label{subsec:Predictions}
The idea of applying percolation to social and market contagion \cite{Solomon 2000}, \cite{Goldenberg 2000}, introduced in the previous section, has since been exploited in several different contexts (see \cite{Zeppini 2013} for a review): immunization of communication networks \cite{Goldenberg 2005}, advertising  and product acceptance \cite{Yaari 2006},  interacting networks (optimism-pessimism influences vs. saving ratios interactions) \cite{Erez 2005}, negative word of mouth \cite{Erez 2004}. Moreover, as in \cite{Weisbuch 2000} and \cite{Vega-Redondo 2007}, one can consider the epidemiological case in which the contaminated, in this case infected, nodes recover automatically after some time (SIR). It was shown in \cite{Weisbuch 2000} that, in the presence of top-down reaction, such SIR systems can reach a self-organized critical phase with collective phenomena that lead in time to macroscopic long range correlations and fractal spatial structures. Again, while the numerical simulation of such systems is possible and useful, the understanding and ascertaining of the phase transition and scaling properties of the models is highly dependent on an analytical treatment of the type we demonstrate below (Eqs. \ref{eq5}--\ref{eq8}) in the simplest of settings.

We will start with the simplest model and gradually introduce additional feedback loops and study their effects. We will then use, in the following section, these iterative contagion feedback loops to explain the propagation of defaults during an economic crisis following a Minsky moment.

Let's consider, as the first example, an inter-company network as shown in Figure \ref{fig:net1}, where each node has exactly $K$ connections and there are no loops. The ponzi units are shown as full circles. The fragility of the ponzi companies is represented in this simplest model by the fact that they fail by `contagion' as soon as one of their connections fails. Starting from this rule, one can predict how failures propagate across the network. More sophisticated hypotheses concerning network geometry may be developed. For instance, it is likely that in order to destabilize a company, a significant share of its trade has to be affected rather than just one of the partners being in default. A model that captures this has been studied theoretically in \cite{Kindler 2013}. For the sake of clarity, we discuss here the simplest model.

Assume that at time 1 one starts with just one node, marked by 1 in Figure \ref{fig:net1}, failing. This could happen exogenously or by contamination from its neighbor (from `above' as shown in Figure \ref{fig:net1}). This brings the number of new failed companies at time $t=1$ to $N(1)=1$.  How many failures will follow as a consequence of this trigger? An illustration of the process is given in Figure \ref{fig:net1}.  The ponzis which fail are marked by the time step of their failure.  After eight time steps, eight companies failed. Note that not all the ponzi companies fail by the end of the process, only the ones belonging to the same cluster in which the trigger occurred. Also note that in the Figure \ref{fig:net1} the rate of contagion is one ponzi per time step.

Statistically, if the density of ponzi companies is $\rho_{ponzi}$, and out of the $K$ neighbors of the first failed node, one neighbor has already failed, then the expected number of new failed companies in the next time step $t=2$ will be:
\begin{equation}
\label{eq5}
N(2)=(K-1)\rho_{ponzi}
\end{equation}
In the next time step, each of these $(K-1)\rho_{ponzi}$ companies will contaminate another $(K-1)\rho_{ponzi}$ number of companies.  Extending this iterative process to $t$ time steps, the total number of failed companies by time $t$ becomes  $N_{failed}=N(1)+N(2)+ \cdots +N(t)$, which is the series:
\begin{equation}
\label{eq6}
N_{failed}(t)=1+(K-1)\rho_{ponzi}+[(K-1)\rho_{ponzi}]^2+\cdots+[(K-1)\rho_{ponzi} ]^{(t-1)}.
\end{equation}
The number of failed companies after $t$ time steps is, therefore, obtained by summing Eq. \ref{eq6}:
\begin{equation}
\label{eq7}
N_{failed}(t)=([(K-1)\rho_{ponzi}]^{t}-1)/([(K-1)\rho_{ponzi}]-1).
\end{equation}

\begin{figure}
\centering
\includegraphics[scale=0.45]{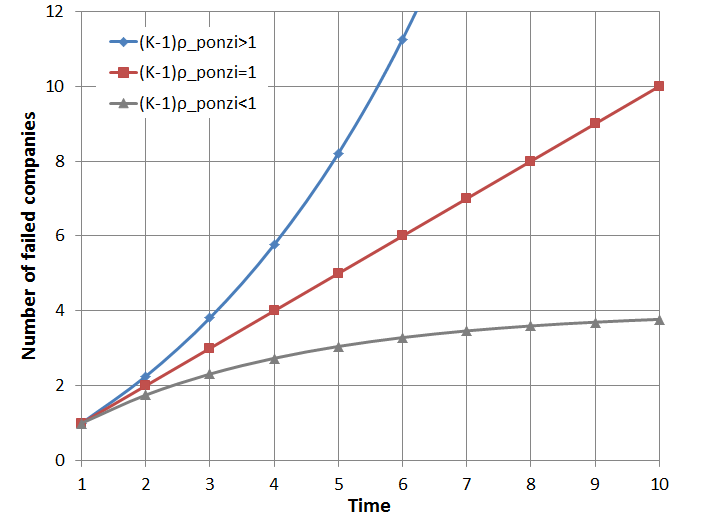}
\caption{\small The number of failed ponzi units as a function of time according to Eq. \ref{eq7}. The red (squares) line is for $(K-1)\rho_{ponzi}=1$, the blue (diamonds) line for $(K-1)\rho_{ponzi}>1$ and the grey (triangles) line for $(K-1)\rho_{ponzi}<1$.}
\label{fig:Krho}
\end{figure}

To understand the `failure avalanche', or domino process, that the iterative contagion of neighbors implies, let us consider first the special case where $(K-1)\rho_{ponzi}=1$. Then, each term in Eq. \ref{eq6} equals 1 and at each time step there is one more ponzi that fails.  This means that the number of failed companies grows linearly in time. This is a limiting case represented by the red straight line in Figure \ref{fig:Krho}.

If $(K-1)\rho_{ponzi}>1$, then the number of failed companies will grow exponentially to infinity (blue (diamonds) line in Figure \ref{fig:Krho}). If however $(K-1)\rho_{ponzi}<1$, then the process saturates (grey (triangles)  line in Figure \ref{fig:Krho}) and the amount in Eq. \ref{eq7} is a sum of a convergent series even for time $t \rightarrow \infty$:
\begin{equation}
\label{eq8}
N_{failed}(t=\infty)=[1-(K-1)\rho_{ponzi}]^{-1}
\end{equation}
This is plotted for a few values of $\rho_{ponzi}$  in Figure \ref{fig:Krho} and its economic implications are discussed below. This happens in the simple geometry of our network, where only one neighbor that fails leads to the failure of all its susceptible connected peers.

More generally speaking, when the network is more realistic, the critical value of the ponzi companies density is not  $\rho_{C}=(K-1)^{-1}$  as in Eq. \ref{eq8}, but another value which depends on the geometry of the network. Moreover the divergence `critical exponent' in Eq. \ref{eq8} is, in general, not one but another value, $\gamma$,  that depends on the network.  The generic formula for a very large class of networks (including all regular lattices, random Erdos-Renyi networks with any average number of neighbors, small-world networks etc.) which have finite average number of neighbors per node,  $\rho_{ponzi}$ close to $\rho_C$, is:
\begin{equation}
\label{Nfailed of rho}
N_{failed}= S \left[1-\frac{\rho_{ponzi}}{\rho_{C}}\right]^{-\gamma}.
\end{equation}
$S$ is a constant close to $1$ depending on the details of the geometry and of the initial conditions.
When the density of ponzi companies, $\rho_{ponzi}$, reaches the critical density, $\rho_{C}$, this formula implies that the number of failed companies jumps from a finite to an `infinite' value. This, in physics, is called a (percolation) phase transition: from the regime where the contagion causes only `microscopic' localized disruptions involving small clusters, the system switches to a regime where the contagion affects clusters of the size of the system itself. 
This result, however, holds only in the infinite system limit when the total number of agents $N_{total}$ is assumed infinite.
Of course for a finite system, the number of failed companies cannot be larger then the number of ponzies:
\begin{equation}
\label{Nponzi of rho}
N_{failed} \le  N_{ponzi} = \rho_{ponzi}N_{total}.
\end{equation}
In particular, at the critical density, $\rho_{C}$, the number of failures cannot exceed the critical number of ponzis:
\begin{equation}
\label{Ncrit of rhocrit}
N_{failed} (\rho_{C} )  \le N_{C}=\rho_{C}N_{total}.
\end{equation}
In general, the number of failed companies, $N_{failed}$, is bounded from above by the value:
\begin{equation}
\label{Nfailed above percolation}
N_{failed} < \rho_{ponzi} N_{total}.
\end{equation}
In fact this upper bound becomes a good approximation once the largest cluster is large enough to be neighbor to most of the nodes that it doesn't contain. Given its fractal properties this happens very soon after $\rho_{ponzi}$ exceeds $\rho_{C}$
and we will use this approximation in the next section:
\begin{equation}
\label{Nfail of rho}
N_{failed}= \min   \left\{ S [1- {\rho_{ponzi}} / {\rho_{C}}]^{-\gamma}, \: \rho_{ponzi} N_{total} \right\}
\end{equation}

The  analytic evaluation of the number of contaminated nodes as a function of the ponzi density is not the only percolation theoretical result that one can extract about contagion propagation among agents interacting through a network of connections. The details of the process by which the clusters coalesce as the density of ponzi increases can provide a vast amount of additional information about systemic crises.  In particular it determines the fractal geometrical properties of the clusters and impinges on the dynamics of the system in crucial ways. The intermittent fluctuations of the propagation process, both in time and among different realizations of the process, have been studied in detail theoretically and by Monte Carlo simulation [Solomon 2000, Goldenberg 2000, Hohnisch 2007]. Their implications for real market systems have been discussed in [Cantono 2010] (Figures 3--4 and 7--8). 

In fact the formula 
\ref{Nfail of rho} characterising the size of the susceptible clusters, Figure \ref{fig:net1}, can be interpreted in its wider implications: it characterizes the size of the expected fluctuations of the system in space and time, as well as between samples, as a function of the proximity of the density to the transition point $\rho_C$ 
\cite{Herrmann 1992}, \cite{Derrida 1994}, \cite{Parisi 2004}.
Thus, the size of the percolation clusters are a measure, not only of the extent of network regions affected by the contagion avalanches, but also of their space, time and inter-sample variability.

In the rest of the paper we will repeatedly use the fact that around the critical state $\rho \sim \rho_C$ 
 the system displays extreme fluctuations which are manifested in:
\begin{itemize}
\item[-] the fractal spatial distribution of the failed units, 
\item[-] the intermittent time series of the number of contagions, 
\item[-] the non-self averaging variability between different stochastic realizations of the same system.
\end{itemize}
These characteristics will provide tools for diagnostic, prediction and steering of the Minsky crisis propagation process in real systems.

Transcending the average formula \ref{Nfail of rho} and studying the details of the percolation model at the agent level is of great theoretical and practical importance;  it gives the means to evaluate properties that until now were outside the grasp of a quantitative or even conceptual treatment: 
\begin{itemize}
\item[-] what are the departures of the individual empirical realizations from the statistical average, Eq. \ref{Nfail of rho}?
\item[-] what are the time delays between the moment that the first individual agent in the giant cluster is contaminated and until the entire cluster fails?
\item[-] are there moments in the propagation process where a local intervention can stop, delay or accelerate the propagation of distress?
\item[-] what is the role of individual agents and events in determining the macroscopic fate of the system?
\item[-] can one identify those individual agents and events and prescribe an appropriate action for them?
\end{itemize}
Solomon and \cite{Weisbuch 2000}, \cite{Hohnisch 2005}, \cite{Cantono 2010} have provided examples of how to answer such questions in a few specific and generic market contexts. 
The present paper opens the way for applying those techniques also to the crisis percolation transition.

\subsection{Diversification increases Fragility }
\label{subsec:Fragility}

The analytic formula \ref{Nfail of rho}  is an example of how an agent model can be treated theoretically. While it is common throughout much of the literature to consider agents models as identical with computer simulations, we see here an example in which a large set of connected agents (nodes) have collective properties that can be evaluated analytically. Moreover, the analytical formulae allow one to identify transitions between qualitatively different parameter ranges. 

In the present case the Eq. \ref{Nfailed of rho} clearly indicates the transition, at $\rho_{ponzi}=\rho_{C}$, between the range in which the susceptible clusters (and consequently the contagion avalanches) are localized in limited regions of the network and the parameter range in which they extend arbitrarily far across the system. 
Thus, following our assumption that a ponzi company will fail if one of its neighbours fails, one can obtain the result that a number of companies (which could be a significant fraction of the entire system) will fail. Note that not necessarily all the companies will fail: while some companies become ponzi because of the financial system dynamics, some companies among the most resilient may remain non-ponzi. Furthermore, even among the ponzi, not all of them will be reached by the failure avalanche (e.g. as in Figure \ref{fig:net2}).
\begin{figure}
\centering
\includegraphics[scale=0.4]{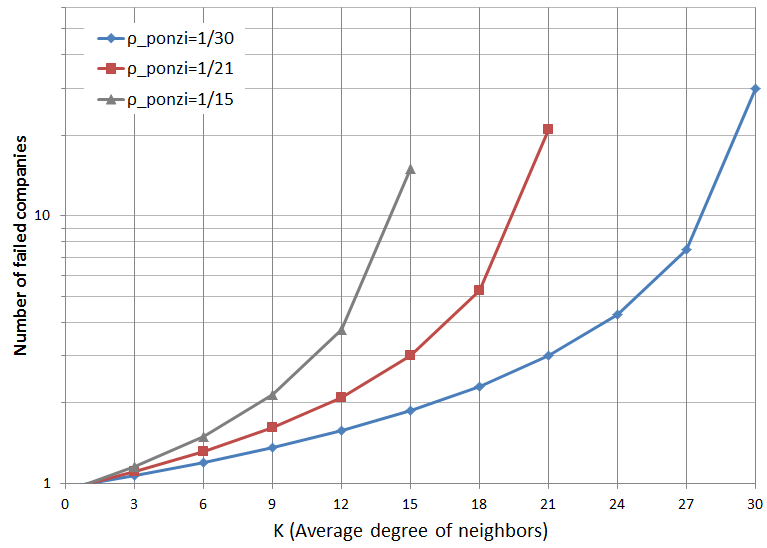}
\label{fig:NK}
\caption{\small Increase in the number of failed companies with an increase in the average degree, $K$, of neighbors in the network, for different densities of ponzi companies, $\rho_{ponzi}$, susceptible to default.}
\end{figure}

Looking at Eq. \ref{eq8} from another point of view, not through the dependence on $\rho_{ponzi}$, but through the dependence on $K$, one sees that the number of failed companies increases with $K$. So, this simple model predicts that a highly connected network (high diversification of trade debt partners) will increase the probability of systemic failure by favoring contagion avalanches and thus facilitating the crisis percolation phase transition. 
This is, of course, at variance with the mainstream economics main theme that diversification is always good \cite{Sharpe 1964}. According to the very simple model above, diversification of itself is neither good nor bad; whether it is or is not depends on the state of the economy and the company's own financial fragility. During a boom, having large $K$ will amplify these positive trends. However, in the situation when the density of ponzi, $\rho_{ponzi}$, is large, a large $K$ in combination with the large number of ponzi companies will imply the possibility that the failure chain reaction might sweep the system from one end to another. In order to discriminate between those situations where diversification is a source of instability from those where it is a stabilizing feature, one has to perform empirical measurements rather than using mere models.

From the theoretical viewpoint, one can only argue that the diversification, $K$, may have opposite effects on the individual and collective economic stability \cite{Battiston 2012}, \cite{Caccioli 2012}. Such a destabilizing effect of diversification and risk-sharing among financial actors has been discussed in the context of securitization \cite{Adrian 2008}, \cite{Stein 2011}. The processes of pooling and securitization have increased leverage which led to the present crisis. Lenders who would not have considered lending using a single subprime mortgage as collateral, when acting as a buyer of the pool of mortgages, actually borrow 70 percent of those pools' collective value \cite{Geneakoplos 2010}, and securitization took this borrowing on pools one step further by converting the loans into long-term loans. Losses by leveraged buyers of assets can cause a chain reaction when a leveraged buyer is forced to sell, which might lower the price of the assets and force another leveraged buyer to sell and so on.  
Concerning credit expansion, Adair Turner, Chairman of UK's Financial Services Authority asked in a fall of 2011 speech:
\begin{quotation}
``how far [can we] rely on traditional policy levers to ensure that either the aggregate amount of credit created or its sectoral allocation is socially optimal?  The answer I will give is not much. \ldots  We need to challenge the idea that financial innovation is axiomatically beneficial in a social as well as private opportunity sense."  Adair Turner, Credit Creation and Social Optimality, speech at Southampton University, 29.09.11. \footnote{\url{http://www.mondovisione.com/_assets/files/Credit-Creation-Social-Optimality-Southampton-Uni-20110929.pdf}}.
\end{quotation}

\section{Minsky accelerator on financial / trade network}
\label{sec:Minsky on network}
\subsection{Definition of the Minsky accelerator percolation model}
\label{subsec:Minsky Percolation def}
The simple Minsky accelerator idea of Section \ref{sec:Minsky accelerator} will in this section be combined with the simple market and social percolation idea  as applied to economic and  financial systems in Section \ref{sec:Percolation}. 

Already in \cite{Solomon 2000} the idea of making the susceptibility of individual nodes dependent on the current extent of the contagion was used to obtain the self-organization of a system in a critical state.
However the difference is that in the original social percolation paper the top-down feedback was negative, or translated in the terminology of the present article: the next interest rate $i_{t+1}$ was a decreasing function of the current number of contaminated nodes  $N_t$ i.e.:
\begin{equation}
i\left(N_{fail}(t)\right) < i\left(N_{fail}(t-1)\right) \text{  iff  } N_{fail}(t) > N_{fail}(t-1)  
\label{old}
\end{equation}

This lead to a \emph{\textbf{self-regulating feedback loop}}. Using the terminology from  the Minsky accelerator model, the self-regulating feedback loop consists of the following sequence of steps:
\begin{itemize}
\item[-] suppose some exogenous event or a endogenous fluctuation {\bf increases the interest rate from $i_{t-1}$ to $i_t > i_{t-1}$};
\item[-] this transforms into ponzi all the nodes $n$ with resiliences in the range 
$i_{t} > r(n) > i_{t-1}$;
\item[-] thus  the number of ponzi increases from 
$N_{ponzi} (t-1)$ to $N_{ponzi}(t)>N_{ponzi} (t-1)$;
\item[-] those of the new ponzi that have failed neighbors get contaminated and fail too thereby increasing the number of failures from
$N_{fail} (t-1)$ to $N_{fail} (t) >N_{fail} (t-1)$,
\item[-] but according to the assumption of model Eq. \ref{old} this implies  $i(N_{fail}(t)) < i(N_{fail}(t-1))$ 
i.e. {\bf the original increase in the interest rate  $i_t > i_{t-1}$ causes a decrease in the interest rate: $i_{t+1} < i_t$ closing the self-regulating loop}.  
\end{itemize}

This negative feedback loop caused the network system in \cite{Solomon 2000} to self-organize in the critical state: 
 $N_{ponzi}(t) \sim N_{C}$.
To see this assume that:
\begin{itemize}
\item[-] due to some fluctuation or perturbation the number of ponzi $N_{ponzi}$ decreases
 from a value above the percolation threshold $N_{ponzi}(t-1) > N_{C}$ to a number below the percolation thershold $N_{ponzi}(t) < N_{C}$.
\item[-] Then, according Eq. \ref{Nfailed of rho}, the number of failures will decrease dramatically
 from a giant cluster of the size of the system (Eq. \ref{Nfailed above percolation}) to a localized small cluster:
$N_{fail}(t) << N_{fail}(t-1)$.
\item[-] According to Eq. \ref{old}  this will cause the interest rate to increase
$ i(N_{fail}(t)) > i(N_{fail}(t-1))$,
\item[-] which would increase the number of ponzi nodes according Eq. \ref{default of i}.
\item[-] Thus, in  \cite{Solomon 2000} as long as the ponzi density was below the percolation threshold the number of 
ponzi nodes increased and vice versa: as long as the ponzi density was above the percolation threshold the number of ponzi nodes decreased.
\item[-] Therefore the regulatory loop kept the  system fluctuating around the critical point $N_{ponzi}(t) \sim N_{C}$.
\end{itemize}

In the present paper rather then the self-regulating feedback Eq. \ref{old},
we have a \emph{\textbf{positive feedback loop}} [Hohnisch 2007, Cantono 2010, Cantono 2012] between the number of failures / defaults and the interest rate, according to  Eqs. \ref{default of i} and \ref{i of default}.
Thus, the effect will not be self-organized convergence to the critical point $N_{ponzi}(t) \sim N_{C}$ but rather the possibility for instability: the Minsky instability.

In the present section we will study in detail the conditions in which this destabilizing scenario can take place.

Suppose that due to some internal fluctuation or some external perturbation one or more clusters of companies fail.
This will have two effects corresponding respectively to the phenomena introduced in Section \ref{sec:Minsky accelerator} and in Section \ref{sec:Percolation}:
\begin{itemize}
\item[-] the interest rate will increase cf. Eq. \ref{i of default}:
\begin{equation}
\label{i of N perc}
i =i_{0} (N_{fail})^{\alpha} 
\end{equation}
and consequently all the nodes $n$ with resilience less then the new interest rate:
$$r(n) <i$$
will become ponzi.
According to Eq. \ref{default of i} their number will be
\begin{equation}
\label{Nponzi of i perc}
N_{ponzi}=({i} / {k})^{\beta}.
\end{equation}
\item[-] Among those ponzi companies, cf. Section \ref{sec:Percolation}, only those ponzi nodes which are connected to the failed cluster(s) will become contaminated and fail too. According to  Eq. \ref{Nfailed of rho} their number will be:
\begin{equation}
\label{Nfail of rho perc}
N_{fail}= \min   \left\{ S [1- {\rho_{ponzi}} / {\rho_{C}}]^{-\gamma}     , \:   \rho_{ponzi} N_{total} \right\}.
\end{equation}
\end{itemize}

\begin{figure}
\centering
\includegraphics[scale=0.30]{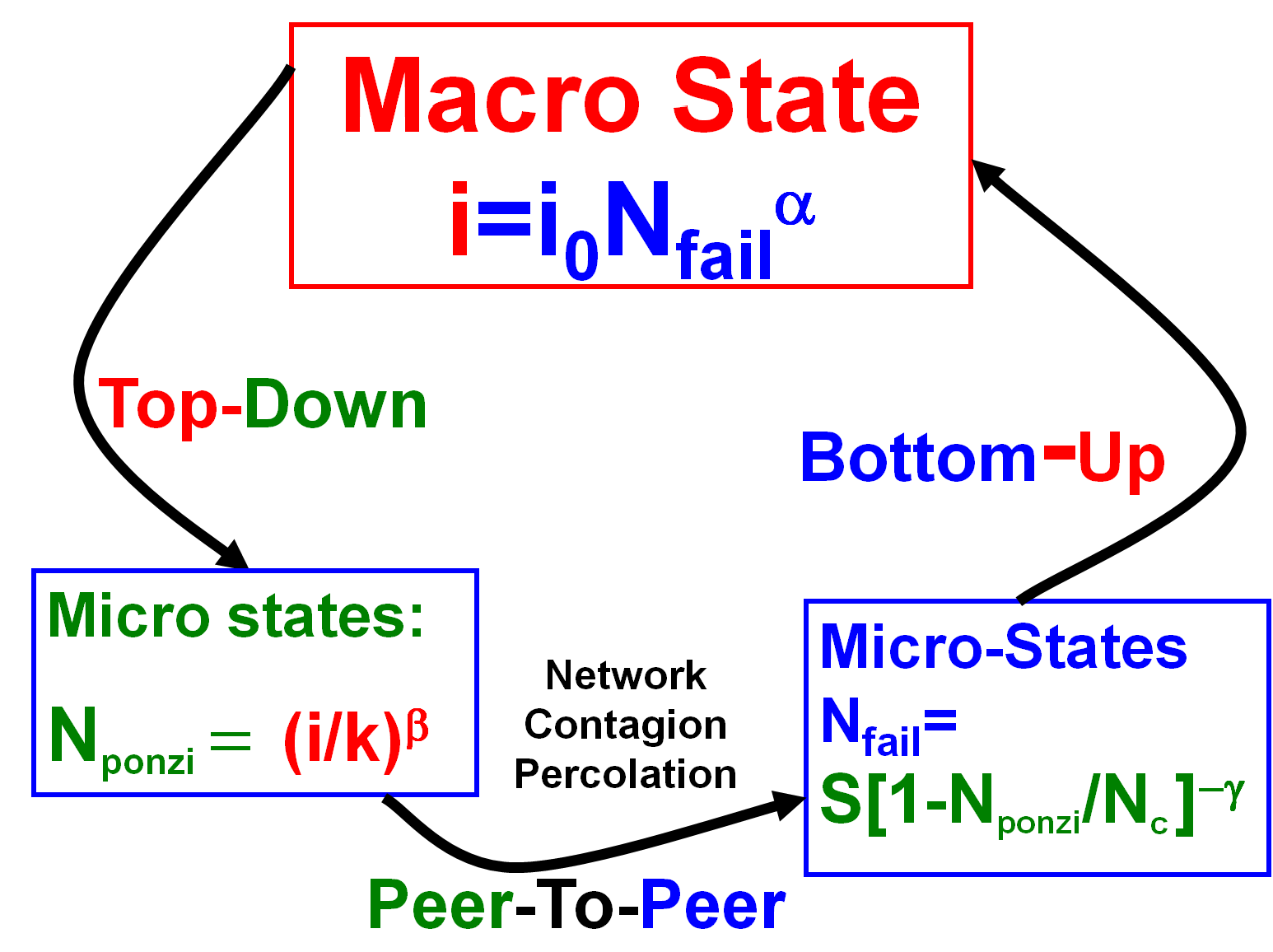}
\caption{\small The more complicated but more realistic case (defined in \ref{subsec:Minsky Percolation def}) where not all the ponzi companies fail, but only those ones that have a partner that failed them. Because of this, the feedback loop acquires an additional link: the peer-to-peer interaction through which the failure of a trade partner causes a ponzi company to actually fail and be recognized as such by the other peers and by the entire system. Only at that stage, when its status switches from ponzi ('susceptible to failure') to 'failed', does the ponzi company contribute to the increase in the interest rate through the bottom-up interaction Eq. \ref{default of i}.}
\label{fig:Loops}
\end{figure}

The conceptual structure of this {\bf top-down (Eq. \ref{Nponzi of i perc}) $\rightarrow$ peer-to-peer (Eq. \ref{Nfail of rho perc}) $\rightarrow$ bottom-up (Eq. \ref{i of N perc})} Minsky crisis accelerator is visualized in Figure \ref{fig:Loops}.

We use Eqs. \ref{Nponzi of rho}, \ref{Ncrit of rhocrit} to express $\rho_{ponzi} / \rho_{C}$ as $N_{ponzi} / N_{C}$ and
to rewrite Eq. \ref{Nfail of rho perc} as
\begin{equation*}
N_{fail}= \min   \left\{ S [1- {N_{ponzi}} / {N_{C}}]^{-\gamma}     , \:   N_{ponzi}  \right\}
\end{equation*}
This allows us to eliminate $N_{ponzi}$ using Eq. \ref{Nponzi of i perc} and reduce the system of Eqs. \ref{i of N perc},  \ref{Nponzi of i perc} and \ref{Nfail of rho perc}
 to a format similar to the one used in the previous sections of two equations iterating only the time evolution of $N_{fail}$ and $i$:
\begin{equation}
\label{N final}
N_{t}= \min \left\{ S [1- ({i_{t}} / {i_{C}})^{\beta} ]^{-\gamma}, \: ({i_t} / {k})^{\beta} \right\}
\end{equation}
and Eq. \ref{i of default}:
\begin{equation}
\label{i final}
i_{t+1}=i_{0}N_t^{\alpha}
\end{equation}
where $i_{C}$ is defined by $ N_{C}=({i_{C}} / {k})^{\beta}$. We also assume here $\alpha \beta <1$. The discussion of the less empirically relevant case $\alpha \beta>1$ is in the Appendix \ref{sec:appendix4}.

Eqs. \ref{N final} and \ref{i final} close a new autocatalytic positive feedback loop similar to Eqs. \ref{default of i} and \ref{i of default} and leading to a iterative process similar to 
\ref{process} and \ref{process increase} :

\begin{equation}
\label{process Nfail i}
N_{0}\xrightarrow{Eq. \ref{i final}} i_1 \xrightarrow{Eq. \ref{N final}} N_1 \xrightarrow{Eq. \ref{i final}} i_{2} \cdots N_t  \xrightarrow{Eq. \ref{i final}} i_{t+1} \xrightarrow{Eq. \ref{N final}} N_{t+1} \cdots
\end{equation}

\begin{figure}
\centering
\includegraphics[scale=0.30]{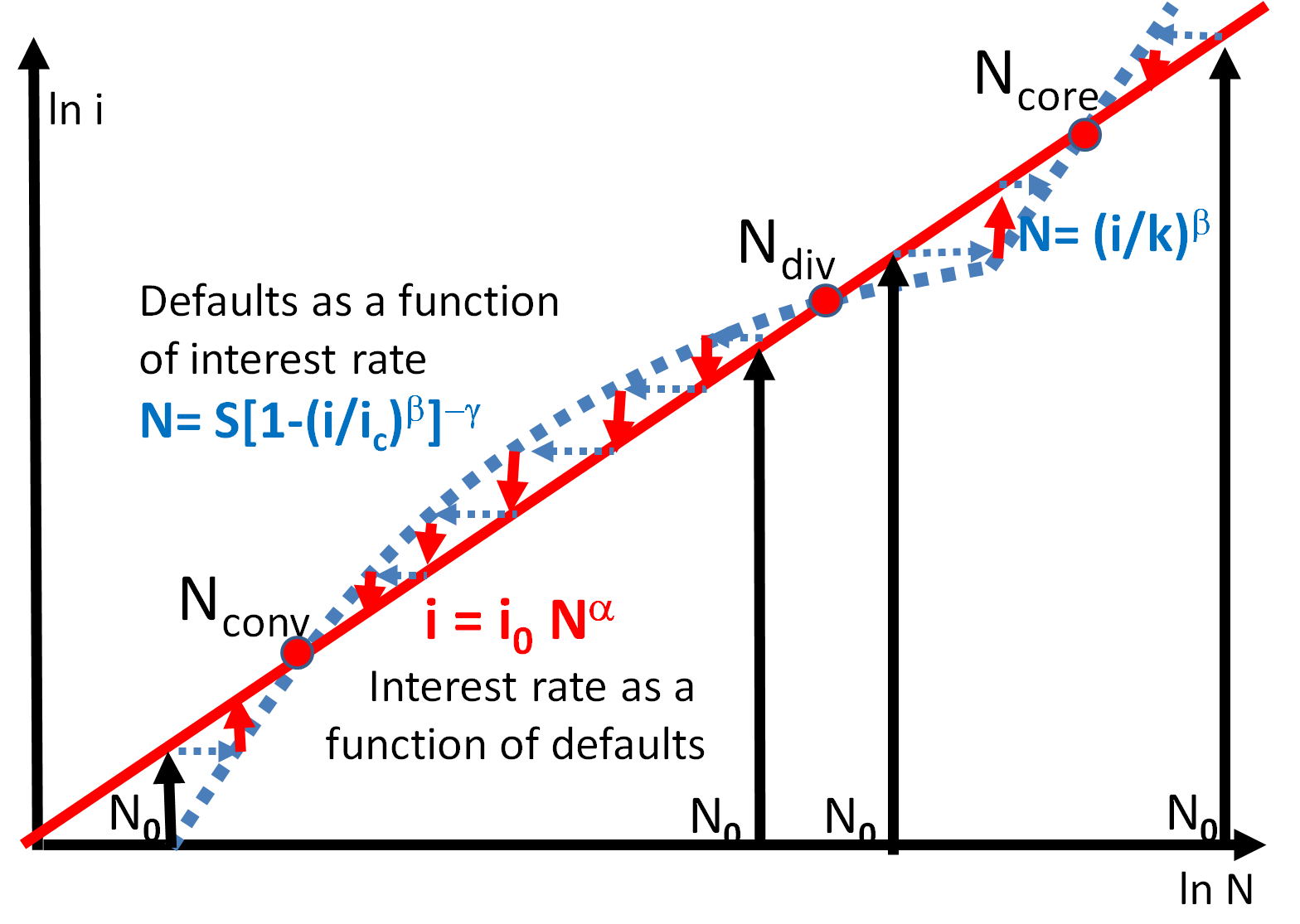}
\caption{\small 
\\ There are three possible fixed points resulting from the intersections of the curves $N(i)$ Eq. \ref{N final} and $i(N)$ Eq. \ref{i final}: $N_{conv}$, $N_{div}$, $N_{core}$. The fixed points separate 4 phases   characterized by different prospects of the iterative process \ref{process Nfail i}:
\\ $ \ \ { \ \ \ \ \ \ \ \ \ \ \ - }  $ $N_0<N_{conv}$: direction of the arrows shows an increase of failed companies. After a limited crisis, the iterative process converges to the point $N_{core}$.
\\ $ \ \ { \ \ \ \ \ \ \ \ \ \ \ - }  $ $N_{conv}<N_0<N_{div}$: the initial number of failed companies decreases during the iterative process, attracted by the convergent point $N_{conv}$, which means that the initial crisis heals up.
\\ $ \ \ { \ \ \ \ \ \ \ \ \ \ \ - }  $ $N_{div}<N_0<N_{core}$: the arrows show that the number of failed companies grows towards the $N_{core}$ convergent point. This phase is very dangerous as it implies that the contaminated cluster grows beyond the percolation transition critical point until it reaches the non-network finite size effects.
\\ $ \ \ { \ \ \ \ \ \ \ \ \ \ \ - }  $ $N_0>N_{core}$: in this range, the initial crisis is large, but the number of failed companies decreases with the iterative process towards the convergent point $N_{core}$.\\
}
\label{fig:solutions}
\end{figure}

This is represented by Figure \ref{fig:solutions}, similarly to Figures \ref{fig:convergence} and \ref{fig:divergence} except that this time the line Eq. \ref{N final} is not straight, not even on a double logarithmic scale. This opens the way to the possibility of having more then one fixed point and to more dynamical phases (regimes)  as detailed in the next sub-section.

\subsection{Analysis of the dynamics of the Minsky accelerator percolation model}
\label{subsec:Dynamicsf}

As detailed in the Appendix \ref{sec:appendix4} and represented in Figure \ref{fig:solutions}, there are {\bf 3 possible fixed points} corresponding to common solutions of  \ref{N final} and \ref{i final}:
\begin{itemize}
\item[-] {\bf two fixed points $N_{conv}$ and $N_{div}$} due to the possible intersections of the curve
 $N=S [1- (i / i_{C})^\beta ]^{-\gamma}$ with $i=i_{0}N^{\alpha}$. The names of the points imply that one of those two points is convergent and the other divergent.
\item[-] {\bf one fixed point $N_{core}$} identical with the fixed point 
 that we obtained in the previous sections as the common solutions of $N=(i / {k})^{\beta}$ and $i=i_{0}N^{\alpha}$. This point is convergent under the assumption that $\alpha \beta <1$.
\end{itemize} 

The fixed points  $\mathbf N_{conv}$, $\mathbf N_{div}$ and  $\mathbf N_{core}$  are convergent (stable, attactive) or divergent (unstable, repulsive) according to the criterion  Eq. \ref{divergence condition}. Stability can also be graphically determined by following the arrows in Figure \ref{fig:solutions} in the same way as described in Figure \ref{fig:convergence}. The exact formulae for  $N_{conv}, N_{div}, N_{core}$ are deduced and listed in the Appendix \ref{sec:appendix4}. These 3 fixed points divide the range of the possible initial number of failures $N_0$ into four regions /phases which we characterize below:

\begin{itemize}
\item[$ N_0 < N_{conv} $] \

 If the process \ref{process Nfail i} starts with an initial shock characterized by a number of defaults $N_0 < N_{conv}$ then $N_t$ initially increases with $t$ but eventually converges to $N_{conv}$. We say we have a limited crisis which consumes itself at the microscopic level by failing a limited number of ponzi companies in the neighborhood of the initial shock. 

\item[ $ N_{conv}<N_0 < N_{div} $] \

This region is still in the attraction range of the fixed point $ N_{conv}$ so a crisis starting with a shock in this range will not propagate at all and it might eventually heal itself by converging to $N_{conv}$ and its corresponding low interest rate. 

\item[$ N_{div}<N_0 < N_{core} $] \

In this region the contaminated cluster grows beyond the percolation transition critical point until it reaches the non-network finite size effects described in Section \ref{sec:Minsky accelerator} and stops at $N_{core}$. 

\item[$ N_{core}<N_0 $]  \

This case is characterized by a catastrophic number of initially failed companies. Following such a catastrophe, a limited core of very solid companies will recover endogenously. The rest will remain in the failed state but no additional failures are expected. Thus the process will converge to $ N_{core}$.

\end{itemize}
\begin{figure}
\centering
\includegraphics[scale=0.45]{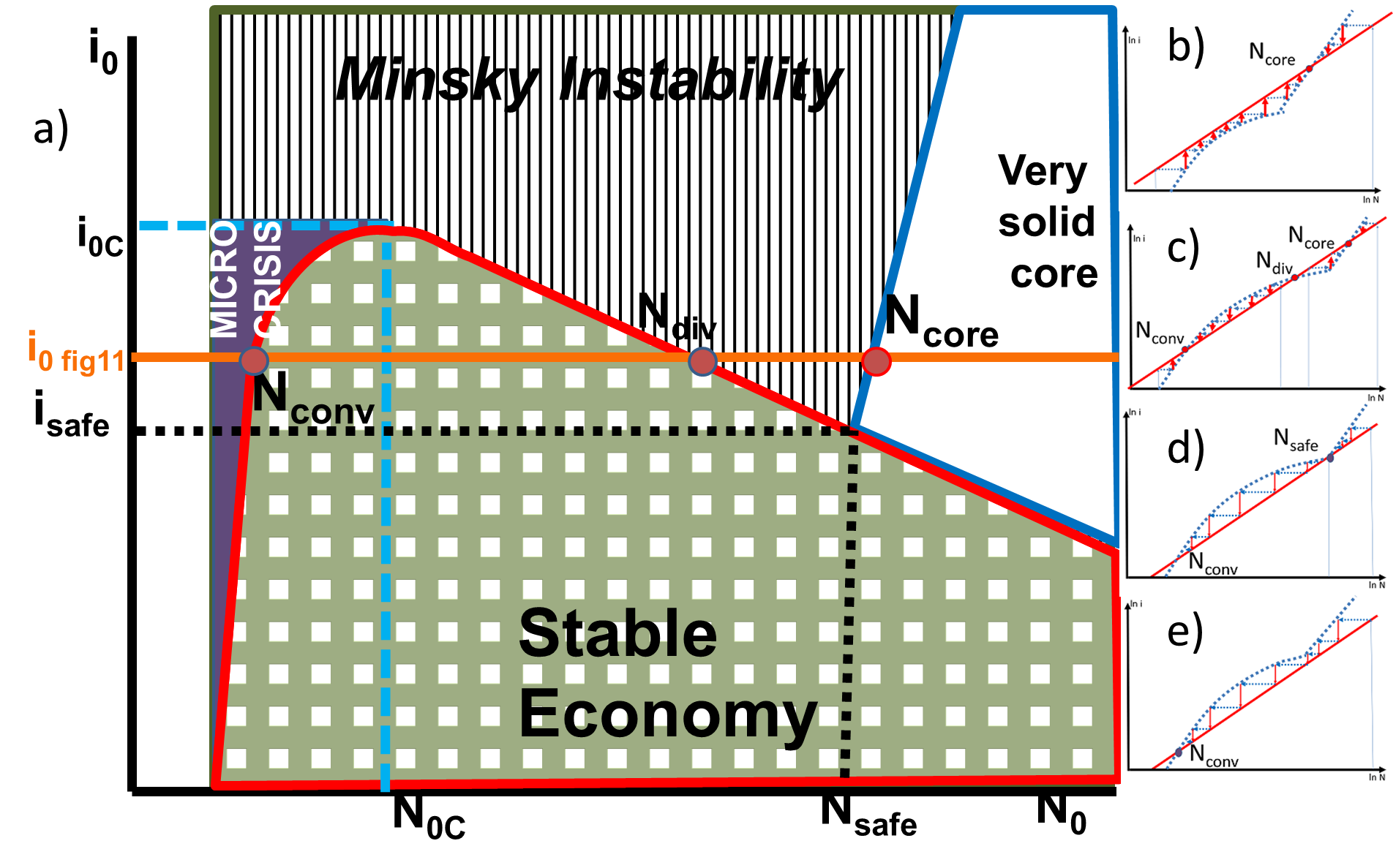}
\caption{\small
In Figure \ref{fig:phasediag}(a),
the horizontal axis represents the initial number of failed companies $N_0$, while the vertical axis represents the initial interest rate $i_0$.  On the right side of the phase diagram \ref{fig:phasediag}(a) are the auxiliary graphs describing the various dynamics of the iterative process Eq. \ref{process Nfail i}, similar to Figure \ref{fig:solutions} (which itself is represented as subfigure (c)).  
Each of these graphs correspond to a particular $i_0$ value and are placed with respect to the map \ref{fig:phasediag}(a) such that they are roughly at a hight equal to the $y$ axis value $i_0$ to which they correspond.\\
Note that while each of the Subfigures \ref{fig:phasediag}(b)-(e) displays explicitly through the arrows the evolution in time of the system for a fixed $i_0$ value and given $N_0$ initial values, by contrast, the Figure  \ref{fig:phasediag}(a) is a static map summarizing the 
final state of the system corresponding to the entire range of initial conditions ($i_0$, $N_0$). Thus, even though it is tempting, it is not correct to imagine the time evolution of the system as a trajectory in the 
($ i_0, N_0$) \ref{fig:phasediag}(a) map since of course the initial conditions $i_0 , N_0$ do not vary during the process. However, one can follow the evolution of $N_t$ and $i_t$ on the $x$ and $y$ axes of 
 \ref{fig:phasediag}(b)-(e) for given $i_0$ and $N_0$ initial conditions. \\
 The phase where the system is stable is square patterned,  the Micro Crisis phase is colored uniform in lilac;  the phase of Minsky instability is marked with strips, the solid core phase is left blank.\\
In the subfigures (b)-(e), which are on the logarithmic scale, the variation of $i_0$ is represented simply by rising or lowering the red line $i= i_0 N^\alpha$ while maintaining its slope $\alpha$ unchanged.  Accordingly, by varying $i_0$ one can have 3 cases:
 \\
$ (c) \ { \ \ }  $   $i_{0C} < i_0 < i_{safe}$ \\ all 3 intersections $N_{conv}$,  $N_{div}$, $N_{core}$  exist. 
An instance is shown in subfigure \ref{fig:phasediag}(a) as the orange horizontal line $i_{0 \ Fig \ 10}$  and the three small filled circles, corresponding to the three fixed points $N_{conv}$, $N_{div}$, $N_{core}$ in Figure \ref{fig:solutions}. 
Depending on the strength of the initial shock $N_0$ the process may belong to one of the 4 phases: Micro-Crisis (uniform lilac color) for $N_0<N_{conv}$, Stable (square pattern) for $N_{conv}<N_0<N_{div}$, Minsky instability (strips) for $N_{div}<N_0<N_{core}$, solid core (blank) $N_0>N_{core}$. \\
$ (b) \ { \ \ }  $ $i_0 > i_{0C}$ \\ the line $i= i_0 N^\alpha$ is raised above 
$ N=S\left[1- (i / i_{C} )^{\beta} \right]^{-\gamma}$ and consequently the intersections $N_{conv}$ and $N_{div}$ do not exist. 
As a result, the iterative process Eq. \ref{process Nfail i} always reaches the fixed point $N_{core}$ for all initial shocks $N_0$. 
 This corresponds the region of the diagram Figure \ref{fig:phasediag}(a) above the horizontal line $i_{0C}$ and to the subfigure \ref{fig:phasediag}(b). It contains most of the Minsky instability phase.  In this range, the effect of the network is to delay the crisis but it does not affect the ultimate outcome: all of the companies outside the very solid core fail.\\
$ (e) \ { \ \ }  $ $i_0 < i_{safe}$ \\ the line $i= i_0 N^\alpha$ is lowered below the junction of 
$ N=S\left[1- (i / i_{C} )^{\beta} \right]^{-\gamma}$
with
$N= ({i_t} / {k})^{\beta}$
and consequently the fixed points $N_{core}$ and $N_{div}$ do not exist. 
As a result  the iterative process $N_t$ Eq. \ref{process Nfail i} always converges to $N_{conv}$ for all initial shocks $N_0$.  This corresponds to subfigure \ref{fig:phasediag}(e) and the part of the phase diagram \ref{fig:phasediag}(a) below the horizontal line $i_{safe}$, where the stable economy phase is contiguous with the solid core phase. The network effects are protecting almost completely the range of parameters left in danger by the non-network case and the only danger are local limited Micro-Crises (lilac region to the left of $N_{conv}$).\\
$ (d) \ { \ \ }  $ $i_0 = i_{safe}$ \\ describes the situation in which  $N_{core}$ and $N_{div}$ meet and form the point $N_{safe}$.
}
\label{fig:phasediag}
\end{figure}

While Figure \ref{fig:solutions} depicts the behavior of the system for a given initial interest rate $i_0$, 
chosen such that all three fixed points $N_{conv}$, $N_{div}$, $N_{core}$ (and the phases that they separate) exist, Figure \ref{fig:phasediag}(a) contains the fixed points configurations for {\bf the entire range of values of $i_0$}. So the Figure \ref{fig:solutions}, which miniature version is added to Figure \ref{fig:phasediag} as subfigure(c), corresponds to just one horizontal section $i_{0 fig 11}$ in Figure \ref{fig:phasediag}(a). In exchange for this more comprehensive display of the information (joint dependence on $N_0$ and $i_0$ of the process outcome) Figure \ref{fig:phasediag}(a) misses the time evolution of the system as displayed in Figure \ref{fig:solutions} and Figures \ref{fig:phasediag} b,c,d,e.

The connection between theses complementary views is better understood if one notices that in Figure \ref{fig:solutions}, which is on the logarithmic scale, the variation of $i_0$ corresponds simply to rising or lowering of the red line $i= i_0 N^\alpha$ while maintaining its slope $\alpha$ unchanged. Varying $i_{0}$, it may occur that some of the three possible fixed points will not exist. Indeed, this is what subfigures \ref{fig:phasediag}(b)-(e) represent. Similarly to the subfigure (c) which illustrate the horizontal section $i_{0 fig 11}$, subfigures (b), (d), (e) have been added to illustrate the horizontal cross sections in the other relevant areas of the phase diagram (a). In fact, the Figures \ref{fig:phasediag}(b)-(e) are placed with respect to  \ref{fig:phasediag}(a)
such that their vertical position corresponds to the height on the $y$ axis in \ref{fig:phasediag}(a) that marks the value of $i_0$ for which they are plotted. 

The intervals between fixed points in Figure \ref{fig:solutions} correspond to entire areas in the phase diagram  \ref{fig:phasediag}(a)
and each of the fixed points of Figure \ref{fig:solutions}, marked by little circles on $\i_{0 fig 11}$ is part of the boundaries separating the phases in  \ref{fig:phasediag}(a). 
We use in Figure \ref{fig:phasediag}(a) different patterns to mark the four phases that the iterative process may belong to, depending on the strength of the initial shock $ N_0$ and the initial state of the economy $i_0$:
\begin{description}
\item[-- Micro-Crisis] for $N_0<N_{conv}$, uniform dark lilac color;
\item[-- Stable] for $N_{conv}<N_0<N_{div}$, square pattern; 
\item[-- Minsky instability] for $N_{div}<N_0<N_{core}$, striped pattern;
\item[-- Solid core] for $N_0>N_{core}$, blank area. 
\end{description}

The structure of Figure \ref{fig:phasediag}(a) is determined by the way in which the initial value $i_0$ affects the existence and position of the the 3 fixed points $N_{conv}$, $N_{div}$, $N_{core}$. 
Thus according to the computations in the Appendix \ref{sec:appendix4} we distinguish three main cases according to the position and intersections of the line $i= i_0 N^\alpha$ with respect to the line $N= \min \{ S\left[1- (i/ i_{C})^{\beta}  \right]^{-\gamma}, ({i_t} / {k})^{\beta} \}$:
\begin{itemize}

\item[$i_0 > i_{0C}$  subfigure \ref{fig:phasediag}(b)] \

In this case the line $i=i_{0}N^{\alpha}$ is positioned above the   $S\left[1- (i / i_{C})^{\beta}  \right]^{-\gamma}$ curve and their intersections $N_{conv}$, $N_{div}$ do not exist nor does the "stable phase". 
In the absence of the fixed points $N_{conv}$ and $N_{div}$, the failures contagion process (Eq. \ref{process Nfail i}) continues unhindered beyond the percolation threshold.
This is the meaning of the part of the {\bf Minsky instability} (which is indicated by striped pattern) phase  which, as seen in the diagram \ref{fig:phasediag}(a), extends from $S$ to the only left fixed point: $N_{core}= (i_{0} / k)^{\beta / (1- \alpha \beta )}$
The $N_{core}$ convergent fixed point insures the survival of only the most resilient companies and defines the {\bf “very solid core”} (blank area) phase in the diagram \ref{fig:phasediag}(a).

The “very solid core” depends on the initial state of the system before the shock $i_0$ and on the shape on the resiliences distribution. In particular it contains those companies that have such a large resilience that exceeds even the values of $i$ corresponding to all the other companies in the economy having failed. 
Of course, not all the companies that were "hedge" for $i=i_0$ belong to the "very solid core".
This is a crucial point of the Minsky instability idea: many companies that look safe before the Minsky moment, are likely to fail in the crisis following it.

\item[$i_{safe} \le i_0 \le i_{0C}$ subfigure \ref{fig:phasediag}(c) ] \

This range corresponds to Figure \ref{fig:solutions} and the horizontal line $i_{0  fig 11}$
from Figure \ref{fig:phasediag}(a).  All three fixed points  $ N_{conv}<N_{div} < N_{core} $ exist and so do the four phases above.

For $i_0 =  i_{safe}$, subfigure \ref{fig:phasediag} (d), the fixed points $N_{core}$ and $N_{div}$ coincide (cf. Appendix \ref{sec:appendix4}) and the "Minsky instability phase" shrinks to one point.

For $i_0=i_{0C}$ the two solutions $N_{conv}$ and $N_{div}$ coincide
and the stable phase shrinks to one point.

\item[$i_0 < i_{safe}$ subfigure \ref{fig:phasediag}(e) ] \

In this case the line $i=i_{0}N^{\alpha}$ is below the meeting point of 
$N=  S [1- (i / i_{C})^{\beta}  ]^{-\gamma} $ and $ ({i_t} / {k})^{\beta}$
and thus does not intersect 
$N= \min \{ S [1- (i / i_{C})^{\beta} ]^{-\gamma}, (i / k)^{\beta} \}$.
Consequently $N_{conv}$, $N_{div}$ do not exist nor does the 'Minsky instability phase'. 
In the absence of  $N_{div}$ and $N_{core}$  the only fixed point remains $N_{conv}$ and the process Eq. \ref{process Nfail i} converges to it.
\end{itemize}

This ends up the analytic quantitative description of the phase diagram of the percolation Minsky model.
 We turn now to the discussion of the empirical meaning and policy implications of the model.

\section{Towards a microscopic view of economic monitoring, regulation and intervention}
\label{sec:Policy }

\subsection{Using the Phase Diagram as a regulatory tool}
\begin{figure}
\centering
\includegraphics[scale=0.50]{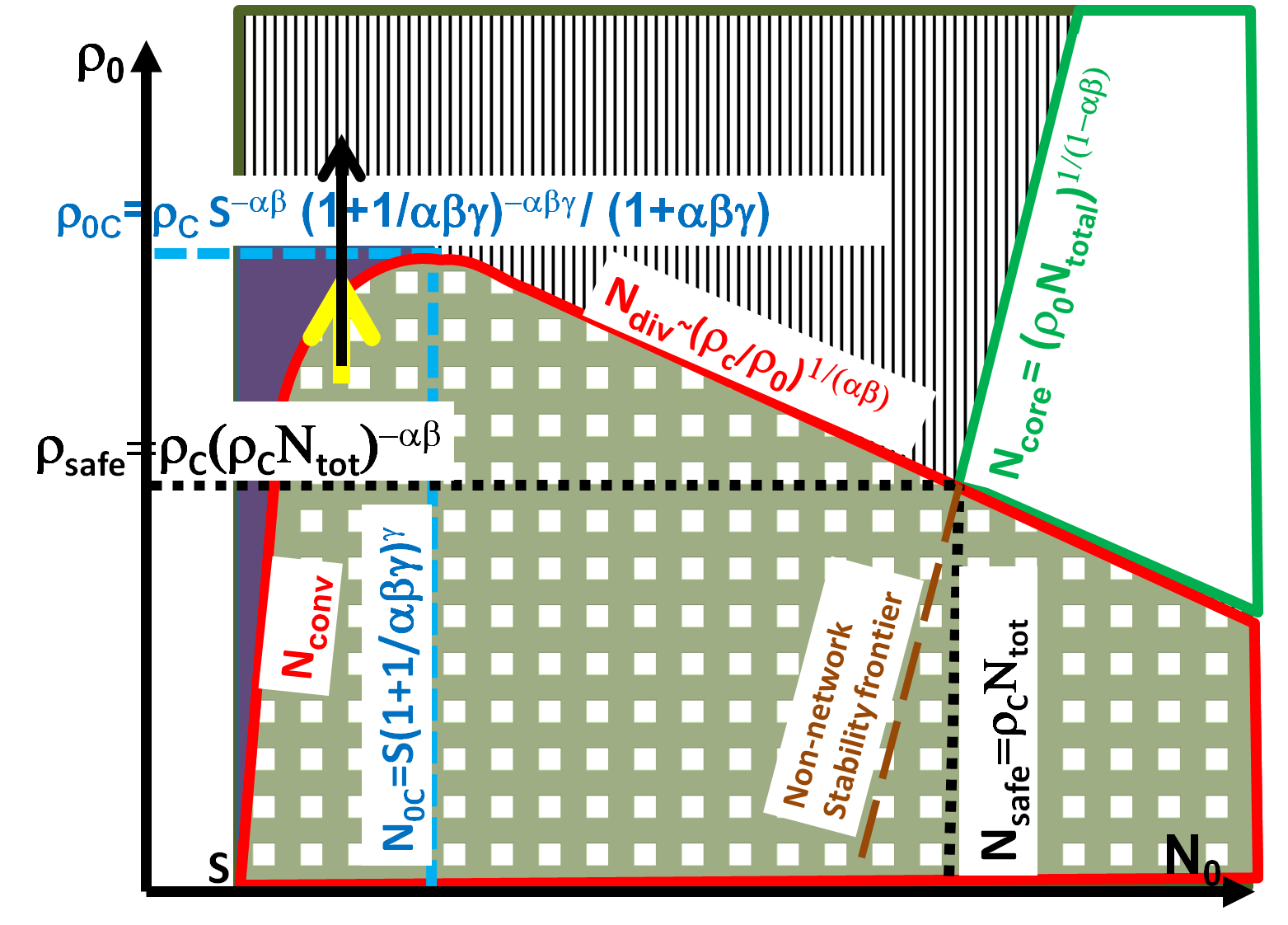}
\caption{\small 
The figure illustrates the crucial effect of a network in limiting the spread of a crises.\\
For $\rho_0 > \rho_{0C}$ the effect of the network is to delay the crisis but it does not affect the ultimate outcome (unless the delay is used by the policy makers to intervene, e.g. by lowering the interest rate 
or supporting key susceptible companies).\\
For $\rho_0 < \rho_{safe}$ the network effects are protecting almost completely the range of parameters
left in danger by the non-network case and the only danger are local limited Micro-Crises.\\
In the case in which there are no network obstructions, the entire region at the left of the line 
`Non-network Stability frontier' would be unstable and any process starting with $\rho_0$, $N_0$ in that 
range would cause a positive feedback loop that would leave a large number of companies to default. 
In fact only the very resilient companies in the solid core would survive. \\
It is tempting to interpret the Figure \ref{fig:phaserho} as a map of the evolution of the crisis in the $(i,N)$ plane. Unfortunately this is not exact. All this figure gives is the outcome of the process for given {\bf initial} values  $(i_0,N_0)$. For details of the process one has to refer to Figure \ref{fig:solutions}.}
\label{fig:phaserho}
\end{figure}

The phase diagram presented in Figure \ref{fig:phasediag}(a) was formulated in terms of the initial shock to the number of ponzi companies, $N_0$, their resilience (parametrized by $k$ and $\beta$) and the initial interest rate, $i_0$. In discussing the empirical applications, it might be easier to use the phase diagram Figure \ref{fig:phaserho}, where the system dynamics is expressed in terms also of the initial failures shock, $N_0$, but of the initial ponzi density, $\rho_0$, rather than $i_0$.
 The relation connecting $\rho_0$ to $i_0$ is $(i_0/k)^\beta = \rho_0 N_{total}$; see Eq. \ref{i of rho}  in Appendix \ref{sec:appendix4}. 
One of the advantages that makes Figure \ref{fig:phaserho} more natural is that in all the formulae therein, the exponent $\alpha$, characterizing the inter-scale feedback Eq. \ref{i of default}, and the coefficient $\beta$, characterizing the companies heterogeneous resilience Eq. \ref{default of i}, appear in the combination $\alpha \beta$ showing that agents' resilience heterogeneity, bottom-up feedback (and network geometry) are combined together by the Minsky accelerator scenario. 
In fact, the parameters tend to appear in the combination $\alpha \beta \gamma$ which lumps together the geometrical properties of the economic network, the structure of the agents heterogeneity and the interscale feedback. This is likely to ease the empirical calibration of the model by reducing the effective number of free parameters.

As seen in the log-log phase diagram Figure \ref{fig:phaserho}, and the previous Section, the effect of the network is that the borders between the different phases become non-linear.

Moreover, as opposed to the non-network case where the fixed point was attractive (for $\alpha \beta <1$), in the network case one has also unstable fixed points. These points generate a frontier between the phases that can be very abrupt.  For instance, the percolation transition at Section \ref{sec:Percolation}, which was continuous in the absence of the Minsky accelerator, becomes discontinuous. At $N_0=N_{div}$ the slightest increase in the density of ponzi to $N_0 > N_{div}$ can throw the system into a process of accelerated crisis, while for a slight decrease to an $N_0 < N_{div}$ the system is self-healing. 

 A crucial network effect is the shift of the crisis contagion range from quite low values of $\rho_0$ to a relative large ponzi density threshold $\rho_{safe}$.
This is because  in a network with critical percolation transition density $\rho_C$, for $\rho_t < \rho_C$
the ponzi units are isolated in localized mutually disconnected clusters. Thus the failure contagion chains are quite short and one has a rather extended stable range of parameters. 
In the absence of network effects, the entire square patterned stable area at the left of the $N=( \rho_0 N_{total} )^{1/(1-\alpha \beta)}$ line would be unstable, with $N$ increasing to $N_{core}$.

 In the range $\rho_{safe}< \rho_0< \rho_{0C}$, the same network effect is partially holding, and  a systemic crisis can take place only following a   shock strong enough to produce spontaneously a cluster of size $N_0 > N_{div}$ failures. Then the number of ponzi failures will increase but only to $N_{core}$.

\begin{figure}
\centering
\subfigure[A crisis triggered by the failure of the node labeled by $1$. The numbers on yellow ponzi nodes correspond to their time of failure. The bottleneck at the time step $ 5$ slows down the propagation of failures. In fact eliminating the 2 nodes that get contaminated at time step $5$,  the network splits in 3 small clusters and the crisis is stopped and confined to only one of them. 
]{\includegraphics[scale=0.40]{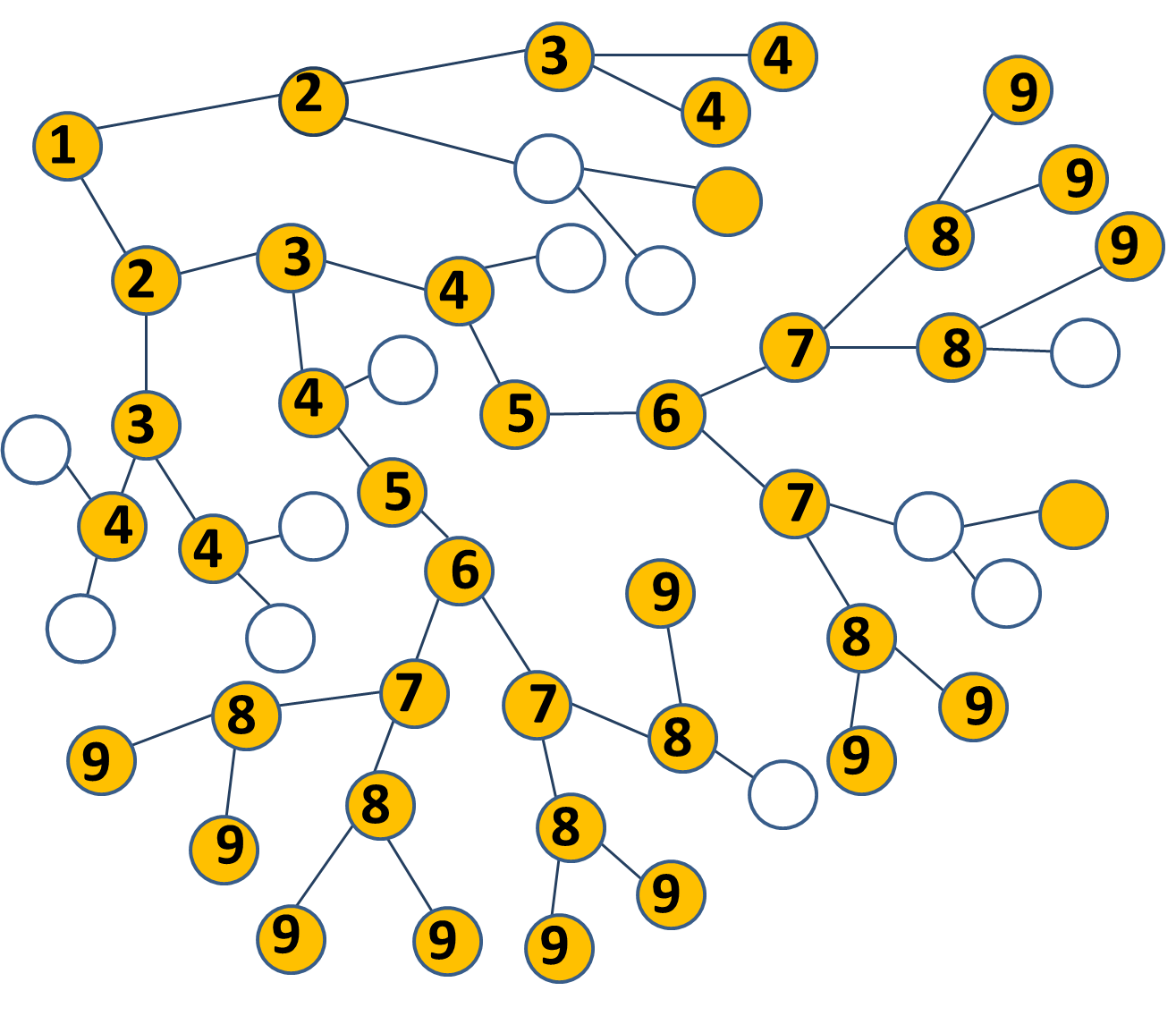} \label{fig:bottleneck1}}
\\
\subfigure[The plot of the number of new failures at each time corresponding to the process in Figure \ref{fig:bottleneck1}. One sees that there is a bottleneck at time $5$. This means that one can stop the crisis if one immunizes the 2 nodes marked by $5$ in Figure \ref{fig:bottleneck1}. Guaranteeing their debts to the 2 points labeld $6$ would also suffice in stopping the crisis. ]{\includegraphics[scale=0.25]{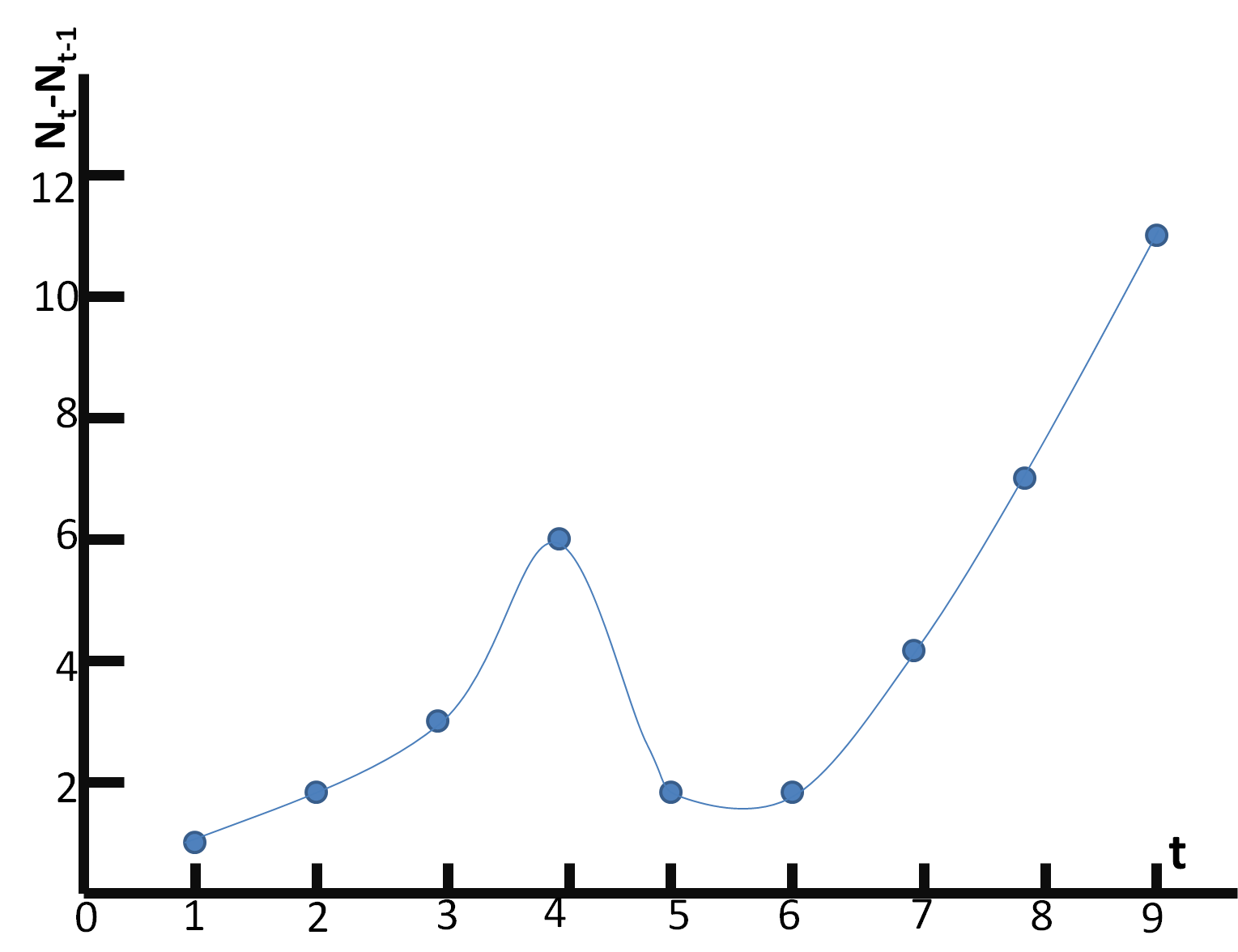} \label{fig:bottleneck2}}
\hfil
\subfigure[Rate of failures corresponding to the graph \ref{fig:phasediag}(b) when the initial ponzi density is above the bottleneck around $\rho_{0C}$. One observes a long crisis slow-down period.]{\includegraphics[scale=0.30]{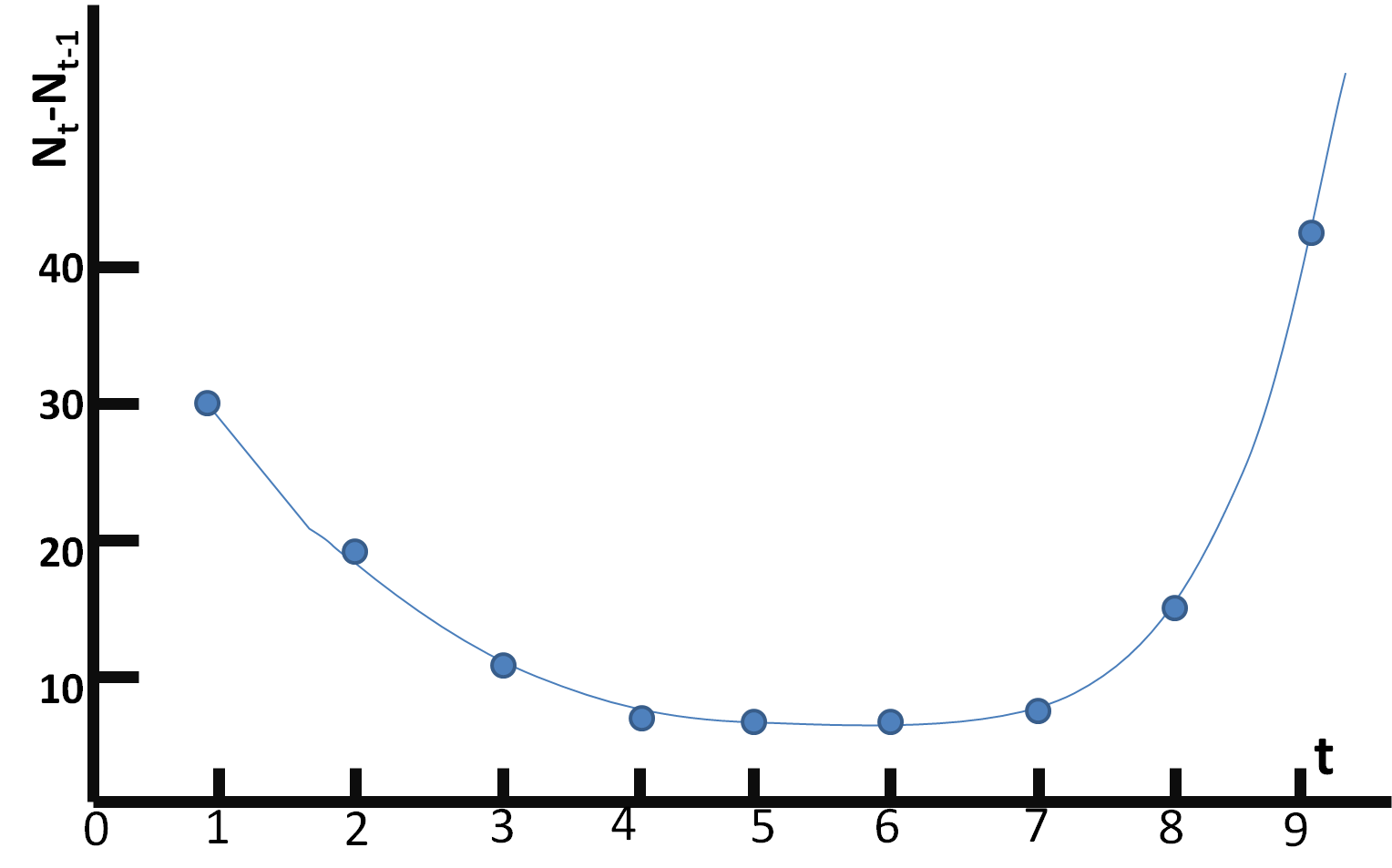}
\label{fig:bottleneck3}}
\caption{\small The failure dynamics in the critical region. The giant cluster in Figure \ref{fig:bottleneck1} depends one rather weak bridges between the sub-clusters. This leads to the existence of bottlenecks in the propagation rate, to large fluctuations in the contagion rate and to opportunities of cheply stopping the crisis. 
Figure \ref{fig:bottleneck3} corresponds to the region in Figure \ref{fig:phasediag} where the initial number of failures is in the vicinity of the critical point $(N_{0C}, \rho_{0C})$.}
\label{fig:bottle}
\end{figure}

Even for  $\rho_0> \rho_{0C}$, the network has significant effect in the planning and implementation of the policy interventions: it delays the advancement of the crisis in the stage $N_t  \sim N_{0C}$ when the flow of failures  is very weak and can be cut completely by targeting for support a very small and specific set of nodes. 
For instance in order to stop the crisis propagation in Figure \ref{fig:bottleneck1} it is enough to intervene and prevent the failures of the points marked by $5$ in Figure \ref{fig:bottleneck2}.
These are the nodes belonging to the bottleneck that connects the already contaminated cluster to new clusters: if this bottleneck is identified and severed, the crisis propagation is stopped. 
 Alternatively, one may lower the interest rate temporarily to stop their contagion. 

\subsubsection{Further elaboration of the dynamics of $N$ in the vicinity of the critical points $(N_{0C},\rho_{0C})$ and $(N_{C},\rho_{C})$.}
\label{bottleneck}
It is important to note that in addition to the three $\rho_0$ regimes above, in the ranges around their borders, the process  \ref{process Nfail i} may show features that are very sensitive to the agent based nature of system.

There are two parameter ranges that are particularly sensitive to it: 
\begin{enumerate}
\item{
\begin{itemize}
\item [a)] For $\rho_0=\rho_{0C}$, where  the two solutions $N_{conv} = N_{div}= N_{0C}$ coincide
and the stable phase shrinks to one point, the `decision' of the system whether to stop 
the propagation of failures locally at $N_{0C}$  or to continue to the macroscopic $N_{core}$ 
depends on microscopic details. 
Sometimes the fate of the process depends on one or two agents 
that have their resilience above or below the value of the current $\rho_t$ when the $N_t$ approaches $N_{0C}$. Such situation is illustrated in Figures \ref{fig:bottleneck1} and \ref{fig:bottleneck2}.
For this parameter range, obviously the different realizations of the dynamics, even with identical macroscopic parameters, may have very different outcomes: microcrisis vs systemic crisis.
In order to provide a precise prediction one needs to know the details of the resiliences and connections of the agents failing around $N_t = N_{0C}$.
\item[b)]  Moreover, even if $\rho_0>\rho_{0C}$ there is a pronounced slowing down [Hohnisch 2007, Cantono 2010] in the advance of $N_t$ as one passes the `bottleneck' range around $N_{0C}$.
This is also seen in the subfigure \ref{fig:phasediag}(b)  where it takes many iterations for the process to advance through the region where the two curves $N(i)$ and $i(N)$ are very close one to the another, see Figure \ref{fig:bottleneck3}.  If the advancement of the process $N_{t+1} - N_t$ is of the order of a few units, the agent granular structure of the system becomes macroscopically evident and the continuous approach is informative only on average and the predictive power can be restored only by detailed knowledge of  the connections and resiliences of the agents involved at this phase of the process.
\end{itemize}}

\item{
As $\rho_t$ increases towards $\rho_C$, the growth of the contaminated cluster is not based on the addition to it of individual agents. Rather, the mechanism of cluster growth is by cluster coalescence: a ponzi cluster that is already contaminated is connected by a new ponzi agent (may be generated by a recent increase in $\rho_t$) to an until then uncontaminated ponzi cluster. An example is in Figure \ref{fig:bottleneck1} where the points marked by $5$ connect between the 3 subclusters composing the giant cluster. This generates small avalanches which stop or slow down when the ponzi (sub-)cluster is entirely contaminated. Thus the propagation is dominated by the successive failure of clusters or cluster of clusters. In Figure \ref{fig:bottleneck2} one sees the slowing down around time $5$. This clusters structure leads both to a fractal rate and a fractal geometry of failures. In statistical mechanics terms these are called critical fluctuations. In the context of product adoption their presence was used to predict the micro or macro outcome of the propagation process.}
\end{enumerate}

\subsection{Predicting unpredictable}
\label{Interventions}
\subsubsection{Coping with the crisis}

For averting systemic crisis it is crucial to prevent the coalescence of the ponzi clusters into a giant cluster.
Thus it is important for the regulator to intervene very energetically and lower $i$ as soon 
as $\rho_t$ approaches $\rho_C$.  This is not difficult to recognize since at this stage $N_t$ is growing at increasing speed.

A more difficult but rewarding measure is to keep $\rho_0$ below $\rho_{safe}$, but the diagnosis for this condition is more difficult to obtain. However, as discussed in the previous section, there are signs that indicate whether $\rho_t$ approaches 
the limits of the stable phase: the dynamics becomes intermittent and the chains of contagion become longer.
A rule of thumb indicating the proximity of the percolation phase transition would be whether the failure of a company would affect typically not only its suppliers but also their own suppliers. This technique of measuring the clustering of new contagions within the network has been shown to be effective in predicting contagion in marketing \cite{Solomon 2000}, \cite{Goldenberg 2005}.

As mentioned at the end of Subsection \ref{subsec:Dynamicsf} and seen in Figure \ref{fig:AppFig5},
even if one is in the divergent phase, if the parameters are close enough to the boundary of the stable phase,
the regulator has a lot of advanced warning signs and intervention options to stop the crisis: 
\begin{enumerate}
\item As the trajectory of the process approaches the stable phase the number failing ponzis per unit time slows down (Figure \ref{fig:bottleneck3} around time 3-7): there is time to react.
\item To stop the process it is enough to prevent from failing those ponzis that are in danger of being contaminated in a given time interval (Figure \ref{fig:bottleneck2} around time 5-6).
\item To stop the process in the slow, narrow bottelneck stage, a slight lowering in the interest rate is sufficient. 
\item Supporting, in a targeted way, the very few ponzies (e.g. the couple of $5$'s in Figure \ref{fig:bottleneck1}) that are actively failing during the slow-down of the process is relatively cheap. 
\item Supporting the partners of those ponzis might be even cheaper in as far as they might only need guarantees as means to lower their interest rate risk premium included in interest rate $i$. E.g. guaranteeing the debts of the nodes $5$ to the nodes $6$ in Figure \ref{fig:bottleneck1}.
\end{enumerate}
Thus, if the mere lowering of $i$ does not help, one may consider stopping the propagation by more targeted regulator intervention: one should guarantee to the second neighbors  that the payments of the distressed first neighbors are going to be met.
This intervention is helped by the fact that the slow contagion period might allow quite a long time for regulators' action.
In fact one may envisage a set of rules for intervention that can be automatically enforced by computer (including credit lines opened electronically to the units targeted for support by the regulator's policy).


\subsubsection{Treating a Minsky loan accelerator with a interest rate hike: Doing the right thing at the wrong time}
\label{sec:rate}

The yellow and the black vertical arrows in Figure \ref{fig:phaserho} are indicating two realizations of the same policy intervention, performed with a different intensity/timing. It is due to the effect of the network that these two interventions can have diametrically opposite effects. 

If one wishes to deflate the debt created during an `irrationally exuberant' Minsky Loan Accelerator period, one can try to increase the interest rate $i_0$. This, however, will increase the fraction of ponzis, $\rho_0$, in the system. On the map, this amounts to moving the point representing the state of the system towards the top of the figure. If this move brings the system from the stable (squared pattern) region to the MicroCrisis (uniform lilac) region, this leads to a successful bubble deflation with a limited price in terms of the number of failures. Those failures are mainly the most exposed ponzi and thus constitute a limited Schumpeter creative destruction. If however, the vertical move takes the system from the stable region to the (striped) Minsky instability region the result is that the regulator himself triggered a major catastrophe by his very trial to avoid it. Such instances are often encountered in the historical accounts.
In fact the wikipedia description of the beginning of the Lost Decade(s) in Japan has a striking similarity with the scenario above.\footnote{
``Recognizing that this bubble was unsustainable, the Finance Ministry sharply raised interest rates and the exchange rate with the US $ \$ $ (a geopolitical decision triggered by bilateral negotiations) in late 1989. This sharp policy caused the bursting of the bubble, and the stock market crashed. A debt crisis followed and the Japanese banks and insurances were now loaded with bad debts. The financial institutions were bailed out through capital infusions from the government, loans from the central bank and the ability to postpone the recognition of losses" \url{http://en.wikipedia.org/wiki/Lost_Decade_(Japan)}.} 

The policy solution is then to act locally by applying direct intervention that supports the agents more likely to trigger, transmit or amplify a failure avalanche. As mentioned above, one rule of thumb criterion could be to see if the second neighbors  of an initially failed company (the $3$'s in Figure \ref{fig:bottleneck1}) are in trouble and to provide regulator guarantees for the first neighbors (the $2$'s in Figure \ref{fig:bottleneck1}) debt. This combines the minimal amount of expenses with the maximal protection from percolation of the avalanche contagion systemic crisis. 
Such an algorithm has been developed in the context of network immunization against electronic viruses \cite{Goldenberg 2005}. 

Of course there are no perfect solutions: the intervention of the regulators in an ostensibly free market has always problems of fairness, moral hazards, not to speak about the ideological reluctance to give `bureaucrats' 
power over the free market or to help banks with the taxpayer's money. 
The agent based models can claim no superiority from this point of view, but they may have superior capabilities to offer
methods for massive searches within the trade and balance sheet data in order to select particular types of 
companies to be singled out as eligible for particular surgical interventions rather then the coarse macro-economic measures.
To recurr to a medical analogy, they might help fixing the knee rather then amputating the leg.

\subsubsection{Is counter-cyclically lowering the interest  stabilizing?}

As seen above, the interest rate is one of the most available and controllable instruments that the regulators have and which they may and will use whenever they feel the necessity.

\begin{figure}
\centering
\includegraphics[scale=0.30]{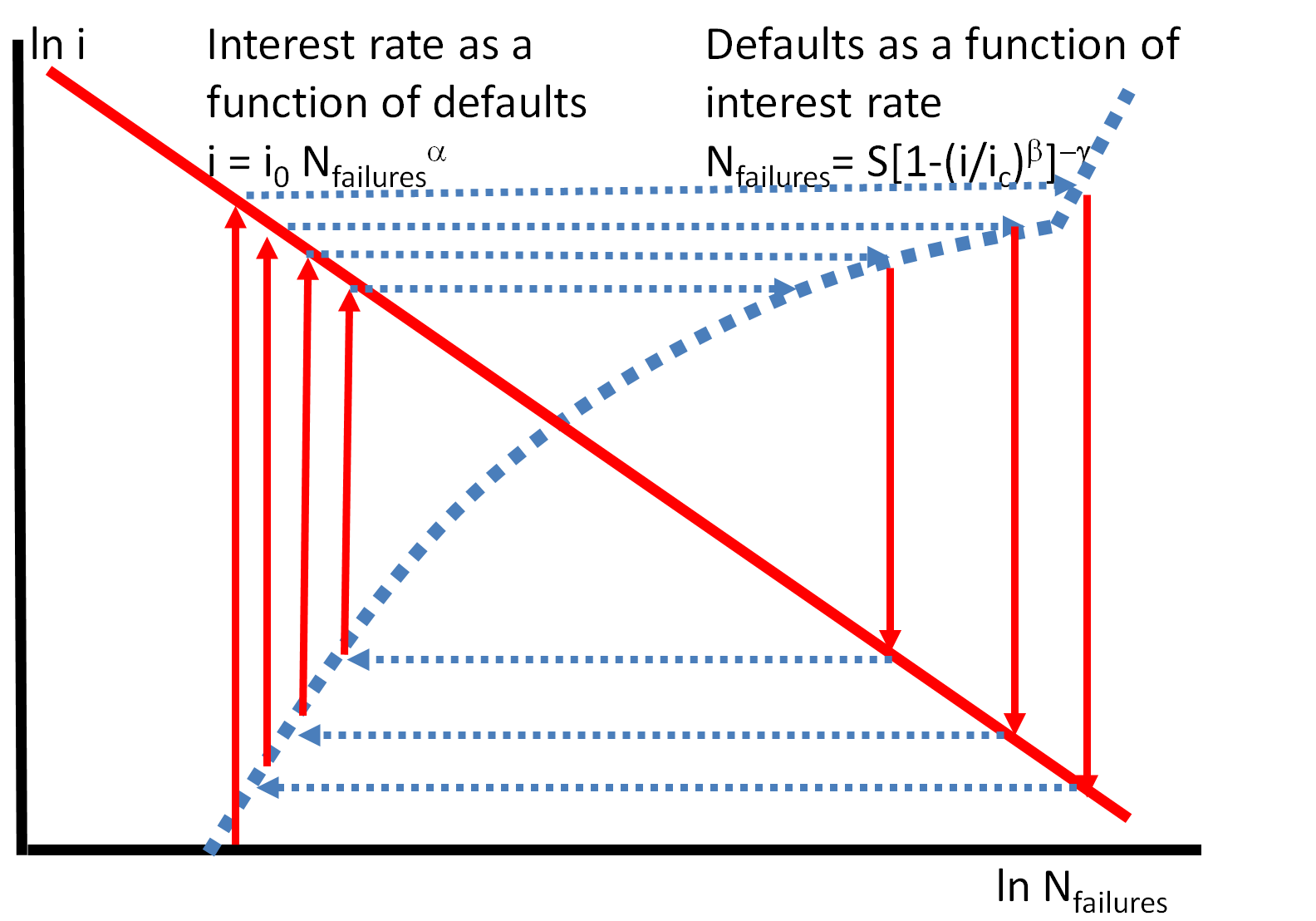}
\caption{\small A hypothetical economy where the regulators' policy imposes a decrease of the interest rate with the number of defaults.
The problem with this is that it might destabilize the graph of loans (trigger a new Minsky loans accelerator) Figure \ref{fig:fast increasing returns}. 
However there might exist a window where both Figure \ref{fig:increasing returns converge} and the present figure hold: both the loans and the defaults loops will be stabilized.}
\label{fig:stabilization}
\end{figure}

In particular: should a regulator tune $i$ opposite to $N$, e.g. by imposing a negative exponent, $\alpha$, in Eq. \ref{i of default} (which the Fed and the ECB eventually did, following the spread of the great recession), one would avoid (of course at the price of other risks and distortions) the run-away crisis.  An example of a convergence (admittedly quite  bumpy) imposed  on the loans market by an exogenously imposed negative $\alpha$ is shown in Figure \ref{fig:stabilization}.

One can observe that the overshooting in the convergence process shown in Figure \ref{fig:stabilization} is milder if the slope of the 
interest rate counter-cyclical exponent is small enough : $0< \alpha << 1$.
Thus a more stabilizing policy could be to use the central bank rate such as to keep the interest rate to the non-financial sector constant. 

However, there are changes in the  inflation rate, foreign exchange, production costs, assets prices,
that are likely to throw some companies into a ponzi position by breaking the balance between the 
resilience of companies and the interest rate. In fact some of these exogenous factors may amount to an effective 
change in the real interest rate.  The regulator may intervene in the first instance by lowering the interest rate.
However, in the case of a very dramatic deflation even an interest rate of 0 would mean a high real interest rate, as 
was the case of Japan and arguably of the great recession. Some people dubbed it `liquidity trap' (though Keynes probably meant something different when the introduced the term). 
In such situations when the nominal  interest rate cannot be lowered anymore, the regulator may resort to intervene by guaranteeing the continuation of rolling loans at the same real interest rate. This would insure that the credit lines that the real sector needs to continue its production would function as before the financial troubles. 

This is partially what the Japanese regulators did throughout the 1990's in order to protect the real production sector following the bubble burst ({\it hokai}) mentioned above and the ensuing massive deflation.
The same policy may be applied to households (especially in the real estate mortgages context):
 the regulator may guarantee the debt that was served for a long enough period before the crisis moment (and thus proved solvent in normal economic conditions).

\subsection{Assumptions underlying our model}

Until a fully computerized system to administrate policy can be supported by a `real time' data acquisition system, and until a software policy enforcing platform can administrate it in a localized targeted way, one is limited to use the diagram in Figure \ref{fig:phaserho} as a static and macro tool to analyse the short term prospects of the system. 
 
The assumptions underlying such a use are the short term (on a scale of weeks) reliability of the parameters:
\begin{itemize}
\item[$\gamma$ :] At the level of weeks, the financial ties are stable, the resilience of the individual agents may change but the system as a network maintains macroscopically the same critical density $\rho_C$  and a critical exponent $\gamma$. 
\item[$\alpha$ :]  The coefficient $\alpha$ governing the interest rate dependence on the defaults is also static at this time scale: in Europe, the policy decisions on the key interest rate are chosen once per month at the ECB during the Governing Council meeting, when the ECB assesses the economic situation and the stance of the monetary policy.
\item[$\beta$ :]  The density of ponzi companies in the system, $\rho_t$, is a variable that microscopically can change on a daily basis (when the payments are due then the ponzi agents can default). However, the stochastic heterogeneity exponent $\beta$ is likely to remain stable. 
\end{itemize}
Some of these parameters are under the regulator's influence and could be targeted within the policy guided by the diagram:
 For example, the exponent $\beta$ can be affected by changing the tax policy, with different regulations for the agents with large and those  with small incomes. Under such a policy, the distribution of agents' resilience will change considerably and the diagram in Figure \ref{fig:phaserho} would have to be reproduced. 

In a more developed data acquisition environment, the diagram \ref{fig:phaserho} can also be used in conjunction with dynamic simulations into which the real time parameters/events (for example the redesigned network after a large shock such as bankruptcy of a major company, natural disaster etc.) would be fed and the phases of the system would have to be re-established as frequently as the data allows. In the absence of genuinely micro-resolution tools, the diagram \ref{fig:phaserho} can be used as an  additional tool to the already existing procedures for the ECB council to simulate which changes in the phases of the system would happen with a planned change of the interest rate.

\subsection{Dealing with uncertainty of the global stochastic parameters}
\label{uncertainty}

The statistical methods used in the present article circumvent the necessity of microscopic knowledge of each company and each trade connection, and reduce it to the knowledge of global parameters such as $\rho_0$, $\alpha$, $\beta$, $\gamma$ etc.
Even if one stops here,  the evaluation of those parameters (or operationally equivalent ones) will demand 
at least the kind of effort and will face similar problems to the neoclassical macro-econometric approach.
For difficulties to extract reliably quantitative connections between macro-economic measurables see \cite{Rogoff 2010} and their criticism by \cite{Herndon 2013}, and \cite{Alesina 2010} and their criticism by \cite{Guajardo 2010}.
 
Moreover, the famous complaint of decision makers against ambidextrous economists
(Harry Truman:``Give me a one-handed economist"), would be expanded to complaints about models that depend on yet a larger (and unfamiliar!) number of empirically difficult to determine new quantities:

\begin{itemize}
\item[$\beta$] : The resilience heterogeneity parametrized by the exponent $\beta$ is not easy to establish.
Not  only that it depends on companies' balance sheet data that are not always available or reliable, but 
also it is likely to depend on the financial structure of the economy and the details of the conditions in which the debt was acquired, more specifically, the debt structure (long vs short term, with risk premium or not, with strong or weak collaterals, old debt or rolling credit etc).
\item[$\alpha$] : The influence of the individual failures on the interest rate is parametrized by $\alpha$ and it is even more elusive and dependent on circumstances. The same event (a company failure) can be perceived in a more severe or more light perspective depending on media coverage, other contemporary events, etc. 
Moreover different banks may impose different rates to different borrowers, transforming the global parameter $\alpha$, too, into a microscopically inhomogenous one.
Of course one can still organize a systematic measurement connecting the volume of  failures in a time span with the interest rate and extract an average $\alpha$.  
\item[$\gamma$] : The parameter $\gamma$ describes the influence of the network geometry on the size of the contagion avalanches and in particular the way their sizes diverge to systemic dimensions as one increases the ponzi density.
Calibrating $\gamma$ it is not less difficult than the other parameters: 
\begin{enumerate}
\item As mentioned previously, it is not even clear that  the simple percolation (rather than the Ising or bootstrap percolation [Kindler 2013]) picture is the correct one.
\item the links of the network have in reality very different weights.
\item The access to the details of the trade connections between companies is even more 
limited by privacy rules then the balance sheet data. The debt connections between financial institutions are somewhat easier to deal with. These connections were monitored by central banks under the previous regulatory regimes.
\item The entire network is never complete: even having data for an entire country leaves the imports and export data for individual companies undefined to a large extent.
\item The network of trade connections is not stable: every year a fraction ($10\%$) of the companies go bankrupt and are replaced by other trade partners in the network connections. 
\end{enumerate}
And one can go on with those qualifications indefinitely.
\item[$\rho_0$] : To create a diagram as in Figure \ref{fig:phaserho}, one needs to know the values of $\rho_C$ and $\rho_0$.
For this, in turn, one has to monitor the financial ties between the agents, as well as their `ponziness'. 
However, the network  links can denote different types of ties between the economic agents, depending on the domain of analysis.  It is of particular interest to focus on credit relationships, on exposures between banks and on liquidity flows in the
interbank payment system, as such network connections add on the ponzi status in a dynamic way. However, relevant ties are also created between the agents that share common assets \cite{Battiston 2012} or by companies residing in the same geographic location (sharing the same households as labor and$/$or as consumers).
\end{itemize}

The hope of the simplifications leading to the phase diagram \ref{fig:phaserho} is that the existence of the sharp phase frontiers between such different phases may ensure that much of the dramatic transitions based on the above abstractions may remain, even after introducing the entire host of realistic empirical details.
The strategical message of the agent based model is that one should organize the monitoring capabilities 
of the regulators to a much higher resolution level: 
by analogy, one would never expect the traffic jams in a city to be resolved only on the basis of 
global parameters, such as the number of cars per meter or on the average speed.

Until then we may congratulate ourselves with the fact that according to our model, the network, top-down and bottom-up parameters appear often in the combination $\alpha \beta \gamma$ which reduces greatly the number of effective empirical parameters to calibrate and reduces drastically the sensitivity to their errors.

\subsubsection{Examples of models relaxing / modifying assumptions of the present model}

In the above we assumed statistical macroscopic homogeneity within the microscopically inhomogenous network of agents.
By this we mean that while microscopically companies have different rank (numbers of partners), different resiliences, different clustering coefficients, at the macroscopic scale, the probability to find a certain rank, resilience or clustering coefficient does not change significantly between various regions of the network.  

This, of course, is not necessarily the case in the real world:
and one may consider systems that are formed by connected but different sub-networks  \cite{Erez 2005} or other localized departures from macroscopic statistical uniformity \cite{Cohen 2003}. 
However in that case, rather then depending on a finite number of global geometrical parameters, the empirical predictions  will depend on the additional microscopic or mesoscopic information as it was feared by Hayek long ago \cite{Hayek 1948}. 

Moreover, in the case in which each company has many contacts it becomes non-realistic to assume that 
it can be failed by the fall of just one partner and one has to take into account the weighted influence of all its connections as described in \cite{Kindler 2013}. The relevant formalism for that case has been shown in \cite{Kindler 2013} to be {\bf `bootstrap percolation'} which is outside the scope of the present exposition. 

Another issue is the reaction of the system to the crisis percolation attack.
Of course in this case the network macroscopic uniformity assumption is broken: the reaction causes the system to behave differently at the locations under attack.  Examples of such `anti-viral' reaction affecting the system properties have been studied in \cite{Hershberg 2001} and \cite{Goldenberg 2005}.
Singling out the regions with more contagion activity and characterizing their geometry has been the very engine of 
predicting very early and very precisely the future of real propagation campaigns \cite{Goldenberg 2005}.

\subsubsection{`Noise' as information}
\label{noise LLS}

As mentioned a few times above, the increase in the fractality of the clusters and in the intermittency of the time evolution are valuable as information that identifies the state of the system and can be used for predicting its future behavior:
 the agent based analysis in this paper connects the amplitude and the extent of the fractal clusters,  intermittent fluctuations and non-self averaging deviations to the proximity of a percolation transition.

 The presence of large fluctuations indicates that the current density of ponzis, $\rho_t$,  is close to the critical value, $\rho_{C}$, and most likely advancing in the Minsky instability region. It was often been observed [Sornette 2002], [Louzoun 2002] that the proximity of a crisis is signalled by large fluctuations in the time evolution of the system. In such an instance the regulator has to intervene energetically to prevent failures and to reduce, at least temporarily, the interest rate: the prospects are that in the absence of such an intervention, even without an external shock $N_0$, a macro crisis may spontaneously emerge which would wipe out the entire population of ponzi companies.  Such a shock would likely transmit to the rest of the economy.

\section{Conclusions:  Seeing `it' as THEM}

A crucial aspect emerging from the present study of the Minsky accelerator on economic networks is the irrelevance of statistical averages. While one can define an average over different realizations of the system, this average will often be irrelevant for any one individual realization; once specific individual rare events can trigger systemic changes, they cause the system to take dramatically different time trajectories. 
The  most significant fact about the predictions is that the most singular effects or most singular features in space and time are the ones which are ultimately responsible for the macroscopic behavior of the entire system. This is in sharp contrast with the usual procedure of considering the most likely or average behavior of the whole system. Taking averages over the various realizations would make as much sense as taking the average between a dead cat and a cat alive in the classical Schroedinger's Gedankenexperiment \cite{Schroedinger 1935}.

Wondering what Minsky would have said about the origins of the current crisis \cite{Wray 2012}, \cite{Wray 2013} one may reach a general thought:
 Since the causality of an emergent macro-phenomenon cannot be pinned on a particular micro-individual or event \cite{LLS 2000}, insisting on finding a single cause to a systemic crisis would be like insisting on searching for the first (or last) individual of a crowd that trampled over the victim of a stampede. Of course one should rather be looking rather for the organizers of the occasion that created the conditions that allowed the stampede to happen. And possibly look for an alternative organization making those conditions unlikely to happen again.
 
Microscopic heterogeneity, autocataliticity and granularity are key factors for the description and prediction of the non-equilibrium reality. We refer to the words of Phillip W. Anderson:
\begin{quotation}
``Much of the real world is controlled as much by the `tails' of distributions as by means or averages: by the exceptional, not the mean; by the catastrophe, not the steady drip; by the very rich, not the middle class. We need to free ourselves from `average' thinking." \cite{Anderson 1997}
\end{quotation}

Thus, in order to catch, in real time, the beginning of a macroscopic process, one has to keep an eye on each of the microeconomic  agents involved.

The first step is to establish the {\bf study of the economic network properties} as one of the intrinsic tasks of the economic regulators and watchdogs. Resources of the size presently invested in collecting and processing data for constructing DSGE models should be invested in collecting and processing data for constructing network (agent based) characterizations of the economy.

Such an effort can be made along a few directions:
\begin{itemize}
\item[-] measuring directly the properties of the economic units and their connections;  
\item[-] probing the system continuously for its failure propagation parameters: 
in spite of the Minsky mechanism, failure clusters do not jump directly from size 1 to the system dimensions; 
\item[-]watching the degree of propagation of micro-crises to alert the regulator to the imminence of larger ones;
\item[-] learning about the network properties not only from the propagation of failures but also from the propagation 
of milder signals (growth, shrink, crisies of particular instruments etc);
\item[-] identifying clusters and nodes connecting those clusters;
\item[-] putting in place `sensors' in the accounting system that can evaluate the changes in the resilience of units and in their interactions;
\item[-] putting in place simulation systems that can evaluate the effects of those changes. 
\end{itemize}

As is often the case, the contribution of a new approach is not only in helping to address old questions from a new angle but also, and foremost, to ask new questions that would not even have been able to be formulated in the old approach.
The present agent-based approach, structured around the geometrical network obstructions to failure contagion, allows a list of new concepts  that can effectively assist the debate around optimal economic policies, both at the macro and at the individual agent level:  the local fluctuations, the propagation delays, the fixed points, the phase transition boundaries are all related to interscale interactions that amplify  the micro-events to macroscopic processes which trigger or block systemic changes.
It is thus an appropriate conclusion of this paper to propose an interdisciplinary research program where the detailed properties of the companies, their trade and their growth are documented at the greatest possible resolution.

\section*{\small Acknowledgements}
{\small We thank the Institute for New Economic Thinking (grant ID INO1100017) and the Israeli Institute for Advanced Studies (Patterns and Processes in Organizational Networks) for support and hospitality. We thank Yuri Biondi and Moshe Levy for very many instructive comments and very helpful suggestions,  Michael Golosovsky for careful critical reviewing the manuscript and Leanne Ussher, Marco Lamieri and Guy Kelman for discussions. Special thanks are due to David Br\'{e}e which helped extensively with editing and clarifying large sections of the paper. }


\appendix
\section*{Appendices}
The appendices provide  derivations and additional examples  to describe the models introduced in the main text. 
The models include both the microscopic and the macroscopic properties of the system and also their interrelatedness. 
The analysis provides predictions of the response of the system to the changes in system parameters and thus to the system's robustness.
There are four appendices providing details of systems described in the four Sections of the paper: 
\ref{sec:Marshall Walras}, \ref{sec:Minsky accelerator}, \ref{sec:Percolation}, \ref{sec:Minsky on network}.

\section{Evolution of the loans quantities during the autocatalytic Marshall-Walras equalization process in loans market with decreasing returns}
\label{sec:appendix decreasing returns}
This appendix gives some additional mathematical derivation of the iterative processes  in which the supplier (here a lender) reacts to an excess demand by modifying their price (the interest rate) while the customer (debtor) reacts to the new price by modifying the quantity they demand (the amount borrowed), leading to the Marshall-Walras equilibrium, as described  in Section \ref{subsec:decreasing}. 
The impact of such changes in the interest rate on the amount borrowed in the economy
and vice-versa, are  treated under a ceterus paribus assumption. 
This loan market is based on the following two assumptions:
\begin{itemize}
\item[-] The total amount of loans, $N_{loans}$, demanded by the debtors at any one time $t$ is a 
	\emph{\textbf {decreasing}} power law function of the interest rate, $i_t$,  that they have to pay at that time:
	\begin{equation}
		\label{eqMWA4}
		N_{loans}(t)=N_t=\left( \frac{i_t}{k} \right)^{-\mu}
	\end{equation}
	where $\{k, \mu\}>0$.
\item[-] The interest rate, $i_{t+1}$, that the banks are charging in the following period is 
	an \emph{\textbf {increasing}} power law function of the amount of loans  set in the previous period, $N_t$:
	\begin{equation}
		\label{eqMWA5}
		i_{t+1}=i_0 N_t^{\alpha}
	\end{equation}
	where $\alpha>0$.
\end{itemize}
These equations determine the interplay between the amount of the loans outstanding,  $N_t$, 
and the interest rate, $i_t$, as follows:
\begin{equation}
\label{eqMWA chain reaction}
N_0  \xrightarrow{Eq. \ref{eqMWA5}} i_1 \xrightarrow{Eq. \ref{eqMWA4}} N_1 \xrightarrow{Eq. \ref{eqMWA5}} i_{2} \cdots N_t  \xrightarrow{Eq. \ref{eqMWA5}} i_{t+1} \xrightarrow{Eq. \ref{eqMWA4}} N_{t+1} \cdots
\end{equation}
Introducing the equillibrium condition $i_{t+1}$=$i_t$ and $N_t=N_{t-1}$ in Eqs. \ref {eqMWA4} and \ref{eqMWA5}:
\begin{equation}
\label{eqAMW18}
i=i_0 \left({\frac{i}{k}}\right)^{-\alpha \mu}
\end{equation}
\begin{equation}
\label{eqAMW19}
N=\left( \frac{i_0}{k} \right)^{-\mu}N^{-\alpha \mu}
\end{equation}
one can express the fixed point $(N_{fix}, i_{fix})$ in terms of the initial interest rate $i_0$ (note that the point does not depend on the initial value of $N_0$).
So by rewriting Eqs. \ref{eqAMW18} and \ref{eqAMW19}, 
it can be expressed in terms of just the initial interest rate $i_0$:
\begin{equation}
\label{eqMWA fixed point}
(N_{fix}, i_{fix})=\left( \left(\frac{i_0}{k} \right)^{\frac{-\mu}{1+\alpha\mu}}, \left( \frac{i_0}{k^{\alpha \mu}} \right)^\frac{1}{1+\alpha \mu} \right)
\end{equation}

\begin{figure}
\centering
\subfigure[Visualisation of the law of decreasing returns, Eq. \ref{eqMWA chain reaction}, in which the red line, Eq. \ref{eqMWA5}, represents the increasing price of  loans as their demand increases, 
while the blue line represents the decreasing demand for loans as their price increases, Eq.  \ref{eqMWA4}. Here $\alpha \mu <1$, so the loan market, $N_{loan}$ and $i$, converges to a fixed  equilibrium point 
irrespective of their initial values (in this case $(N_0,i_0)=(2,0.2\%)$).]{\includegraphics[scale=0.44]{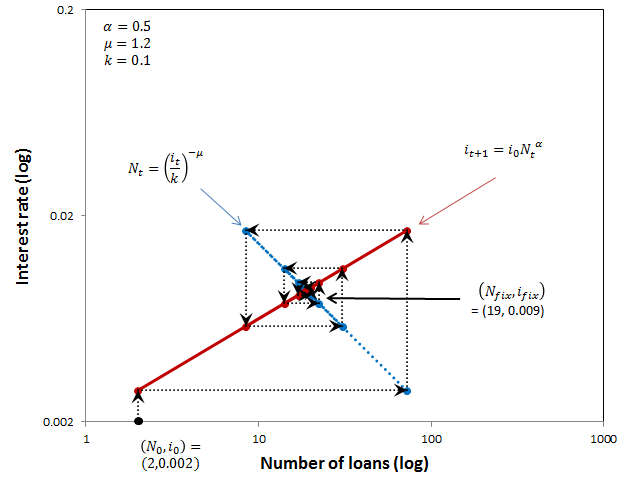} \label{App 15a}}
\hfil
\subfigure[Similar to Figure \ref{App 15a} with the difference that now $\alpha\mu >1$. 
This entails either that the banks react to a decrease of demand for loans by decreasing the interest rate very strongly, 
or that the borrowers react to an increase in the interest rate by decreasing their demand very strongly, or both.
This brings the system into a state of a temporary disequilibrium until it reaches the minimum value of demand. 
Here the initial values are $(N_0, i_0)=(12, 0,2\%)$.]{\includegraphics[scale=0.44]{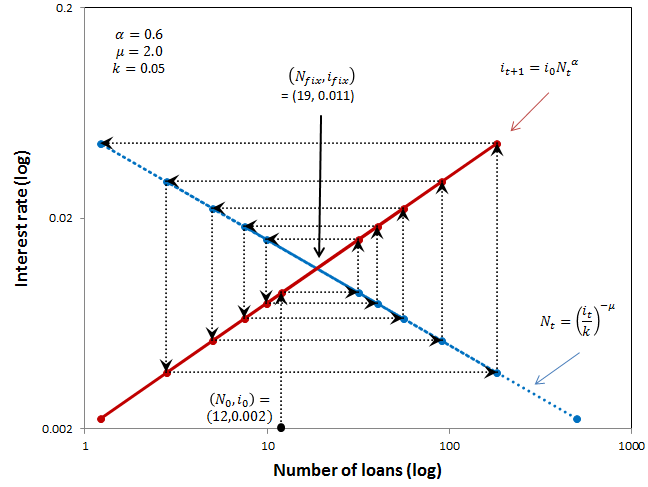} \label{App 15b}}
\caption{\small Illustration of the Marshall-Walras iterative procedure of equilibrating the number of loans and the interest rate, in a market with decreasing returns, Eq. \ref{eqMWA chain reaction}. 
If the decrease of returns (coefficient $\mu$) is faster then the demand for loans (coefficient $1/\alpha$) then the fluctuations in the system become larger, with $N_{loans} \rightarrow 0$, $i \rightarrow 0$.}
\label{fig:App MW Convergence}
\end{figure}

This iterative proces is visually represented in Figure \ref{fig:App MW Convergence}, for both $\alpha\mu<1$ and $\alpha\mu>1$.

By repeated application of Eqs. \ref{eqMWA4} and \ref{eqMWA5} and then using Eq. \ref{eqMWA fixed point}, one can express the number of loans in terms of the number of loans at the fixed point $N_{fix}$:
\begin{multline}
\label{Nloan as f of Nfix A}
N_t=\left( \frac{i_t}{k} \right)^{-\mu}=
\left( \frac{i_0} {k}\right)^{-\mu}N_{t-1}^{-\alpha \mu}=
\left( \frac{i_0} {k}\right)^{-\mu(1-\alpha\mu)}N_{t-2}^{-(\alpha \mu)^2}=\\
\left( \frac{i_0} {k}\right)^{-\mu(1-\alpha\mu-(\alpha\mu)^2)}N_{t-3}^{-(\alpha \mu)^3}=\cdots=
\left( \frac{i_0} {k}\right)^{-\mu(1-\alpha\mu-(\alpha\mu)^2- \cdots-(\alpha\mu)^{(t-1)})}N_{0}^{-(\alpha \mu)^t} =\\
\left( \frac{i_0}{k} \right)^{-\mu \frac{1+(\alpha \mu)^t}{1+\alpha \mu}}N_{0}^{-(\alpha \mu)^t} =
N_{fix}^{1+ (\alpha \mu)^t} N_0^{-(\alpha \mu)^t}=
N_{fix}\left[\frac{N_0}{N_{fix}} \right]^{-(\alpha \mu)^t}
\end{multline}
This fixed point can be attractive, in which case the system converges to it, or repulsive. Iff $|\alpha \mu| < 1$ the fixed point is stable. A proof of that in given in the Appendix \ref{sec:appendix2}. A more elaborated and complete proof of convergence can be found in \cite{Galor 2007}.

\section{Evolution of the loans quantities during the autocatalytic Marshall-Walras equalization process in loans market with increasing returns}
\label{sec:appendix increasing returns}
\begin{figure}
\centering
\hfil
\subfigure[When $\alpha \mu <1$, the iterative process Eq. \ref{eqMWAD chain reaction} converges to the fixed point irrespective of the initial values $N_0$ and $i_0$ (shown here are two cases $(N_0,i_0)=(2,0.18\%)$) and $(N_0,i_0)=(900,0.18\%)$)]{\includegraphics[scale=0.44]{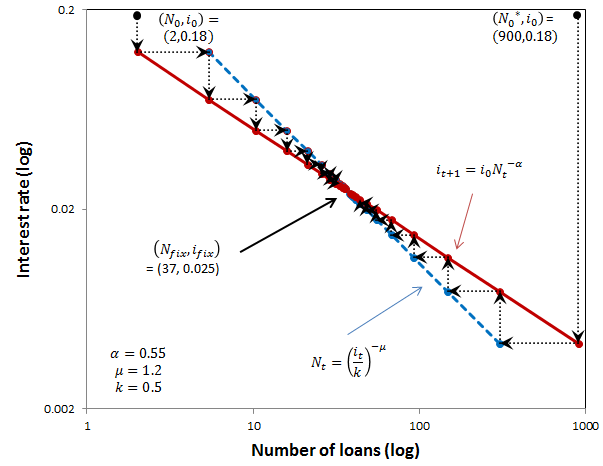}}
\hfil
\subfigure[When $\alpha \mu >1$, the iterative process Eq. \ref{eqMWAD chain reaction} diverges from the fixed point irrespective of the initial values $N_0$ and $i_0$ (shown here are two cases $(N_0,i_0)=(2,0.18\%)$) and $(N_0,i_0)=(900,0.18\%)$)]{\includegraphics[scale=0.44]{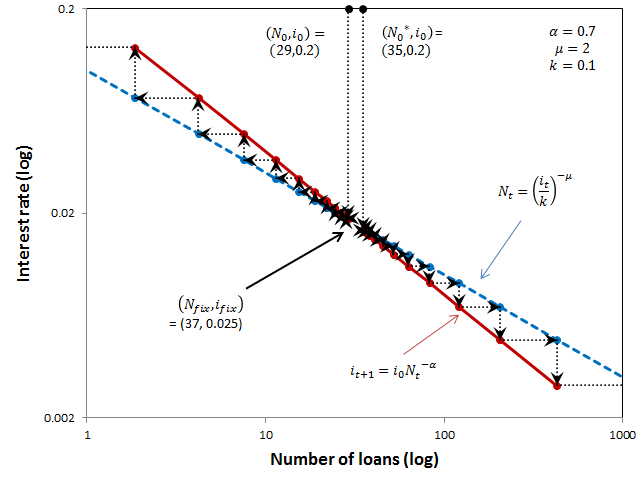}}
\caption{\small Coevolution of the loan quantity and the interest rate during the autocatalytic Marshall-Walras equalization process in a loan market with increasing returns for the lenders.}
\label{fig:App MWI Visualization}
\end{figure}

This appendix gives a mathematical derivation of the   coevolution of loans $N_{loans}$ and the interest rate $i$  introduced in Section \ref{subsec:increasing}. 
This loan market differs from that described in Appendix \ref{sec:appendix decreasing returns} 
in the reaction of lenders to an increase in the credit demand.
Instead of an increase in borrowing leading to an increase in the interest rate, it here leads to a decrease.
This is motivated by increasing returns to scale; the more that is lent, the cheaper is the process associated with arranging and monitoring loans.
So this process is based on the following two assumptions:
\begin{itemize}
\item[-] As before, the amount of loans $N_{loans}$ demanded by the debtors is a \emph{\textbf{decreasing}} power law function of the interest rate that they have to pay:
\begin{equation}
\label{eqMWAD4}
N_{loans}(t)=N_t=\left( \frac{i_t}{k} \right)^{-\mu}
\end{equation}
where $\{k, \mu\}>0$.
\item[-] But now  the assumption concerning the interest rate, $i_{t+1}$,  that the banks are charging is that
	it is a \emph{\textbf{decreasing}} power law function of the amount of loans $N_t$ borrowed
	in the previous period:
\begin{equation}
\label{eqMWAD5}
i_{t+1}=i_0 N_t^{-\alpha}
\end{equation}
where $\alpha>0$.
\end{itemize}
Again these equations determine the interplay between the amount of loans, $N_t$, and the interest rate, $i_t$, as follows:
\begin{equation}
\label{eqMWAD chain reaction}
N_0  \xrightarrow{Eq. \ref{eqMWAD5}} i_1 \xrightarrow{Eq. \ref{eqMWAD4}} N_1 \xrightarrow{Eq. \ref{eqMWAD5}} i_{2} \cdots N_t  \xrightarrow{Eq. \ref{eqMWAD5}} i_{t+1} \xrightarrow{Eq. \ref{eqMWAD4}} N_{t+1} \cdots
\end{equation}
This iterative process is visually represented in Figure \ref{fig:App MWI Visualization}.

Introducing the equilibrium condition $i_{t+1}$=$i_t$ and $N_t=N_{t-1}$ in Eqs. \ref {eqMWAD4} and \ref{eqMWAD5} gives:
\begin{equation}
\label{eqAMWD18}
i=i_0 \left({\frac{i}{k}}\right)^{\alpha \mu}
\end{equation}
\begin{equation}
\label{eqAMWD19}
N=\left( \frac{i_0}{k} \right)^{-\mu}N^{\alpha \mu}
\end{equation}
The resulting fixed point $(N_{fix}, i_{fix})$ depends only the initial interest rate $i_0$, not on the initial value of the amount of loans, $N_0$.
The fixed point is therefore equal to:
\begin{equation}
	\label{eqMWAD fixed point}
	(N_{fix}, i_{fix})=\left( \left(\frac{i_0}{k} \right)^{\frac{-\mu}{1-\alpha\mu}}, \left( \frac{i_0}{k^{\alpha \mu}} \right)^\frac{1}{1-\alpha \mu} \right).
\end{equation}

The fixed point can be either attractive, in which case the system converges to it, or repulsive. 
Iff $|\alpha \mu| < 1$ the fixed point is stable. A proof of this is given in  Appendix \ref{sec:appendix2}. 
A more elaborated and complete proof of convergence can be found in \cite{Galor 2007}

By regrouping Eqs. \ref{eqMWAD4}, \ref{eqMWAD5} and \ref{eqMWAD fixed point} one can express the number of loans at any time in terms of the number of loans at the fixed point $N_{fix}$:
\begin{multline}
\label{Nloan as f of Nfix B}
N_t=\left( \frac{i_t}{k} \right)^{-\mu}=
\left( \frac{i_0} {k}\right)^{-\mu}N_{t-1}^{\alpha \mu}=
\left( \frac{i_0} {k}\right)^{-\mu(1+\alpha\mu)}N_{t-2}^{(\alpha \mu)^2}=\\
\left( \frac{i_0} {k}\right)^{-\mu(1+\alpha\mu+(\alpha\mu)^2)}N_{t-3}^{(\alpha \mu)^3}=\cdots=
\left( \frac{i_0} {k}\right)^{-\mu(1+\alpha\mu+(\alpha\mu)^2+ \cdots+(\alpha\mu)^{(t-1)})}N_{0}^{(\alpha \mu)^t} =\\
\left( \frac{i_0}{k} \right)^{-\mu \frac{1-(\alpha \mu)^t}{1-\alpha \mu}}N_{0}^{(\alpha \mu)^t} =
N_{fix}^{1- (\alpha \mu)^t} N_0^{(\alpha \mu)^t}=
N_{fix}\left[\frac{N_0}{N_{fix}} \right]^{(\alpha \mu)^t}
\end{multline}

\section{Evolution of the autocatalytic Minsky crisis accelerator process based on the non-spatial model}
\label{sec:appendix2}
The objective of the present Appendix is to explain the details of the method of analysis of the autocatalytic processes introduced in the Section \ref{sec:Minsky accelerator}, in terms of finding the fixed point of the iterative processes generated by the interactions and feedbacks between the number of ponzi companies $N_{ponzi}$ and the interest rate $i$. 

This model is non-spatial (it ignores the credit/debt network or other contagion effects, such as holding common assets). It assumes that the interest rate, $i$, determines, through the heterogeneous distribution of the individual resilience, $r(n)$, of the companies, the fraction of units in the system that are ponzi.
The resilience of a company $n$ is defined as the ratio of its earnings to its debt:
\begin{equation}
\label{eqA1}
r(n)=\frac{earnings(n)}{debt(n)}.
\end{equation}
Assuming that the earnings and the nominal debts of each company do not change in the short term, 
e.g. during the financial crisis, its ability to pay its interest charge would depend only on the variation of the nominal interest rate $i$. More precisely, the company $n$ would become ponzi if the interest rate $i$ exceeds its resilience: $i>r(n)$. 
Thus the resilience, $r(n)$, of company $n$ is the lowest level of the interest rate, $i$, which renders the company $n$ ponzi. In other words, the resilience $r(n)$ is the largest value of the interest rate which still allows company $n$ to pay the interest on its debt without incurring further debt.

We consider a set of $N_{tot}$ companies with heterogeneous level of resilience. For definiteness (and based on empirical data \cite{Takayasu 2000}), we take the distribution of company resiliences to be a Pareto or Zipf power law with a heterogeneity exponent $\beta$. That means that the probability that a company will have ponzi status (its resilience $r(n)$ lower than a given interest rate $i$) is:
\begin{equation}
\label{eqA2}
Prob[r(n)<i]=\left(\frac{i}{r_{max}}\right)^{\beta}
\end{equation}
where $r_{max}$ is the maximum value of resilience found in the given set of $N_{tot}$ companies. The heterogeneity exponent $\beta$ is typically found to take a value between 1 and 2.

Consequently, given the total number of companies in the system, $N_{tot}$, the number of ponzi companies, $N(i)$, depends as a power law on the interest rate $i$:
\begin{equation}
\label{eqA3}
N(i)=N_{tot}P[r(n)<i]=N_{tot} \times \left( \frac{i}{r_{max}} \right)^{\beta}=\left( \frac{i}{k} \right)^{\beta}
\end{equation}
where we have introduced a new constant $k=r_{max}/N_{tot}^{1/\beta}>0$ for the sake of simplicity.
So, in our financial accelerator model, an increase in the  interest rate at certain moment $t$, $i_t$, leads to an increase in the number of ponzi companies, $N_t$, according to Eq. \ref{eqA3}: 
\begin{equation}
\label{eqA4}
N_t{=N_{ponzi}(t)}=\left( \frac{i_t}{k} \right)^{\beta}
\end{equation}
An increase in the number of ponzi, all of which are seen to have failed, increases the interest rate; 
the banks increase interest rate not only if the demand for credit is increased but also if there is an increase in the number of their clients that are not fulfilling their debt obligations. The increase on the interest rate reduces resilience endogenously. The mechanism that the banks apply when adjusting interest rate may be modeled in different ways. In our current analysis we will use a specific power law relation between the number of ponzi (or distressed) companies at a given time $t$, $N_t$, and the interest rate, $i_{t+1}$,  that they induce in the next period.
\begin{equation}
\label{eqA5}
i_{t+1}=i_{0}N_{t}^{\alpha} 
\end{equation}

The Eqs. \ref{eqA4} and \ref{eqA5} define an iterative process where, starting from a certain initial interest rate, $i_0$, and an exogenous shock consisting of the failure of  $N_0$ companies, a chain reaction is triggered:
\begin{equation}
\label{eqA chain reaction}
N_0  \xrightarrow{Eq. \ref{eqA5}} i_1 \xrightarrow{Eq. \ref{eqA4}} N_1 \xrightarrow{Eq. \ref{eqA5}} i_{2} \cdots N_t  \xrightarrow{Eq. \ref{eqA5}} i_{t+1} \xrightarrow{Eq. \ref{eqA4}} N_{t+1} \cdots
\end{equation}
The analysis below is exactly exploring the properties of the iterative process Eqs. \ref{eqA4}, \ref{eqA5} and preparing the tools for the similar analysis of the network model where the system has more than one fixed point. 

Eqs. \ref{eqA4} and \ref{eqA5} specify the reciprocal influences (feedbacks) between the number of ponzi companies (that are assumed to have failed) and the interest rate in the next time period. 
To see the iterative process resulting from such equations, assume that one starts  in an initial state with interest rate $i_0$. Assume a shock occurs that makes a number, $N_0$, of companies ponzi, i.e. leads to their failure. 
According Eq. \ref{eqA5} this event will jump the interest rate in the next period to:
\begin{equation}
\label{eqA6}
i_1=i_0N_0^{\alpha}.
\end{equation}
Following this rise in the interest rate, some of the companies which until now were viable (speculative: with the old interest rate,  $i_0$, they were able to pay interest on their own debts $(r(n)>i_0)$ will become ponzi. More precisely, all the companies with resilience  $r(n)<i_1=i_0 N_0^{\alpha}$  will now be ponzi. According to Eq. \ref{eqA4}, the total number of ponzi companies will be: 
\begin{equation}
\label{eqA7}
N_1=\left( \frac{i_1}{k} \right)^{\beta}=\left( \frac{i_0}{k}N_{0}^{\alpha} \right)^{\beta}=\left( \frac{i_0}{k} \right)^{\beta}N_{0}^{\alpha \beta}.
\end{equation}
Let's assume for definiteness (we will study all the cases below) that the parameters are such that $N_1>N_0$. 
The new, increased number of ponzi companies, $N_1$,  induces in turn, according to Eq. \ref{eqA5} a new increased interest rate:
\begin{equation}
\label{eqA8}
i_2=i_0N_1^{\alpha}=\left( \frac{i_0}{k} \right)^{1+\alpha \beta} kN_0^{\alpha^2 \beta}
\end{equation}
which in turn leads to a larger number of ponzi companies:
\begin{equation}
\label{eqA9}
N_2=\left( \frac{i_2}{k} \right)^{\beta}=\left( \frac{i_0}{k}N_{1}^{\alpha} \right)^{\beta}=\left( \frac{i_0}{k} \right)^{\beta}N_{1}^{\alpha \beta}=\left( \frac{i_0}{k} \right)^{(1+\alpha \beta)\beta} N_{0}^{\alpha^2 \beta^2}.
\end{equation}
The new number of ponzi companies has systemic consequences and it raises the interest rate to the value:
\begin{equation}
\label{eqA10}
i_3=i_0N_2^{\alpha}=\left( \frac{i_0}{k} \right)^{1+\alpha \beta+\alpha^2 \beta^2} kN_0^{\alpha^3 \beta^2}.
\end{equation}
which in turn leads to a higher number of ponzi companies:
\begin{equation}
\label{eqA11}
N_3=\left( \frac{i_3}{k} \right)^{\beta}=\left( \frac{i_0}{k} \right)^{(1+\alpha \beta+ \alpha^2 \beta^2)\beta} N_{0}^{\alpha^3 \beta^3}.
\end{equation}
In general, after a number of iterations, $t$, the interest rate will be:
\begin{equation}
\label{eqA12}
i_t=\left( \frac{i_0}{k} \right)^{1+\alpha \beta+\alpha^2 \beta^2+ \cdots +\alpha^{-1}\beta^{-1}} kN_0^{\alpha^t \beta^{t-1}}
\end{equation}
and the number of ponzi companies will be:
\begin{equation}
\label{eqA13}
N_t=\left( \frac{i_0}{k} \right)^{(1+\alpha \beta+ \alpha^2 \beta^2)\beta+ \cdots+ \alpha^{t-1} \beta^{t-1}} N_{0}^{\alpha^t \beta^t}.
\end{equation}
For $\alpha \beta \neq 1$, the geometric series in the exponents of Eqs. \ref{eqA12} and \ref{eqA13} can be summed, which brings the equations into their compact forms:
\begin{equation}
\label{eqA14}
i_t=\left( \frac{i_0}{k} \right)^{\frac{1-\alpha^t \beta^t}{1-\alpha \beta}} kN_0^{\alpha^t \beta^{t-1}}
\end{equation}
and
\begin{equation}
\label{eqA15}
N_t=\left( \frac{i_0}{k} \right)^{\frac{1-\alpha^t \beta^t}{1-\alpha \beta}} N_{0}^{\alpha^t \beta^t}.
\end{equation}

From these equations one can easily discuss the outcome of the process. 
More precisely, one may compute under which conditions the process will continue after the initial Minsky moment, under which conditions it will diverge into a crisis and under which conditions it will converge and stop at a fixed point.  Moreover, if the ponzi status is reversible, the system can also develop towards a recovery. 

To clarify the similarity to the Walrasian way of computing the equilibrium state, note that if such an equilibrium / static state $(i_{t=\infty},N_{t=\infty})$ exists for Eqs. \ref{eqA14}, \ref{eqA15}, this state will be a fixed point solution, $(N_{fix}, i_{fix})$, of the system Eqs. \ref{eqA4}, \ref{eqA5}:
\begin{equation}
\label{eqA16}
i_{t+1}=i_{t}=i_0 N_t^{\alpha}
\end{equation}
\begin{equation}
\label{eqA17}
N_{t+1}=N_t=\left( \frac{i_t}{k} \right)^{\beta}
\end{equation}
which is easily solved by substitution:
\begin{equation}
\label{eqA18}
i=i_0 \left({\frac{i}{k}}\right)^{\alpha \beta}
\end{equation}
\begin{equation}
\label{eqA19}
N=\left( \frac{i_0N^{\alpha}}{k} \right)^{\frac{\beta}{1-\alpha \beta}}
\end{equation}
and gives:
\begin{equation}
\label{eqA20}
i_{fix}=\left(i_0 k^{-\alpha \beta} \right)^{\frac {1} {1-\alpha \beta}}
\end{equation}
\begin{equation}
\label{eqA21}
N_{fix}=\left( \frac{i_0}{k} \right)^{\frac{\beta}{1-\alpha \beta}}
\end{equation}

When $\alpha \beta <1$, this result is indeed the same result that is obtained by taking $t \rightarrow \infty$ limit in the Eqs. \ref{eqA14} and \ref{eqA15}, since then $\alpha \beta ^ \infty=0$, and these equations reduce to Eqs. \ref{eqA20}, \ref{eqA21}, respectively. Note that in this case, for a given $i_0$ the final equilibrium point is $(N_{t=\infty},i_{t=\infty})=(N_{fix},i_{fix})$  and does not depend on the initial value of $N_0$.

Of course the fixed point given by Eqs. \ref{eqA20}, \ref{eqA21}:
\begin{equation}
\label{eqA fixed point}
(N_{fix}, i_{fix})=\left( \left(\frac{i_0}{k} \right)^{\frac{\beta}{1-\alpha\beta}}, \left( \frac{i_0}{k^{\alpha \beta}} \right)^\frac{1}{1-\alpha \beta} \right)
\end{equation}
is the same for $\alpha \beta>1$ but its role in the process \ref{eqA14}, \ref{eqA15} is very different: it is not anymore an attractive, but it is a repulsive point.

Indeed by regrouping Eq. \ref{eqA15} in the form:
\begin{multline}
\label{Nt as f of Nfix}
N_t=\left(\frac{i_0}{k} \right)^{\frac{\beta}{1-\alpha\beta}}
\left( \frac{i_0}{k} \right)^{\frac{-\beta(\alpha^t \beta^t)}{1-\alpha \beta}} N_0^{\alpha^t \beta^t}\\=
\left(\frac{i_0}{k} \right)^{\frac{\beta}{1-\alpha \beta}} 
\left[\left(\frac{i_0}{k}\right)^{\frac{-\beta}{1-\alpha \beta}} 
N_0 \right]^{\alpha^t \beta^t}\\=
\left( \frac{i_0}{k}\right)^{\frac{\beta}{1-\alpha \beta}} 
\left[\frac{N_0}{N_{fix}} \right]^{\alpha^t \beta^t}=
N_{fix}\left[\frac{N_0}{N_{fix}} \right]^{(\alpha \beta)^t}
\end{multline}
one sees that for $\alpha \beta >1$ the exponent $\alpha^t \beta^t$ of $\left[\frac{N_0}{N_{fix}} \right]$ rather then vanishing for $t \rightarrow \infty$, diverges. Thus rather than $(N_{t \rightarrow \infty}, i_{t \rightarrow \infty})$ converging towards $(N_{fix}, i_{fix})$, the opposite happens: if the initial point $N_0$ is even slightly smaller $N_0<N_{fix}$ then $N_{t \rightarrow \infty} \rightarrow 0$ while if $N_0 > N_{fix}$, N increases until the entire system is ponzi. This unrealistic feature is corrected by the network model where even for systemic crises only part of the companies fail.

In conclusion, the dynamics defined by the iterative process Eqs. \ref{eqA4}, \ref{eqA5} has two possible outcomes:
\begin{itemize}
\item[-] when $\alpha \beta <1$ the system converges after a crisis to a fixed interest rate and fixed number of ponzi companies; 
\item[-] when $\alpha \beta >1$ the system either flows in a state with very low interest rate where there are no ponzi companies or in a state where the entire system becomes ponzi and the interest rate diverges. 
\end{itemize}
\begin{paragraph} {Graphical representation of convergence/divergence.}  On a double logarithmic graph Figure \ref{fig:AppMiskyNoNetwork} with the number of ponzi companies, $N$, on the horizontal axis and the interest rate $i$, on the vertical axis, the Eqs. \ref{eqA4} and \ref{eqA5} are straight lines, as shown in Figure \ref{fig:AppMiskyNoNetwork}:
\begin{itemize}
\item[-] the red line gives the interest rate as a function of the number of ponzi companies, determined by  Eq. \ref{eqA5}, as a line with a slope equal to $\alpha$:
\begin{equation}
\label{eqA24}
\ln(i_{t+1})=\alpha \ln(N_t)+\ln(i_0)
\end{equation}
\item[-] the blue dashed line plots how many ponzi companies there are in the system,  as given by using Eq. \ref{eqA4}. 
		 Please note that in the graphical representation Eq. \ref{eqA4} is `inverted', because of the interest rate, $i$, being on the vertical axis, and the number of ponzi companies, $N$, on the horizontal axis:
\begin{equation}
\label{eqA23}
\ln(i_t)=(1/\beta) \ln(N_t)+\ln(k)
\end{equation}
\end{itemize}

When $\alpha \beta <1$, the fixed point, as defined by Eq. \ref{Nt as f of Nfix}, is a stable attracting point. For any initial number of ponzi companies, the system reaches (after sufficient time) the fixed point. This is graphically represented in Figure \ref{fig:alphabeta lt one}. The choice of parameters in this illustration of a basic autocatalytic process is as follows: initial interest rate $i_0=0.4\%$, initial number of ponzi companies $N_0=2$ or $N_0^{*}=900$, $\alpha=0.5$, $\beta=1.3$ and $k=0.0015$. 
The figure shows that the position of the two lines is such that left of the fixed point the blue dashed line is above the red solid line and therefore $i_{t+1}>i_{t}$, so the interest rate and the number of ponzi companies grow with every new iteration towards the fixed point. In the area which is on the right-hand side of the fixed point, the red solid line is above the blue dashed line and therefore $i_{t+1}<i_{t}$ i.e. the values of the interest rate and the number of ponzi companies decreases towards the fixed point.
\begin{figure}
\centering
\subfigure[Number of ponzi companies and interest rate in two possible developments of the system when the fixed point is stable, $\alpha \beta<1$. Here $\beta=1.3$, $\alpha=0.5$, $i_0=0.4\%$ and $k=0.0015$. The initial number of ponzi companies is set either to $N_0=2$ or to $N_0^{*}=900$. In both cases after an initial shock the system converges towards $N_{fix}=38$.] {\includegraphics[scale=0.44]{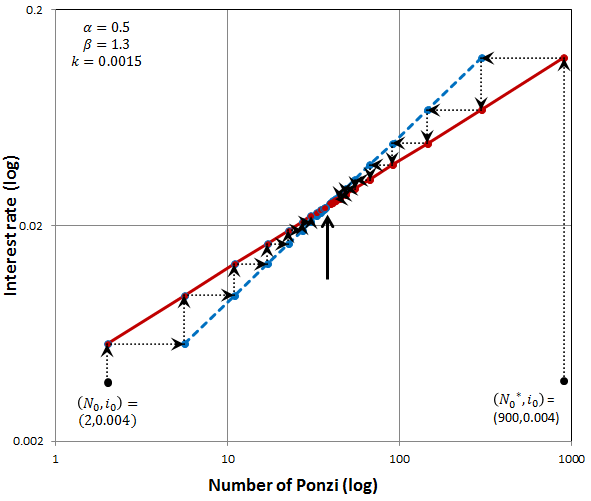} \label{fig:alphabeta lt one}}
\hfil
\subfigure[Number of ponzi companies and interest rate in two possible developments of the system when the fixed point is not stable, $\alpha \beta>1$. Here $\beta=1.8$, $\alpha=0.75$, $i_0=0.003$ and $k=0.005$. The initial number of ponzi companies is set either to $N_0=12$ to $N_0^{*}=16$. In both cases the system diverges; in the first case the number of ponzi companies decreases, while in the second it increases.]{\includegraphics[scale=0.44]{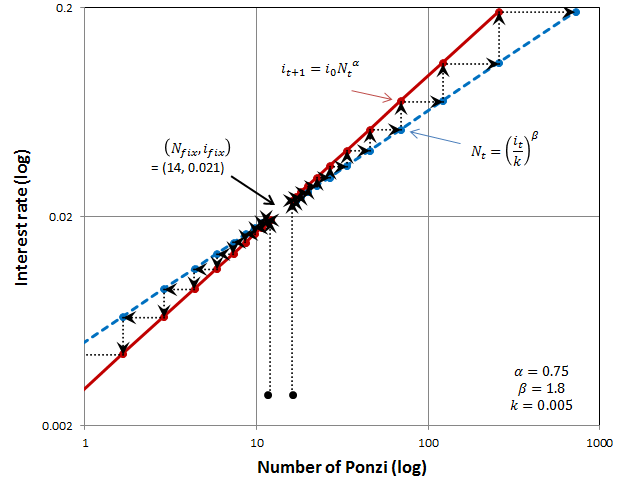} \label{fig:alphabeta gt one}}
\caption{\small Evolution of the autocatalytic Minsky crisis accelerator process based on the non-spatial model. The mechanism determining the number of ponzi companies, Eq. \ref{eqA4}, is given by the red line.
The mechanism determining the interest rate, Eq. \ref{eqA5}, is given by the blue line.
\ref{fig:alphabeta lt one}: $\alpha \beta <1$ so that the  fixed point is attractive;  \ref{fig:alphabeta gt one}: $\alpha \beta >1$ the fixed point is repulsive.}
\label{fig:AppMiskyNoNetwork}
\end{figure}

When $\alpha \beta >1$ the fixed point is unstable and the dynamics brings the system away from $(N_{fix}, i_{fix})$. This is shown in the following example and Figure \ref{fig:alphabeta gt one}. We chose: $\beta=1.8$, $\alpha=0.75$, $k=0.005$, and $i_0=0.3\%$. The fixed point becomes $(N_{fix}, i_{fix})$=$(14, 2.1\%)$. From Eq. \ref{Nt as f of Nfix} one sees that starting from below $N_{fix}$, given by Eq. \ref{eqA21}, the process goes to lower $N$; while starting above $N_{fix}$ the process diverges to larger $N$. Therefore, if one starts at $N_0=12$ then $N_0 < N_{fix}$ and the number of ponzi companies goes down with each iteration; in the graphical representation, this means that the red solid curve is above the blue dashed curve. 
If one starts with $N_0=16$ which is larger then the $N_{fix}=14$ then the number of ponzi increases and this leads to a crisis.
\end{paragraph} 
\begin{paragraph} {Fixed point stability.} Eqs. \ref{eqA24} and \ref{eqA23} represent a system of two linear difference equations:
\begin{equation*}
\ln(i_{t+1})=\alpha \ln(N_t)+\ln(i_0)
\end{equation*}
\begin{equation}
\label{eqA Nt+1}
\ln(N_{t+1})=\beta \ln(i_{t+1}) -\beta \ln(k)=
\alpha \beta \ln(N_t)+ \beta \ln(i_0/k).
\end{equation}
This system can be conveniently rewritten in matrix form as:
\begin{equation}
\label{A difference}
\left[ \begin{array}{c} \ln(i_{t+1}) \\
\ln(N_{t+1}) \end{array} \right] =
\left[ \begin{array}{cc} 0 & \alpha \\
0 & \alpha \beta \end{array} \right] 
\left[ \begin{array}{c} \ln(i_{t}) \\
\ln(N_{t}) \end{array} \right] +
\left[ \begin{array}{c} \ln(i_{0}) \\
\beta \ln(i_0/k) \end{array} \right].
\end{equation}
Let's denote the matrix A as:
\begin{equation}
\label{A matrix}
A= \left[ \begin{array}{cc} 0 & \alpha \\
0 & \alpha \beta \end{array} \right] .
\end{equation}

Stability theorem: The first-order matrix difference equation \ref{A difference} is stable -- that is,  converges asymptotically to the steady state $\left[ \begin{array}{c} \ln(i_{fix}) \\ \ln(N_{fix}) \end{array} \right]$ -- if and only if all eigenvalues of the transition matrix A (whether real or complex) have an absolute value which is less than 1. For its proof see one of the textbooks on linear difference equations, such as \cite{Galor 2007}.

The eigenvalues of the given matrix A are the solutions of the equation:
\begin{equation}
\label{A determinant}
det\left(\lambda I - A \right)=\lambda_1(\lambda_2-\alpha \beta)-0=0\\
\end{equation}
\begin{equation}
\label{A eigen values}
\lambda_1=0 \text{ and }\lambda_2=\alpha \beta.
\end{equation}
Therefore, the condition for asymptotic stability of the system \ref{eqA Nt+1} is:
\begin{equation}
\label{stability}
|\alpha \beta|<1.
\end{equation}
\end{paragraph}

\section{Evolution of the Minsky accelerator process based on the network model}
\label{sec:appendix4}
The non-spatial model from the previous Appendix doesn't take into account the credit/debt relationships between companies, their common assets nor any other network effects. The network effect in the context of debt is crucially important, since the direct link between a failed debtor and its creditor has immediate consequences to the creditor. 

Let us consider that we have a network of companies. The network may be directed or undirected, with either a fixed or a variable number of neighbors and with arbitrary clustering properties. The analysis given in the present Appendix depends on only two parameters of the network: its critical density, $\rho_C$, and the critical exponent, $\gamma$, which we now define.

The companies represent the nodes of a network. In general, given a set of nodes (also called sites), a network is defined by the links (or edges) which connect these nodes. Nodes that are connected by a link are `neighbors'. A link may be either directed or undirected.
If a link is `directed' then one of the two nodes made neighbors by it, is named the origin of the link and the other the target of the link. The concept of percolation is introduced by declaring some nodes as `susceptible', i.e. capable of becoming `contaminated' if only one of their neighbors (origins if the links are directed) is already `contaminated'. 
Of course this contamination rule provides the possibility of contamination avalanches: the contaminated nodes can contaminate new nodes (their susceptible neighbors) which in turn may contaminate additional nodes (their own susceptible neighbors) and so on. The crucial concept in percolation is the concept of cluster; imagine one starts a contamination avalanche by contaminating one node. The avalanche triggered by this will go on until there are no susceptible sites (nodes) neighboring any of the currently contaminated nodes.  Such a set of susceptible nodes that can contaminate one another (directly or indirectly through the contamination of intermediate susceptible nodes) but not any additional susceptible nodes, is said to form a cluster.

The crucial property of many networks is that in the limit of infinite size system, the size of the largest cluster, $N$, diverges once the fraction, $\rho$, of (randomly distributed) susceptible nodes approaches a certain critical value, $\rho_C$. In particular, for many types of networks the divergence follows a universal `scaling' law:
\begin{equation}
\label{eqA25}
N_t=S \left[ 1- \frac{\rho_t}{\rho_C} \right]^{-\gamma}
\end{equation}
with $\rho_C$ and $\gamma$ independent of the details of the particular assignment of the links or of the susceptible nodes in the network \cite{Stauffer 1979}. $S>0$ is a constant related with the exact way the contagion process is initialized (see \cite{Cantono 2010}). 

As it turns out, this property will extend the analysis of Appendix \ref{sec:appendix2} to a more realistic system. In Appendix \ref{sec:appendix2} we assumed that as soon as a company, $n$, is ponzi (i.e. $r(n)<i$) the system recognizes it as such and the global dynamics is affected through an increase in the interest rate according to Eq. \ref{eqA5}: $i_{t+1}=i_0 N_t^{\alpha}$.
In reality, the exact financial status of each company is not known, not even to its trading partners; thus, without additional external events, a ponzi company would be treated as a normal one. It is only when the distress of the company is highlighted by negative events in its immediate environment that the company is recognized and treated as such by the system. The conditions required for a company's distress to become public may vary. For instance in \cite{Kindler 2013} one considered the cumulative influence of its neighbors.  

\begin{figure}[h]
\centering
\hfil
\subfigure[When the two functions, the blue dashed line for Eq. \ref{eqA30} and the red line Eq. \ref{eqA31}, do not intersect, this system of equations has no solution. The stability of the system is fully determined by the non-network model given in the Appendix \ref{sec:appendix2}.  The network then has the effect of delaying the divergence by allowing only a very low rate of contagion in the region where the two functions are getting very close one to one another.]{\includegraphics[scale=0.44]{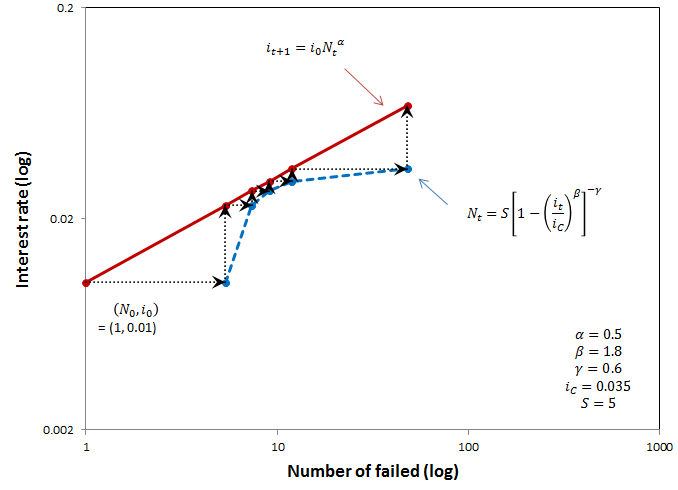} \label{fig:app no solution}}
\hfil
\subfigure[By lowering the initial interest rate, $i_0$, from  $1\%$ in \ref{fig:app no solution} to $0.5\%$ here,
the red solid line of Eq. \ref{eqA31} is lowered to intersect the  blue dashed line of Eq. \ref{eqA30}. The intersections of these lines give  the two solutions of the system. Clearly, looking at the direction of the arrows on the black dotted evolution curve, one of the solutions, $(N_{conv}, i_{conv}),$ is attractive and the system in the close neighborhood converges to it. 
The other solution, $(N_{div}, i_{div})$, is repulsive and the arrows are directed outwards.]{\includegraphics[scale=0.44]{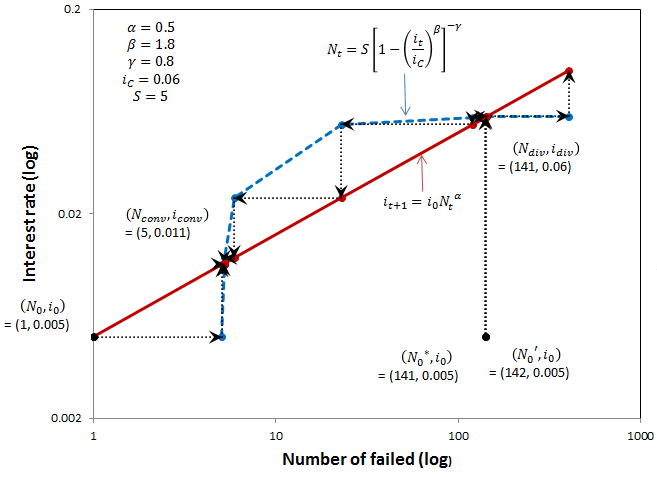} \label{fig:app two solutions}}
\caption{\small The red solid line represents the way the interest rate reacts to the number of companies in distress (Eq. \ref{eqA30}) and the blue dashed line represents the number of failures, Eq. \ref{eqA31}. Depending on the given structure of the network and the way the contagion process is initialized, these two lines either intersect (\ref{fig:app two solutions}) or they do not (\ref{fig:app no solution}). That means that the system of Eqs. \ref{eqA30} and \ref{eqA31} can either have two solutions, or none.}
\label{fig:AppFig5}
\end{figure}

Here we will use a minimal extension of the model in Appendix \ref{sec:appendix2}: we will assume that for a ponzi company to be recognized as such (`uncovered'), it is enough that one of its ponzi neighbors is uncovered.  One is lead to a network model in which the ponzis are identified with the susceptible nodes and which, through their business or financial links, can contaminate one another into open distress. The difference between the usual percolation problem and the present one is provided by the feedback loop of Appendix \ref{sec:appendix2}: as the size of the cluster of contaminated ponzis, $N_t$, expands, cf. Eq. \ref{eqA25}, the interest rate, $i_t$,  increases, cf. Eq. \ref{eqA5}, and the fraction of ponzi companies, i.e. susceptible nodes, $N_{sus}$, increases too, according to Eq. \ref{eqA3}, which leads to an increase of density of susceptible nodes:  
\begin{equation}
\label{eqA26}
\rho_t=\frac{N_{sus}(t)}{N_{tot}}=\frac{(i_t/k)^\beta}{N_{tot}}=\left(\frac{i_t}{r_{max}}\right)^{\beta}.
\end{equation}
By substituting Eq. \ref{eqA26} in Eq. \ref{eqA25} one obtains the dependence of the number of openly contaminated companies, $N_t$, as a function of the interest rate, $i_t$:
\begin{equation}
\label{eqA27}
N_t=S \left[1-\left( \frac{i_t}{i_C} \right)^{\beta} \right]^{-\gamma}
\end{equation}
where $i_C$, according to Eq. \ref{eqA26}, is defined as $i_C=\rho_C^{1/\beta} r_{max}$.
So, by introducing the contamination by a network neighbor as an additional condition for a ponzi company to be recognized and to be acted upon as such, we just substitute for Eq. \ref{eqA4} in the previously defined iterative process (Eqs. \ref{eqA4}, \ref{eqA5}), the Eq. \ref{eqA27}. While slightly more complicated then system \ref{eqA4}, \ref{eqA5}, the new iterative process has the same structure:
\begin{equation}
\label{eqA chain reaction network}
N_0  \xrightarrow{Eq. \ref{eqA5}} i_1 \xrightarrow{Eq. \ref{eqA27}} N_1 \xrightarrow{Eq. \ref{eqA5}} i_{2} \cdots N_t  \xrightarrow{Eq. \ref{eqA5}} i_{t+1} \xrightarrow{Eq. \ref{eqA27}} N_{t+1} \cdots
\end{equation} 
As in the case of Eqs. \ref{eqA4} and \ref{eqA5}, the key to understanding the phase diagram of the process is to find the fixed points, i.e. the solutions of the system:
\begin{equation}
\label{eqA30}
N_{fix}=S \left[1-\left( \frac{i_{fix}}{i_C} \right)^{\beta} \right]^{-\gamma}
\end{equation}
\begin{equation}
\label{eqA31}
i_{fix}=i_0 N_{fix}^{\alpha}.
\end{equation}
By substituting Eq. \ref{eqA31} into Eq. \ref{eqA30}, the problem is reduced to finding the solutions of the equation:
\begin{equation}
\label{eqA32}
N_{fix}=S \left[1-\left( \frac{i_0 N_{fix}^{\alpha}}{i_C} \right)^{\beta} \right]^{-\gamma}
\end{equation}
which, by replacing $i_C=\rho_C^{1/\beta} r_{max}$ (Eq. \ref{eqA26}), can be rewritten as:
\begin{equation}
\label{eqA29a}
N_{fix}=S\left[1-\frac{\rho_0}{\rho_C} N_{fix}^{\alpha \beta} \right]^{-\gamma}.
\end{equation}
The network based process Eq. \ref{eqA chain reaction network} is visualized in the examples given in Figure \ref{fig:AppFig5}, for various values of the parameters, similarly to the way Figures \ref{fig:alphabeta lt one} and \ref{fig:alphabeta gt one} described various cases of the non-network process. 
One sees in these figures (and will calculate below) that in the network case the function determining the number of uncovered ponzi companies, which we will call `failed', Eq. \ref{eqA30}, is now a curve which intersects the straight line representing the interest rate mechanism, Eq. \ref{eqA31}, either in two points or not at all, corresponding to the number of solutions of Eq. \ref{eqA32}, as shown graphically  in Figures \ref{fig:app two solutions} and \ref{fig:app no solution}, respectively. In Figure \ref{fig:app two solutions}, the two fixed points, one convergent and the other divergent,  are labelled $(N_{conv}, i_{conv})$  and $(N_{div}, i_{div})$, respectively.

In order to evaluate quantitatively the crossing points defined above, we re-write Eq. \ref{eqA32} as:
\begin{equation}
\label{eqA33}
\left( \frac{N_{fix}}{S} \right)^{1/\gamma} \left[ 1- {\left( \frac{i_0}{i_C} \right)}^{\beta} N_{fix}^{\alpha \beta} \right]=1
\end{equation}
and further as:
\begin{equation}
\label{eqA34}
\left( \frac{N_{fix}}{S} \right)^{1/\gamma} \left[ 1- {\left( \frac{i_0}{i_C} \right)}^{\beta} N_{fix}^{\alpha \beta} \right]=1
\end{equation}
i.e.
\begin{equation}
\label{eqA34a}
F\left(N_{fix}\right)= \left( \frac{i_0}{i_C} \right)^{\beta}  N_{fix}^{\alpha \beta + 1/\gamma} - N_{fix}^{1/\gamma} +S^{1/\gamma}=0.
\end{equation}
Assuming $\alpha\beta\gamma=1$, Eq. \ref{eqA34a} can be reduced to a quadratic equation:
\begin{eqnarray*}
	\left(\frac{i_0}{i_c}\right)^{\beta} S^{1/\gamma} \left(\frac{N_{fix}}{S}\right)^{2/\gamma} - \left(\frac{N_{fix}}{S}\right)^{1/\gamma} + 1 	&= 	&0.	
\end{eqnarray*}
Solving this quadratic function for $(N_{fix}/S)^{1/\gamma}$, we get the number of failed companies in the convergent and divergent fixed points:
\begin{equation}
	N_{fix}=N_{conv/div} = S \left[\frac{1 \pm \sqrt{1-4\left(\frac{i_0}{i_C}\right)^{\beta} S^{1/\gamma}}}
				{2 \left(\frac{i_0}{i_C}\right)^{\beta} S^{1/\gamma}}
			\right]^{\gamma}
\label{eqAquad}
\end{equation}
which gives the two curves  in the phase space diagrams shown in  Figure \ref{fig:AppPhaseDiagrams} for $N_{conv}$ (lilac) and $N_{div}$ (red).
Obviously, when $4\left(\frac{i_{0C}}{i_C}\right)^{\beta} S^{1/\gamma}=1$, Eq. \ref{eqAquad} has only one solution $N_{0C}=N_{conv}=N_{div}=2^\gamma S$ . The corresponding critical interest rate is, from Eq. \ref{eqA31}, $i_{0C}= i_0 2^{\alpha\gamma} S^{\gamma}$. 

\begin{figure}
\centering
\subfigure[A realization of the evolution Eq. \ref{eqNA final} for the system in which $\alpha \beta < 1$. The fixed point defined by the non network model, $(N_{core}, i_{core})$, is therefore convergent. Then in the small area under and to the left of this fixed point the number of ponzi companies grows. In the area which is shaded a light gray color, the network (as seen from the blue curve) protects a large number of ponzi companies (susceptible modes) from failing. The number of failed companies in this area is limited by the network.]{\includegraphics[scale=0.41]{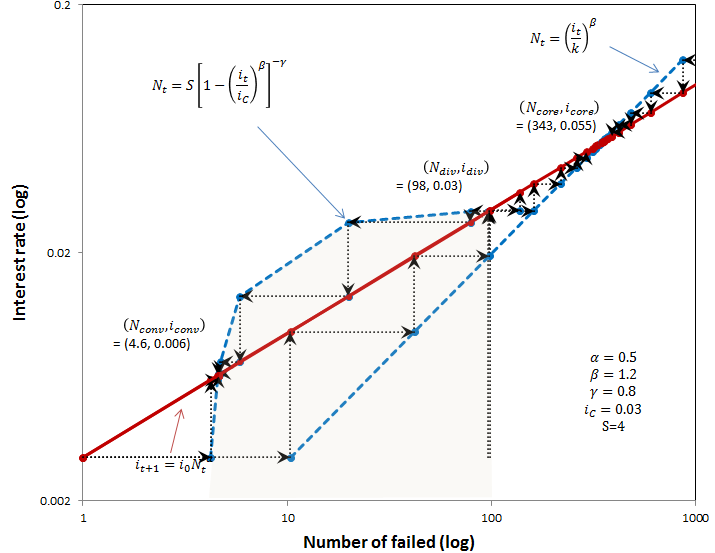}\label{fig:AppMinNetConv}}
\hfil
\subfigure[A visualisation of the two sub-models, the number of failed (blue curve) and the number of ponzi companies (blue line), of the co-evolution given by Eq.\ref{eqNA final}, for the system in which $\alpha \beta > 1$.  In this example, a process that starts with the number of ponzi companies just above the divergent fixed point defined by the non-network model, $(N_{dis}, i_{dis})$, and below the divergent fixed point defined by the network model, ($N_{div}, i_{div}$), is prevented from the contagion/failing of the susceptible nodes (ponzi companies) and the increase of interest rate by the network. ]
{\includegraphics[scale=0.43]{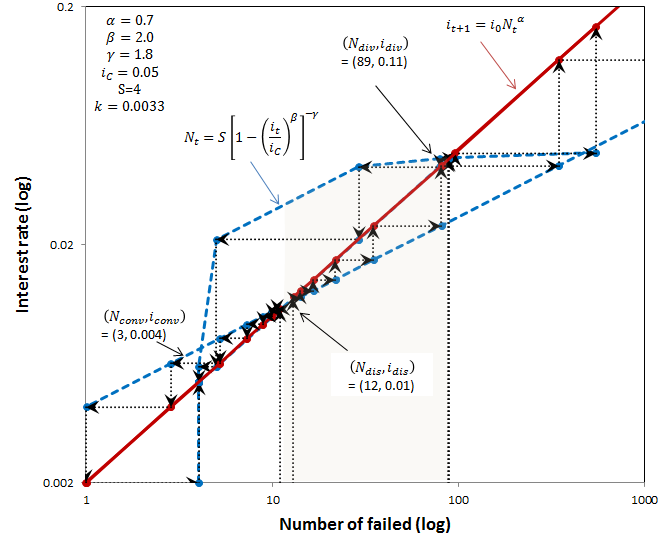} 
\label{fig:AppMinNetDiv}}
\caption{\small The Minsky accelerator percolation model. According to Eq. \ref{eqNA final}, the simultaneous evolution of the number of failed ponzi companies, $N_t$, according the two previously elaborated models (Eqs. \ref{eqA4} and \ref{eqA27}) is necessary to obtain the Minsky accelerator percolation model. These are represented with the blue dashed line and curve, respectively. The red line is used for Eq. \ref{eqA5}. The actual $N_t$ (on the x-axis) for a given $i_t$ (on the y-axis) is the one which is closer to the vertical axis. Therefore it is clear that in the gray shaded areas, the network solution is dominant: instead of following the growing number of ponzi companies (as the non-network model specifies), the number of failed ponzi firms declines. Note that the relevant crossing points on the Figures \ref{fig:AppMinNetConv} and \ref{fig:AppMinNetDiv} are at the same time representing a horizontal cross-section of the phase diagrams shown in Figures \ref{fig:AppPhaseDiag} and \ref{fig:AppPhaseDiagDiv}, respectively. }
\label{fig:network with no network}
\end{figure}

\begin{figure}
\centering
\includegraphics[scale=0.40]{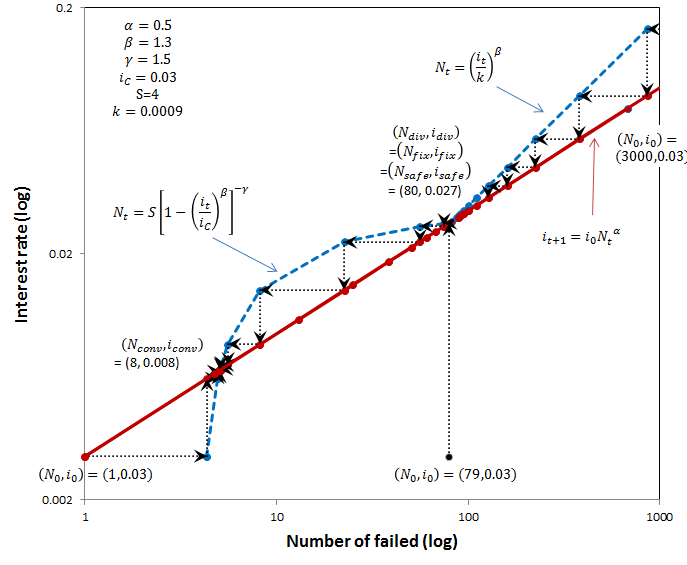}
\caption{\small
Figure for finding one value of $i_{safe}$.
The condition for $ i_{safe}$ is that $N_{core}=N_{div}$, i.e.  that the three lines
representing the three mechanisms --  
the interest rate setting mechanism 
$i_t =i_0 N_t^{\alpha}$ (Eq. \ref{eqA5} red line),
the simple mechanism for setting the number of ponzi companies $N_t =  (i_t /k )^{\beta} $ (Eq. \ref{eqA4})  and 
the network mechanism for determining the number of companies discovered as being ponzi (and so in distress) $N_t=  S [1- ( i_t / {i_{C}} )^{\beta} ]^{-\gamma}$ (Eq. \ref{eqA27}) --
meet at a single point $(N_{safe}, i_{safe})$.
}
\label{fig:app fixed points meet}
\end{figure}

To find the fixed point  $(N_{0C}, i_{0C})$ for the general case, when $\alpha \beta \gamma \neq 1$, we set the derivative of $F(N_{fix})$, as given by Eq. \ref{eqA34a}, with respect to $N_{fix}$ to zero,
since this will give just one solution for $N_{fix}$, being
where $N_{conv}$ and $N_{div}$ coincide 
\begin{eqnarray}
	\label{eqF'}
	\frac{\delta F}{\delta N_{fix}} 
		= \left(\alpha\beta + 1/\gamma\right)
		\left(\frac{i_0}{i_c}\right)^\beta N_{fix}^{\alpha\beta +1/\gamma -1}
		- N_{fix}^{1/\gamma -1}/\gamma &= 	&0	\\
	\label{eqSingular fix}
	\left(\frac{i_0}{i_c}\right)^\beta N_{fix}^{\alpha\beta +1/\gamma}
		&= &\left(\frac{1}{\alpha\beta\gamma + 1}\right)  N_{fix}^{1/\gamma}.
\end{eqnarray}
ubstituting the value of $N_{fix}$ from Eq. \ref{eqSingular fix} into Eq. \ref{eqA34a} gives an equation for $N_{fix}$ of this singular fixed point:
\begin{equation}
	\label{eqA34c}
	\left(\frac{1}{\alpha\beta\gamma + 1}\right)  N_{fix}^{1/\gamma} 
		 - N_{fix}^{1/\gamma} + S^{1/\gamma} 	= 	0.
\end{equation}
Rearranging Eq. \ref{eqA34c} and substituting in Eq. \ref{eqA31} gives the  fixed point $(N_{0C}, i_{0C})$:
\begin{equation}
\label{i0C}
i_{0C}=i_{0}S^{\alpha}\left(1+\alpha \beta \gamma \right)^{1/\beta}\left(1+\frac{1}{\alpha \beta \gamma}\right)^{\alpha \gamma}
\end{equation}
\begin{equation}
\label{N0C}
N_{0C}=S\left(1+1/{\alpha \beta \gamma}\right)^{\gamma}.
\end{equation}

In Figure \ref{fig:network with no network} two examples are given in which one can easily see how the network can have different impacts on the convergence/divergence of the system. The entire system is perceived as a co-evolution of three models: the two microscopic models: the bottom-up model (the blue line) and the peer-to-peer network model (the blue curve) and the macroscopic top-bottom model (the red line). 
According to Eq. \ref{eqAminN}, the actual number of failed ponzi companies in the network case is equal to the lesser of the number of ponzi companies in the system without a network and the number of failed ponzis predicted by the network model. Once the number of failed ponzi is fixed, then the interest rate is determined as the minimal interest rate that the two models (the non-network model Eq. \ref{eqA4} and the infinite network model Eq. \ref{eqA27}) predict.

\begin{equation}
N_{fail}= \min   \{ S [1- {N_{ponzi}} / {N_{C}}]^{-\gamma}     , \\   N_{ponzi}  \}
\label{eqAminN}
\end{equation}
Therefore, eliminating $N_{Ponzi}$ from \ref{eqAminN},  the time evolution of the system is reduced to an iterative process between $N_{fail}$ 
\begin{equation}
\label{eqNA final}
N_{t}= \min \{ S\left[1- (i_{t}/i_{C})^{\beta} \right]^{-\gamma} , ({i_t} / {k})^{\beta} \}
\end{equation}
and $i$, 
Eq. \ref{eqA5} $(i_{t+1}=i_{0}N_t^{\alpha})$.

In Figure  \ref{fig:network with no network}, the system without taking the network into account is represented by the two straight lines: the red line that represents the top-bottom macroscopic effect and the blue straight line that represents the microscopic bottom-up effects. Their intersection is labelled $(N_{core}, i_{core})$ when $\alpha \beta <1$ and $(N_{dis}, i_{dis})$ when $\alpha \beta >1$. The networked system is represented with the red line (macro) and blue curve (micro). Their intersection defines two fixed points: $(N_{conv}, i_{conv})$ and $(N_{div}, i_{div})$, regardless of the values of these parameters. The macroscopic part of the model revises the interest rate according to the number of failed ponzi companies. 

In the visual analysis of Eq. \ref{eqNA final}, Figure \ref{fig:network with no network}, the graphical analysis is not entirely completed: the bottom-up and the peer-to-peer models are plotted as if they were independent of each other and it is now up to the reader to determine the minimum and thus the actual solution for the number of failed ponzi companies, and the final evolution of the process. The end result would contain only one red line, one blue curved line (a combination of the existing blue line and the blue curve) and one evolution curve (an example of a completed procedure is given in Figure \ref{fig:app fixed points meet}).

Figure \ref{fig:AppPhaseDiagrams} gives the final result of the Minsky accelerator model: the prediction of the states of the system, given the initial number of failed ponzi companies and the initial interest rate. As the previous analysis has proved, if the initial conditions are equal to one of the fixed points, the initial state will not change; otherwise, the number of failed ponzi companies and the interest rate will either grow (possibly leading to a crisis) or decay (stable state), depending on the position of the initial state relative to the neighbouring fixed points. The lines representing these positions in Figure \ref{fig:AppPhaseDiagrams} are the lilac curve for $(N_{conv}, i_{conv})$, the red curve for $(N_{div}, i_{div})$, and the blue line  for the convergent fixed point $(N_{core}, i_{core})$ in Figure \ref{fig:AppPhaseDiag} or the divergent fixed point Figure $(N_{dis}, i_{dis})$ in \ref{fig:AppPhaseDiagDiv}; they have been used to construct the phase diagrams.
As can be seen in Eq. \ref{eqAquad} with a negative of the square root, as $i_0$ approaches zero the lilac curve approaches $S$. 
The lines are calculated under the assumption: $\alpha\beta\gamma=1$. A detailed and descriptive interpretation of the phase diagram when $\alpha \beta <1$ (Figure \ref{fig:AppPhaseDiag}) is given in the main text and will be partially and quite technically repeated in the paragraphs that follow. The interpretation of the diagram when $\alpha \beta >1$ (Figure \ref{fig:AppPhaseDiagDiv}) is left to the reader.

\begin{paragraph} {Minsky accelerator model in terms of ponzi density}
The same dynamics can be expressed in terms of the feedback between $N_t$ and $\rho_t$, using the relations derived from reordering Eq. \ref{eqA4} and noting that $N_t=\rho_t/N_{total}$:
\begin{subequations}
\label{i of rho}
\begin{eqnarray}
i_t &=& k N_t^{1/\beta}=k \rho_t^{1/\beta}N_{total}^{1/\beta} \label{eq23a}\\
i_0 &=& k  N_0^{1/\beta}=k \rho_0^{1/\beta}N_{total}^{1/\beta} \label{eq23b}\\
i_{ponzi} &=& k  N_{ponzi}^{1/\beta}=k \rho_{ponzi}^{1/\beta}N_{total}^{1/\beta} \label{eq23c}\\
i_C &=& k  N_C^{1/\beta}=k \rho_C^{1/\beta}N_{total}^{1/\beta}. \label{eq23d}
\end{eqnarray}
\end{subequations}
The system of equations Eqs. \ref{eqNA final} and \ref{eqA5} then becomes:
\begin{equation}
\label{res Nfail of rho}
N_t= \min \{   S\left[1-\frac{\rho_t}{\rho_C}\right]^{-\gamma}, {\rho}_t N_{total} \}
\end{equation}
\begin{equation}
\label{res rho of Nfail}
\rho_{t+1}=\rho_0 N_t^{\alpha \beta}.
\end{equation}
Substituting Eq. \ref{res rho of Nfail} into Eq. \ref{res Nfail of rho}  gives the number of failed companies at the equilibrium fixed points 
when there is a network effect: 
\begin{equation}
\label{res eq29}
N_{fix}=S\left[1-\frac{\rho_0}{\rho_C} N_{fix}^{\alpha \beta} \right]^{-\gamma}.
\end{equation}
\end{paragraph}

\begin{paragraph}
{Deduction of the formulae for $N_{core}$:}
When the network effects are not  felt, the system is given by the two Eqs.  \ref{eqA4} and \ref{eqA5}. 
The two straight lines representing these equations intersect in a fixed point which we have labelled $(N_{core},i_{core})$ when $\alpha\beta<1$ and $(N_{dis},i_{dis})$ when $\alpha\beta<1$. 
The position of these points can be determined by substituting Eq.  \ref{eqA5} into Eq. \ref{eqA4}, as has already be done in Appendix \ref{sec:appendix2}, see Eq. \ref{eqA21}, to give:
\begin{eqnarray}
N_{core} &=& (i_{core}/k)^{\beta}\\
i_{core} &=& i_0 N^{\alpha}.
\label{App derivation Ncore 2}
\end{eqnarray}
Substituting for $i_{core}$ gives:
\begin{eqnarray}
N_{core} &=& (N^{\alpha} i_0/k )^{\beta}\\
N_{core}^{ 1-\alpha \beta } & = & (i_0/k)^{\beta}\\
N_{core} &=& (i_0/k)^{\beta / (1-\alpha \beta) }  .
\label{App derivation Ncore}
\end{eqnarray}
This can be translated in terms of $\rho$, taking into account that $\rho_{total}=1$:
\begin{equation}
(i_0/k)^{\beta}=( i_0 / i_{max})^{\beta} (i_{max} / k)^{\beta}= \rho_0 N_{total} .
\label{App Ncore rho}
\end{equation}
So finally:
\begin{equation}
N_{core}  = \left( \rho_0  N_{total} \right)^{1 / (1-\alpha \beta) }.
\label{App Ncore final}
\end{equation}
Similarly for the point $(N_{dis},i_{dis})$.
\end{paragraph}

\begin{paragraph}
{Deduction of the formulae for $N_{safe}$ and $i_{safe}$:}

In order to find the conditions where the three curves -- the red line, the blue line and the blue curve -- meet, as shown in Figure \ref{fig:app fixed points meet}, 
we assume that $N$ is large enough so that $N=  S [1- ( i / {i_{C} )^{\beta}}]^{-\gamma}$  implies, to a good approximation, $i = i_C$. 
Thus the condition for the three lines to meet becomes, by substitution in Eqs. \ref{eqA4} and \ref{eqA5}:
\begin{equation}
N_{safe} =  (i_C /k )^{\beta}
\label{Nsafe condition}
\end{equation}
(which already determines $N_{safe}$) and:
\begin{equation}
i_C= i_{safe} N_{safe}^{\alpha}
\end{equation}
or
\begin{equation}
N_{safe} =( i_C / i_{safe} )^{1/\alpha}.
\end{equation}
Eliminating $N_{safe}$:
\begin{eqnarray}
(i_C /k )^{\beta} & = & ( i_C / i_{safe} )^{1/\alpha}\\
(i_C /k )^{\alpha \beta } & = & ( i_C / i_{safe})\\
\label{isafe}
 i_{safe} & = & i_C (k / i_C)^{\alpha \beta}.
\end{eqnarray}
Note that this point has been shown in Figure \ref{fig:AppPhaseDiagrams} as the intersection of the blue line with the red curve. In fact the intersection of the blue line with the lilac curve is also a possible position for $(N_{safe},i_{safe})$, all be it with a value of $i_{safe}$ that is approximately the same, but with a smaller value for $N_{safe}$. That only one value is given here is due to the approximation we made that $i=i_c$.

Using the same tricks as above one can translate Eqs. \ref{Nsafe condition} and \ref{isafe} in terms of $\rho$:
\begin{equation}
N_{safe} =  (i_C /k )^{\beta} =( i_C /i_{max})^{\beta}(i_{max} /k)^{\beta}  =\rho_C  N_{total}
\end{equation}
\begin{equation}
i_{safe}  / i_C =  (k / i_C ) ^{\alpha \beta }
\end{equation}
which can be rewritten as:
\begin{equation}
(\rho_{safe} / \rho_C )^{1/\beta} =  (\rho_C  N_{total} ) ^{-\alpha}
\label{App Nsafe rho}
\end{equation}
or
\begin{equation}
\rho_{safe}   = \rho_C (\rho_C  N_{total} ) ^{-\alpha\beta}.
\label{App Nsafe final}
\end{equation}
\end{paragraph}

\begin{paragraph}{Analysis of $i_0$-ranges on the phase diagram in Figure \ref{fig:AppPhaseDiag}:}
As the fixed points do not depend on $N_0$  but, aside from parameter values, only on $i_0$, the boundaries between the different phases are parallel to the x-axis in Figure \ref{fig:AppPhaseDiag}.
There are three ranges of $i_0$ and two borderline values between these ranges which correspond to the different levels of the risk that the system may enter the `Minsky instability phase': 
\begin{itemize}
\item[$i_0 < i_{safe}$] \

In this region the system is safe since then either:
\begin{itemize}
\item The point $(N_{core},i_{core})$ is to the left of the intersection of the blue line, representing the mechanism for setting the number of companies that are ponzi $N_t =(i_t / k)^{\beta}$ (Eq. \ref{eqA4}), and the blue curve, representing the network mechanism for determining how many of them are discovered as being ponzi (and so in distress) $N_t=  S [1- (i_t / i_{C})^{\beta}  ]^{-\gamma}$ (Eq. \ref{eqA27}) in a figure similar to Figure \ref{fig:app fixed points meet} and also the red line, representing the interest rate setting mechanism $i_t=i_{0}N_t^{\alpha}$ (Eq. \ref{eqA31}), is below the meeting point of these blues lines representing the two mechanisms for determining the number of failed ponzi companies. 
\item Or the point $(N_{core},i_{core})$ is to the right of the intersection of the blue line and the red line is above the blue curve.
\end{itemize}
Consequently, $N_{div}$ and $N_{core}$ do not exist and so 
the process Eq. \ref{process Nfail i} converges to $N_{conv}$ (see Figure \ref{fig:app fixed points meet}) and the `Minsky instability phase' does not exist.
\begin{figure}
\centering
\subfigure[Phase diagram-like representation of the possible states of the system with $\alpha\beta <1$. The red line represents the fixed point $(N_{core}, i_{core})$. If the starting point defined by the initial number of ponzi companies and the initial interest rate falls in the area which is below all the lines, the system will remain stable: the number of ponzi companies will decay with each iteration. If the initial number of ponzi companies is greater then $S$ and the $(N_0, i_0)$ point is above all the lines, then the number of ponzi companies with each iteration increases.]{\includegraphics[scale=0.44]{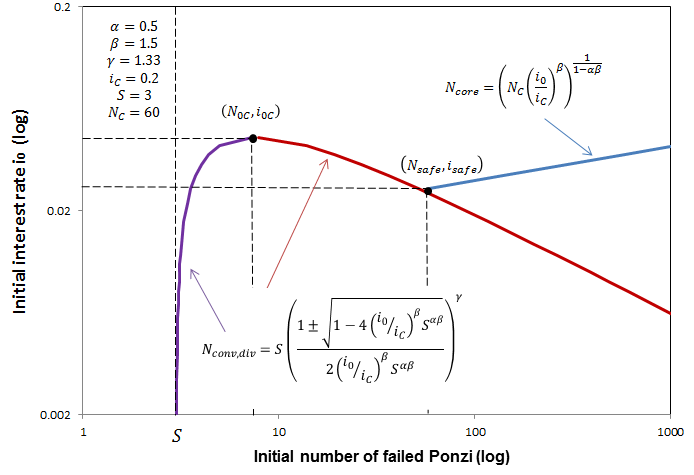} \label{fig:AppPhaseDiag}}
\hfil
\subfigure[The red line represents the fixed points $(N_{dis},i_{dis})$ for $\alpha\beta>1$. $N_{dis}$ uses the same formula as $N_{core}$, but its role is somewhat different: for all starting points $(N_0, i_0)$ which fall below the blue line, the number of failed ponzi companies eventually falls to zero and for all $(N_0, i_0)$ above the blue line, the iteration process develops into a crisis.  If however the starting point falls in the area below the red or the lilac lines, the system is to a great extent protected by the network and the space that would otherwise be above the blue line and would lead to a crisis, remains stable.]{\includegraphics[scale=0.44]{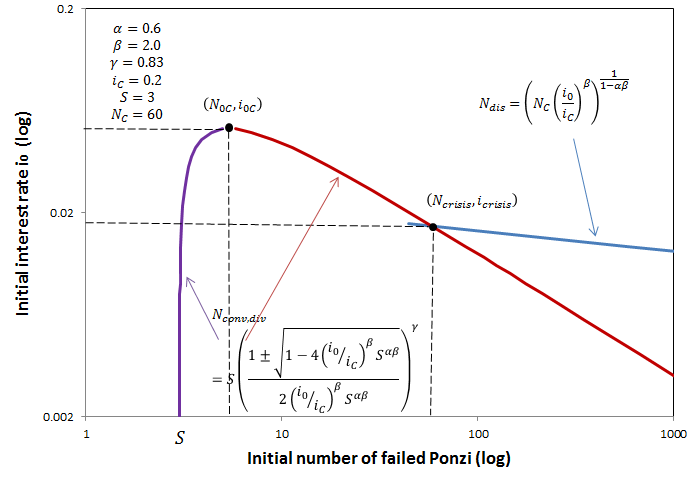} \label{fig:AppPhaseDiagDiv}}
\caption{The plotted lines represent the borderlines of the phase space between the possible phases: crisis acceleration or deceleration following an initial shock $N_0, i_0$. The red and lilac lines represent the fixed points $(N_{conv}, i_{conv})$ and $(N_{div}, i_{div})$, respectively, for the given parameters. They are calculated under the assumption that $\alpha\beta\gamma=1$.}
\label{fig:AppPhaseDiagrams}
\end{figure}

\item[$i_0 =  i_{safe}$] (for definition see Eq.\ref{isafe}) the fixed points $N_{core}$ and $N_{div}$ coincide (previous paragraph):
$$N_{core} =N_{div}=  N_{safe} =(i_C /k )^{\beta}$$ and the `Minsky instability phase' shrinks to one point.

\item[$i_{safe} < i_0 < i_{0C}$] \

In this range all three fixed points  $ N_{conv}<N_{div} < N_{core} $ exist and so do all the phases.

\item[$i_0=i_{0C}$] (for the definition see Eq. \ref{i0C}) the two solutions $N_{conv}$ and $N_{div}$ coincide (Eq. \ref{N0C}): 
\begin{equation}
N_{conv} = N_{div}= N_{0C}=S \left(1+\frac{1}{\alpha \beta \gamma}\right)^{ \gamma}
\end{equation}
and the stable phase shrinks to one point.

\item[$i_0 > i_{0C}$] \

In this range, the risk of entering an accelerated crisis is very high.
The red line, $i=i_{0}N^{\alpha}$, is raised above the blue curve,  
$S\left[1- (i / i_{C})^{\beta}  \right]^{-\gamma}$, so their intersections, $(N_{conv},i_{conv})$ and $(N_{div}, i_{div})$, do not exist, nor does the `stable phase'. 
In the absence of these fixed points the failure contagion process \ref{process Nfail i} continues unhindered beyond the percolation threshold.
This is the meaning of the part of the {\bf Minsky instability} phase  which, as seen in the Figure \ref{fig:AppPhaseDiag}, would extend to the only fixed point that remains:
$$N_{core}= (i_{0} / k)^{\beta / (1- \alpha \beta )}.$$
The $N_{core}$ convergent fixed point insures the survival of only the most resilient companies and defines the very solid core phase in Figure \ref{fig:AppPhaseDiag}. \\
\end{itemize}
\end{paragraph}

\end{document}